\documentclass[a4paper,11pt,twoside,pdftex]{article}
\pdfoutput=1
\usepackage{jheppub} 
\usepackage[T1]{fontenc}
\usepackage[utf8]{inputenc}
\usepackage{defs} 
\usepackage{booktabs,dcolumn}
\newcolumntype{C}{>{$}c<{$}} 
\makeatletter
\def\@xcmidrule{\ifx\@tempa\cmidrule\vskip-\@thisrulewidth
     \global\@lastruleclass=\@ne\else
     \ifx\@tempa\morecmidrules\vskip \cmidrulesep
     \global\@lastruleclass=\@ne\else
     \vskip \belowrulesep\global\@lastruleclass=\z@\fi\fi
     \ifnum0=`{\fi}}
\makeatother
\title{\boldmath%
Non-perturbative tests of continuum HQET through small-volume two-flavour QCD
}
\collaborationImg{\flushleft\includegraphics[height=0.9cm]{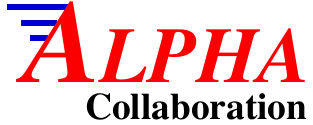}}

\author[a,b,1]{Patrick Fritzsch,%
               \note{Corresponding author. $^{a}$Present address.}}
\author[c]{Nicolas Garron,}
\author[d]{Jochen Heitger}
\affiliation[a]{%
Instituto de Física Te\'orica UAM/CSIC,
Universidad Autónoma de Madrid,\\\hspace{0.5em}  
C/ Nicolás Cabrera 13-15,
Cantoblanco, Madrid 28049, Spain
}
\affiliation[b]{%
University of Southampton, 
School of Physics \& Astronomy,\\\hspace{0.5em}
Highfield, Southampton~SO17~1BJ, United Kingdom
}
\affiliation[c]{%
Plymouth University,
School of Computing \& Mathematics,\\\hspace{0.5em}
Plymouth~PL4~8AA, United Kingdom
}
\affiliation[d]{%
Westf\"alische Wilhelms-Universit\"at M\"unster,
Institut f\"ur Theoretische Physik,\\\hspace{0.5em} 
Wilhelm-Klemm-Stra{\ss}e~9, 48149 M\"unster, Germany
}
\emailAdd{p.fritzsch@csic.es}
\emailAdd{nicolas.garron@plymouth.ac.uk}
\emailAdd{heitger@uni-muenster.de}
\preprint{%
{\flushright
IFT-UAM/CSIC-15-092\\
MS-TP-15-12\\
}
}

\abstract{%
We study the heavy quark mass dependence of selected observables
constructed from heavy-light meson correlation functions in small-volume 
two-flavour lattice QCD after taking the continuum limit. 
The light quark mass is tuned to zero, whereas the range of available heavy 
quark masses $m_{\rm h}$ covers a region extending from around the charm to
beyond the bottom quark mass scale. 
This allows entering the asymptotic mass-scaling regime as $1/m_{\rm h}\to 0$
and performing well-controlled extrapolations to the infinite-mass limit.
Our results are then compared to predictions obtained in the static limit
of continuum Heavy Quark Effective Theory (HQET), in order to verify
non-perturbatively that HQET is an effective theory of QCD.
While in general we observe a nice agreement at the few-\% level, we find it
to be less convincing for the small-volume pseudoscalar decay constant when
perturbative matching is involved.
}

\keywords{%
Non-perturbative Effects, Lattice QCD, Heavy Quark Physics}

\begin{document}
\maketitle
\flushbottom
\section{Introduction}\label{sec:intro}

Heavy quark systems, notably B- and $\Bs$-mesons, provide a unique
opportunity to perform stringent tests of the Standard Model and probe signals
of new physics. To maximize the impact of experiments performed at the Large
Hadron Collider, for example, it is imperative to control various aspects of
the underlying theory. In particular, strong interaction effects must be
understood at the quantitative level, including reliable estimates of
systematic uncertainties.

Although in principle lattice QCD in a large physical volume $L^3$ allows for
\emph{ab initio} computations of hadronic matrix elements and energy levels,
the presence of heavy and light quarks still renders computations very
demanding.
Current state-of-the-art lattice simulations of QCD with $\nf\ge 2$ dynamical
quarks usually reach lattice spacings down to $a\sim 0.05\,\fm$, while
satisfying $\mpi L\gtrsim 4$ in order to keep the finite-size effects under
control.
Due to increasing computer power and algorithmic advances over the last 
decade, it has become possible to simulate close to the physical pion mass
for the figures just given.
However, additionally including relativistic b-quarks with their ``heavy''
physical mass of about $4\,\GeV$
requires very fine lattice spacings $a \ll 1/\mh\sim 0.05\,\fm$ in order to
monitor its discretization effects in the spirit of Symanzik's local effective
theory. Associated with this is the problem of large scale separations, 
$\mB/\mpi=\Or(100)$, to be accommodated in a single simulation, which remains 
out of reach in the near future.
Therefore, one still needs to take a detour to an effective description for
the heavy quark. Here we focus on the Heavy Quark Effective Theory
(HQET)~\cite{stat:eichten,Eichten:1989zv,Georgi:1990um,Grinstein:1990mj} 
which provides a natural framework to study heavy-light mesons through a 
systematic expansion in the inverse heavy quark mass, $1/\mh$.

A non-perturbative implementation of HQET on the lattice~\cite{Heitger:2003nj},
including the next-to-leading order in the $\mhinv$-expansion, has been tested
and applied successfully for the quenched case in the
past~\cite{Blossier2010,Blossier2010a,Blossier:2010mk}.  It requires to solve a
set of matching relations between quantities in continuum QCD and lattice HQET
in a small physical volume. In two-flavour QCD, the resulting non-perturbative 
set of HQET parameters~\cite{Blossier:2012qu} was recently used to 
extract phenomenologically relevant parameters such as the b-quark mass, the 
B-meson decay constant and hyperfine splittings from large-volume 
simulations~\cite{Bernardoni:2013xba,Bernardoni:2014fva,Bernardoni:2015nqa}.  
At present there are efforts to extend the matching strategy to include the 
vector meson channel~\cite{DellaMorte:2013ega,Korcyl:2013ega} in order to 
compute $\fBstar$ and form factors of semi-leptonic 
${\rm B}_{({\rm s})}\to \pi\,({\rm K})$ transitions.

In this paper we probe predictions of HQET by studying the asymptotic behaviour
$\mhinv\to 0$ of continuum-extrapolated lattice QCD observables, computed in
a small volume ($L\approx 0.4\,\fm$) with $\nf=2$ dynamical flavours of
non-perturbatively $\Or(a)$-improved massless Wilson fermions.
Extrapolations to the static limit are performed, which not only gives
numerical evidence of the correctness of static order HQET but also allows
assessing the size of the $\mhinv$-effects. 
As such, the present study constitutes a non-trivial 
\emph{non-perturbative} test of HQET being an effective theory of QCD and 
extends earlier work in quenched QCD~\cite{Heitger:2004gb} to the physically 
more realistic situation with dynamical light quarks. 
In particular, we investigate a considerable set of observables at varied
kinematics, which in the spirit of the general non-perturbative
matching strategy mentioned above are usually employed to define suitable 
matching relations.
Complementary to perturbative
studies~\cite{Hesse:2012hb,DellaMorte:2013ega,Korcyl:2013ega}, this yields 
non-perturbative insights into their heavy quark mass asymptotics and provides
criteria for which of them are to be preferred within the matching strategy
of~\cite{Heitger:2003nj}, such as impact of mass-dependent cutoff effects in
the continuum limit extrapolations, numerical accuracy and magnitude of
higher-order corrections in $\mhinv$. 

In contrast to a non-perturbative matching of (lattice) HQET to continuum
QCD, the perturbative approach relies on a perturbative evaluation of
\emph{matching (resp. conversion) functions}, often called Wilson 
coefficients. 
To properly recover the static limit in HQET as $\mh\to\infty$, also our
QCD observables --- non-perturbatively evaluated at finite quark mass ---
still have to be combined with matching functions of this kind.
These are only \emph{perturbatively} known, up to three-loop order in most
cases.
To disentangle in our tests the genuine non-perturbative properties of the 
theory encoded in the observables from perturbative effects induced by the
conversion functions, we map out their $\mh$-dependence towards the static
limit for conversion functions of different perturbative orders.
This comparison gives a rough idea on the systematic error that is involved,
when the matching between HQET and QCD is performed perturbatively.
As will be exposed by the example of the pseudoscalar decay constant below,
we observe that --- with perturbative matching at work --- the agreement 
between the large-mass QCD asymptotics and the HQET prediction may not be as
good as expected, even if a three-loop expression for the conversion function
is used.
We take this as an indication that in an effective theory framework for
heavy quarks the matching should be done \emph{non-perturbatively},
if one wants to have the systematic errors under control.
In addition, we consider QCD observables, which have a non-trivial static 
limit but do \emph{not} depend on any conversion function.
Hence, their $\mh\to\infty$ limit is much less affected by systematic 
uncertainties such that they are among the cleanest observables in our
non-perturbative tests of the effective theory approach to heavy-light
physics.

Some preliminary results on a smaller subset of observables, data ensembles 
and statistics have already been reported 
in~\cite{DellaMorte:2007qw,thesis:patrickf}.

\section{Observables}\label{sec:obs}

This paper follows up our previous work~\cite{Fritzsch:2010aw} on the
definition of a line of constant physics in $\nf=2$ lattice simulations.  It
allows us to non-perturbatively study the quark mass dependence of relativistic
QCD meson observables in a finite box of extent $L_1\approx
0.4\,\fm$~\cite{Fritzsch:2012wq}.  We explore a wide range of quark masses that
starts below the charm sector and goes beyond the bottom quark region.  This is
done in a partially quenched setup, i.e., the light quark mass is set to the
approximately vanishing mass of a degenerate sea quark doublet and all other
quarks are quenched.  We generically refer to the latter as heavy quarks of
mass $\mh$.
Equivalently, we will assign to them the dimensionless mass parameter
$z=L_1\Mh$ from now on, where $\Mh\equiv M$ denotes some fixed value of the
renormalization group invariant (RGI) heavy quark mass.
In~\cite{Fritzsch:2010aw} we have already shown that in such a small box the
lattice spacing can be chosen small enough so that all heavy quarks up to a
certain value can be simulated relativistically while keeping cutoff effects in
the $\Or(a)$ improved theory well under control.  For any unexplained notation,
the reader may consult~\cite{Fritzsch:2010aw,Heitger:2004gb}.

Since our interest lies in relating predictions made by HQET 
non-perturbatively to the proper counterpart in QCD towards the limit 
$1/\mh\to0$,
we furthermore take into account measurements of HQET observables that
have been done in the framework of a general non-perturbative matching
strategy of HQET and QCD in the very same volume $L_1$, but to a much higher 
statistical accuracy. Additional details can be found in appendices B and C of 
reference~\cite{Blossier:2012qu}. Working in a finite (and small) volume,
all matrix elements and energies become effective quantities which
intrinsically depend on the scale $L_1$. For notational brevity we often
suppress this dependence in the following.

Our main observables are built from Schr\"odinger functional (SF) correlation 
functions~\cite{Luscher:1992an,Sint:1993un} in a $T\times L^{3}$ volume with 
$T=L=L_1$ fixed and periodic boundary conditions in space. The fermion fields 
are taken periodic only up to a phase,
\begin{align}
        \psi(x+\hat{k} L)    &=\rme^{\rmi\theta}\psi(x)\;, &
        \psibar(x+\hat{k} L) &=\psibar(x)\rme^{-\rmi\theta}\;, &
        k &= 1,2,3\;,
        \label{}
\end{align}
where we use $\theta\in\{0,0.5,1\}$. 
In correlation functions, this periodicity angle $\theta$ amounts to a
projection onto quark and antiquark momenta with components $\pm\theta/L$.  In
time direction, Dirichlet boundary conditions are imposed at $x_0=0,\, T$ where
source quark and antiquark fields are separately projected onto vanishing
spatial momentum.  We are interested in the pseudoscalar (PS) and vector (V)
channel using finite-volume heavy-light QCD currents. They are given by the
time component of the axial vector and spatial components of the vector
current, respectively:
\begin{align}
        A_0(x_0) &= \lightb(x_0) \gamma_0\gamma_5 \heavy(x_0) \;,&
        V_k(x_0) &= \lightb(x_0) \gamma_k \heavy(x_0) \;, &
               0 &< x_0 < T \;.
        \label{eq:def_A0_V0}
\end{align}
To be more precise, we use their $\Or$
improved~\cite{DellaMorte:2005se,Sint:1997jx} and non-perturbatively
renormalized~\cite{DellaMorte:2008xb} lattice versions.  Furthermore, we need
the pseudoscalar heavy-light current
\begin{align}
        P(x_0) &= \lightb(x_0) \gamma_5 \heavy(x_0) \;,&
               0 &< x_0 < T \;,
        \label{eq:def_P}
\end{align}
the renormalization factor of which, $\zp(\mu,\gosq)$, has been determined
non-perturbatively in the SF scheme at scale $\mu=1/L_1$ during the production
runs reported in~\cite{Blossier:2012qu}.  Since they have not been quoted in
that reference we list them together with further details in
table~\ref{tab:QCD_run_details}.

\subsection{Definitions}\label{sec:definitions}

How to non-perturbatively set up a line of constant physics in the envisaged
small volume $L_1\approx 0.4\,\fm$ has already been reported
in~\cite{Fritzsch:2010aw}.  In that volume we have four different ensembles
with non-perturbatively $\Or$ improved dynamical Wilson fermions made of a
doublet of massless quarks. The range of lattice spacings used is $0.01\,\fm
\lesssim a \lesssim 0.02\,\fm$.  This ensures feasible continuum limit
extrapolations of the QCD observables to be introduced below, which depend on
the dimensionless RGI heavy quark mass fixed to
\begin{align}
     z &\equiv L_1 \Mh \in \{ 2, 2.7, 3, 3.3, 4, 6, 7, 9, 11, 13, 15, 18, 21 \} \;.
     \label{eq:def_z}
\end{align}
In contrast to our previous work, we added four additional $z$-values at the
lower end to also cover the charm quark region.  Details about the latter,
which were needed to perform the additional measurements, are listed in
table~\ref{tab:QCD_run_details}. 
In the computation of HQET observables we can naturally rely on larger lattice
spacings ($0.025\,\fm\lesssim a\lesssim 0.067\,\fm$). We use the 
``HQET[$L_1$]'' lattices with $T=L=L_1$ as specified in table~C.1 and~C.2 of
ref.~\cite{Blossier:2012qu}. For completeness we list some additional
details in table~\ref{tab:HQET_run_details} that were not published there.

The HQET observables themselves are computed using two different static
actions, referred to as $\hypone$ and $\hyptwo$~\cite{DellaMorte:2005yc}, in
order to have an improved noise-to-signal behaviour compared to the original
Eichten-Hill action.  For the present purpose of testing HQET we want to
express all observables in terms of matrix elements computed in a finite
volume.  The interested reader can find the corresponding notation in terms of
the traditional SF correlation functions in appendix~\ref{app:defs} and
reference~\cite{Heitger:2004gb}.  In passing we just note that only states,
which are eigenstates of spatial momentum with eigenvalue zero, enter them.

\subsection{Effective masses}\label{sec:eff_masses}

The \emph{small-volume effective pseudoscalar} and \emph{vector meson mass} 
in terms of Hilbert space matrix elements read, adopting an operator notation
for the quark bilinear composite fields,
\begin{subequations}      \label{eq:def_eff_mass}
\begin{align}  
   \GamPS(x_0) &= -\drvsym0 \ln\big[ \ketbra{\Om(L)}{\opA}{{\rm B}(L)}    \big]  \;, \label{eq:GamPS}\\
   \GamV (x_0) &= -\drvsym0 \ln\big[ \ketbra{\Om(L)}{\opV}{{\rm B}^*(L)}  \big]  \;, \label{eq:GamV}\\
   \GamP(x_0)  &= -\drvsym0 \ln\big[ \ketbra{\Om(L)}{\opP\,}{{\rm B}(L)}    \big]  \;, \label{eq:GamP}
\end{align}
\end{subequations}
where $\ket{\Om(L)}\equiv \rme^{-x_0\opH}\ket{\varphi_0(L)}$ denotes the vacuum
state given in terms of the Hamiltonian $\opH$ and the SF intrinsic vacuum
boundary state $\ket{\varphi_0(L)}$ at $x_0=0$.  Having in mind a variation of
the heavy quark mass for our non-perturbative tests later on, we denote a
general heavy-light pseudoscalar state by $\ket{{\rm
B}(L)}\equiv\rme^{-x_0\opH}\ket{\varphi_{{\rm B}}(L)}$ and a heavy-light vector
state by $\ket{{\rm B}^*(L)}\equiv\rme^{-x_0\opH}\ket{\varphi_{{\rm B}^*}(L)}$.
Both are again given through the time evolution operator and the well-defined
boundary states $\ket{\varphi_{\rm X}}$ with quantum numbers in the respective
channel, ${\rm X}={\rm B}, {\rm B}^*$.  Since all states naturally depend on
the finite volume (or box) size $L$ and are also considered as functions of
the RGI heavy quark mass $M$ (or $z$),  we drop these dependencies for the
moment to ease notation, as we do so for their additional dependence on the
SF-specific periodicity angle $\theta$ of the fermion fields.  
Here and from now on, we fix $x_0=T/2\equiv L/2$ in the correlation functions
employed to construct the observables above and those to be introduced in the
subsections below.  
It is then worth to emphasize that the time evolution operator $\rme^{-T\opH/2 }$
suppresses high-energy states exponentially such that $\ket{\Om(L)}$ and
$\ket{{\rm B}(L)}$, upon expanding them in terms of eigenstates of $\opH$, are
dominated by contributions from states with energies of at most $\Delta
E=\rmO(1/L)$ above the ground state.  Therefore, as HQET is expected to apply
to correlation functions at large Euclidean time separations, it particularly
describes their large-mass behaviour at large $x_0\geq\rmO(\mhinv)$ in the
present SF setup, too.

Whereas physical masses must be computed in large-volume simulations, 
the definition of our observables is such that they agree with the physical 
ones in the large-volume limit $L\to\infty$.

\subsection{Decay constants and ratios}\label{sec:DecayConst}

Furthermore and analogously, we define the following QCD observables, 
suppressing again their dependence on $M$, $L$ and $\theta$: 
\begin{align}
  \Yps   &\equiv
  + \left[\dfrac{\ketbra{\Om}{\opA}{{\rm B}}}{\norm{\ket{\Om}}\cdot\norm{\ket{\rm B}}}\right]_{\rm R}  \;,& 
  \Yv   &\equiv
  - \left[\dfrac{\ketbra{\Om}{\opV}{{\rm B}^*} }{\norm{\ket{\Om}}\cdot\norm{\ket{\rm B^*}} }\right]_{\rm R} \;, \label{eq:def-Yps-Yv} \\[0.5em]
  \Rpsp&\equiv 
     - \left[\dfrac{ \ketbra{\Om}{\opA}{{\rm B}} }{ \ketbra{\Om}{\opP\,}{{\rm B}} }\right]_{\rm R}    \;, &
  \Rpsv&\equiv 
     - \left[\dfrac{ \ketbra{\Om}{\opA}{{\rm B}} }{ \ketbra{\Om}{\opV}{{\rm B}^*} }\right]_{\rm R}  \;, \label{eq:def-Rpsp-Rpsv}  \\[0.5em]
  \Ypsv   &\equiv
     \dfrac{\Yps}{\Yv}  \;, &
  \Rspin&\equiv  
  \dfrac{3}{4}\ln \dfrac{ \braket{ {\rm B} }{{\rm B}} }{ \braket{ {\rm B}^* }{{\rm B}^*} }  \;,  \label{eq:def-Ypsv-Rspin}
\end{align}
where $\Yps$ and $\Yv$ are the \emph{finite-volume heavy-light pseudoscalar} and
\emph{vector decay constant}, respectively. 
As $L\to\infty$, they become proportional to the physical heavy-light
pseudoscalar and vector meson decay constants.  Accordingly, $\Ypsv$ is the
ratio of the two, which in large volume becomes proportional to $\fB/\fBstar$,
if the heavy quark is set to the b-quark. The ratio $\Rspin$ is proportional to
the spin splitting between the pseudoscalar and vector channel and, as
predicted by HQET, has to vanish in the static limit owing to the heavy-quark
spin symmetry.  These observables also involve boundary-to-boundary SF
correlation functions, see appendix~\ref{app:defs}, which properly cancel the
multiplicative renormalization factors of the boundary quark fields.

All quantities involving $[\bullet]_{\rm R}$ are understood to be renormalized
non-perturbatively, while for others such factors either drop out or are not
needed at all, such that alltogether they thus are \emph{finite} and possess a
well-defined continuum limit.

\subsection{Quantities with different kinematics}\label{sec:otherQuants}

The SF is especially useful, if one wants to probe physics with different
kinematics.  Here we do so by changing the fermionic phase angle $\theta$ as
mentioned earlier. Whereas in the last section all observables were meant to be
evaluated at the same values, i.e., $\theta_0\in\{0,0.5,1\}$, we now turn our
attention to quantities that are made of two matrix elements of heavy-light
composite fields referring to fermionic periodicity phases different from
eachother. With the same notational conventions as before they read
\begin{subequations}\label{eq:all-good-obs}
\begin{align} \label{eq:kinQ-ps}
        R_f   (\theta_1,\theta_2) &=\frac{ \braket{ {\rm B} }{{\rm B}}_{\theta_1} }{ \braket{ {\rm B} }{{\rm B}}_{\theta_2} }\;,  &
        \Rpsps(\theta_1,\theta_2) &=\frac{ \ketbra{\Om}{\opA}{{\rm B}}_{\theta_1} }{ \ketbra{\Om}{\opA}{{\rm B}}_{\theta_2} } \;, \\[0.5em]
              \label{eq:kinQ-v}
        R_k   (\theta_1,\theta_2) &=\frac{ \braket{ {\rm B}^* }{{\rm B}^*}_{\theta_1} }{ \braket{ {\rm B}^* }{{\rm B}^*}_{\theta_2} } \;,   &
        \Rvv  (\theta_1,\theta_2) &=\frac{ \ketbra{\Om}{\opV}{{\rm B}^*}_{\theta_1} }{ \ketbra{\Om}{\opV}{{\rm B}^*}_{\theta_2} } \;,\\[0.5em]
              \label{eq:kinQ-R1}
        \Rone (\theta_1,\theta_2) &=\frac{1}{4}\ln\left[ R_f(\theta_1,\theta_2) R_k(\theta_1,\theta_2)^3 \right]  \;, &
        \Rpp  (\theta_1,\theta_2) &=\frac{ \ketbra{\Om}{\opP\,}{{\rm B}}_{\theta_1} }{ \ketbra{\Om}{\opP\,}{{\rm B}}_{\theta_2} } \;, \\[0.5em]
              \label{eq:kinQ-Y}
        \Ypsps(\theta_1,\theta_2) &=\frac{ \Yps(\theta_1) }{ \Yps(\theta_2) } \;, &
        \Yvv  (\theta_1,\theta_2) &=\frac{ \Yv( \theta_1) }{ \Yv(\theta_2) } \;, 
\end{align}
\end{subequations}
where in our actual calculations we consider the following pairs of phase
angles: $(\theta_1,\theta_2)\in\{(0,0.5),(0.5,1),(0,1)\}$. 

It is important to note that all multiplicative renormalization and 
improvement factors cancel in these ratios.%
\footnote{They differ from 1 (or 0 for $R_1$) only due to
$\theta_1\neq\theta_2$, both at finite $z$ and in the static limit.}
In this respect and because one expects cancellations of cutoff effects at
every fixed value of $z$, these QCD observables (as well as their counterparts
in HQET) may be seen as ``gold plated'' observables for the purpose of testing
the asymptotic behaviour of heavy-light physics as $z\to\infty$ for different
kinematical setups.

\subsection{Observables in HQET}\label{sec:HQETobs}

After the continuum limit of the previously defined QCD observables has been
taken, we aim for an extrapolation to the static limit, $1/z\to 0$, in order to
compare their asymptotic behaviour to the one predicted by HQET in the
continuum. According to the systematic heavy quark expansion, all quantities
approach a well-defined value in that limit.  Classically, the leading
asymptotic behaviour of our effective masses~\eqref{eq:def_eff_mass}, for
instance, is linear in the heavy quark mass $z$, but it receives logarithmic
modifications on the quantum level owing to the scale dependent renormalization
of the effective theory to compare with, viz.
\begin{align}\label{eq:LGam-asympt}
    L\Gamma(L,M)\; &\simas{z\to\infty} \;\Cmass(z)  \cdot z \cdot\Big[1 + \rmO\big(z^{-1}\big) \Big] \;, 
\end{align}
where $\Cmass(z)$ denotes the conversion function that relates the heavy
quark's pole mass to the RGI heavy quark mass $M=z/L_1$. 

A generic conversion function $\Cx(z)$ carries all the logarithmic dependence
of a given quantity $\rmX$ \emph{to some order in perturbation theory} such
that only power corrections in $1/z$ remain in the effective theory.
For more details, we refer to appendix~\ref{sec:ConvFuncs} and appendix B
of~\cite{Heitger:2004gb}.  To avoid a remnant renormalization scheme dependence
in the (static) effective theory, we favour to fully express the asymptotic
behaviour in terms of renormalization group invariants (RGIs).  For the
quantities of section~\ref{sec:DecayConst}, i.e., $\rmX\in\{{\rm PS},{\rm
V},{\rm PS/P},{\rm PS/V},{\rm spin}\}$, this means
\begin{align}
   \label{eq:asym-Yps}      \Yps(L,M)  \;&\simas{z\to\infty}\; \Cps(z)  \cdot \XRGI(L)       \cdot\Big[1 + \rmO\big(z^{-1}\big) \Big] \;,         \\[0.2em]
   \label{eq:asym-Yv}       \Yv(L,M)   \;&\simas{z\to\infty}\; \Cv(z)   \cdot \XRGI(L)       \cdot\Big[1 + \rmO\big(z^{-1}\big) \Big] \;,         \\[0.2em]
   \label{eq:asym-Rap}      \Rpsp(L,M) \;&\simas{z\to\infty}\; \Cpsp(z) \cdot  1             \cdot\Big[1 + \rmO\big(z^{-1}\big) \Big] \;,         \\[0.2em]
   \label{eq:asym-Rav}      \Rpsv(L,M) \;&\simas{z\to\infty}\; \Cpsv(z) \cdot  1             \cdot\Big[1 + \rmO\big(z^{-1}\big) \Big] \;,         \\[0.2em]
   \label{eq:asym-RY}       \Ypsv(L,M) \;&\simas{z\to\infty}\; \Cpsv(z) \cdot  1             \cdot\Big[1 + \rmO\big(z^{-1}\big) \Big] \;,         \\
   \label{eq:asym-spin}     \Rspin(L,M)\;&\simas{z\to\infty}\; \Cspin(z)\cdot\frac{\XRGIspin(L)}{z}\cdot\Big[1 + \rmO\big(z^{-1}\big)\Big]\;. 
\end{align}

The ratios $\Rpsp$, $\Rpsv$ and $\Ypsv$, along with the associated $\Cx$,
approach $1$ in the static limit of HQET, whereas the effective decay
constants $\Yps$ and $\Yv$ both approach the finite-volume RGI static-light
decay constant $\XRGI(L)$ in this limit, as a consequence of the heavy-quark
spin symmetry.
In the two-flavour theory at hand~\cite{DellaMorte:2006sv}, 
the renormalization scale was implicitly fixed by a value for the 
renormalized SF coupling of $\gbsq(\mu)|_{\mu=\lmax^{-1}}\equiv 4.61$.
However, since for the purpose of this study we are working at a slightly
different physical volume $L_1\,\lesssim\,\lmax$, the \emph{universal part} 
$Z^{\rm stat}_{\rm A,RGI}/Z^{\rm stat}_{\rm A}(\mu)$ of the total renormalization
factor, which relates a matrix element of the static axial current 
renormalized at a scale $\mu$, $X_{\rm R}(\mu)$, to the RGI one, had to be 
re-evaluated.
The outcome for $\mu=L_1^{-1}$ based on the data of~\cite{DellaMorte:2006sv} is 
\begin{align}
       \XRGI &= \frac{Z^{\rm stat}_{\rm A,RGI}}{Z^{\rm stat}_{\rm A}(\mu)} \cdot X_{\rm R}(\mu)  \;,  & 
       \left.\frac{Z^{\rm stat}_{\rm A,RGI}}{Z^{\rm stat}_{\rm A}(\mu)}\right|_{\mu=L_1^{-1}} \!\!\!&= 0.875(7) \;, &       
       \left.\gbsq(\mu)\right|_{\mu=L_1^{-1}} &\equiv 4.484 \;,
 \label{eq:Xrgi}
\end{align}
which allows us to directly compute the (renormalization scale and scheme
independent) quantity $\XRGI$ instead of the renormalized (and thus scale
dependent) static-light decay constant $X_{\rm R}(\mu)$ in finite volume. 
Some additional technical details on $\XRGI$ and its error budget are 
postponed to appendix~\ref{app:AxialC}.
 
By contrast, the RGI matrix element of the spin splitting operator,
$\XRGIspin$, is not known, because the corresponding RG running is not
available for the dynamical flavour theory.  Note that in principle it would be
possible to extract it from our QCD data according to the given asymptotic
behaviour, \eqref{eq:asym-spin}, but with the only perturbatively known
conversion function $\Cspin$ we do not expect the result to be particularly
meaningful.

Our most stringent tests of HQET will finally arise from the QCD observables
defined in section~\ref{sec:otherQuants}. In these quantities, the conversion
functions cancel such that no logarithmic corrections are left to all orders in
perturbation theory \emph{and} non-perturbatively.  Hence, all uncertainties
from the matching between QCD and HQET are absent and one just faces the
genuine power corrections in $1/z$ of the effective theory.  This makes the
comparison of continuum QCD in the limit $1/z\to0$ and continuum HQET at static
order entirely non-perturbative and very well controllable, without
encountering any systematic errors induced by the perturbatively evaluated
$\Cx$.  Exploiting again the heavy-quark symmetry, the HQET counterparts of
these QCD observables in the static limit are:
\begin{subequations}
        \label{eq:stat-extrap2}
\begin{align}\label{eq:stat-extrap2-A}
         R_{f}^{\rm stat}(\theta_1,\theta_2) &= \exp\left[ \lim_{1/z\to 0} R_{1}(z,\theta_1,\theta_2) \right] = \lim_{1/z\to 0} R_{\rm i}(z,\theta_1,\theta_2) \;, & \text{for } {\rm i}&=f, k \;, \\
             \label{eq:stat-extrap2-B}
    R_{\rm PS}^{\rm stat}(\theta_1,\theta_2) &=            \lim_{1/z\to 0} R_{\rm i/i}(z,\theta_1,\theta_2) \;,& \text{for } {\rm i}&={\rm PS}, {\rm V}, {\rm P} \;, \\
             \label{eq:stat-extrap2-C}
     R_{X}^{\rm stat}(\theta_1,\theta_2) &=            \lim_{1/z\to 0} Y_{\rm i/i}(z,\theta_1,\theta_2) \;,& \text{for } {\rm i}&={\rm PS}, {\rm V}  \;.
\end{align}
\end{subequations}

\section{Results}\label{sec:results}
In the following subsections we first discuss exemplary continuum
extrapolations for some of our test observables before we turn our attention to
the main results, i.e., their extrapolations as $1/z\to 0$ to the static limit
of HQET.  Additional details about our continuum extrapolations can be found in
appendix~\ref{app:CL}.

\subsection{Representative continuum extrapolations}\label{sec:rep-CL}

For the quantities of
subsections~\ref{sec:eff_masses}~--~\ref{sec:otherQuants},
which as properly renormalized QCD observables are now generically denoted by
$\Omega^{\rm QCD}=\Omega^{\rm QCD}(L,M,a)$, we perform extrapolations to the
continuum limit (CL) using a global fit ansatz in order to have better control
over mass-dependent lattice artefacts.
Due to the latter, we exclude some points at coarsest lattice spacings from
these global QCD continuum extrapolations.

In general we aim at taking the continuum limit of a QCD observable according
to the global fit ansatz
\begin{align}
   \Omega^{\rm QCD}(L,z,a) &=  
   \Omega^{\rm QCD}(L,z) \left[ 1 + (a/L)^2\cdot
   \left\{ \rho_0 + \rho_1 z + \rho_2 z^2 \right\} \right] \;,
   \label{eq:global-CLfit}
\end{align}
which accounts for terms proportional $a/L\times aM$ and $(aM)^2$.
From earlier studies~\cite{Kurth:2001yr,Heitger:2004gb} it is known that an
exclusion limit of $aM > 0.7$ has to be imposed on the data entering in
the extrapolating fits, in order to avoid contaminations which are
potentially dangerous for a reliable continuum limit. 

\begin{figure}[]
  \centering
     \includegraphics[width=0.47\textwidth]{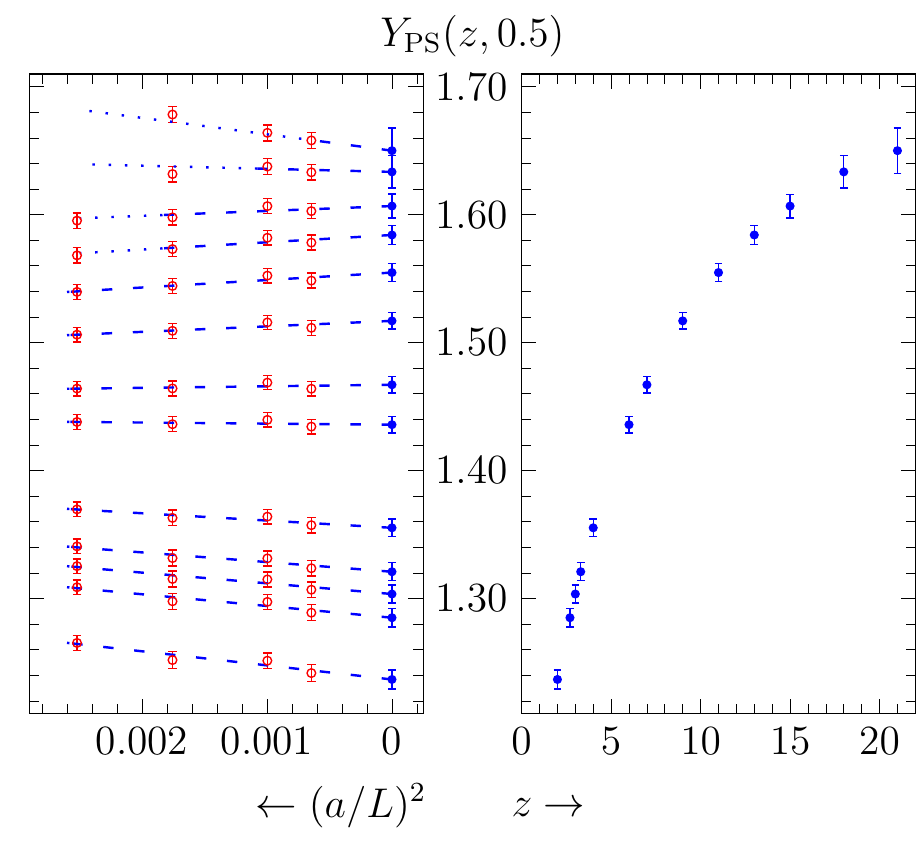}\hfill
     \includegraphics[width=0.47\textwidth]{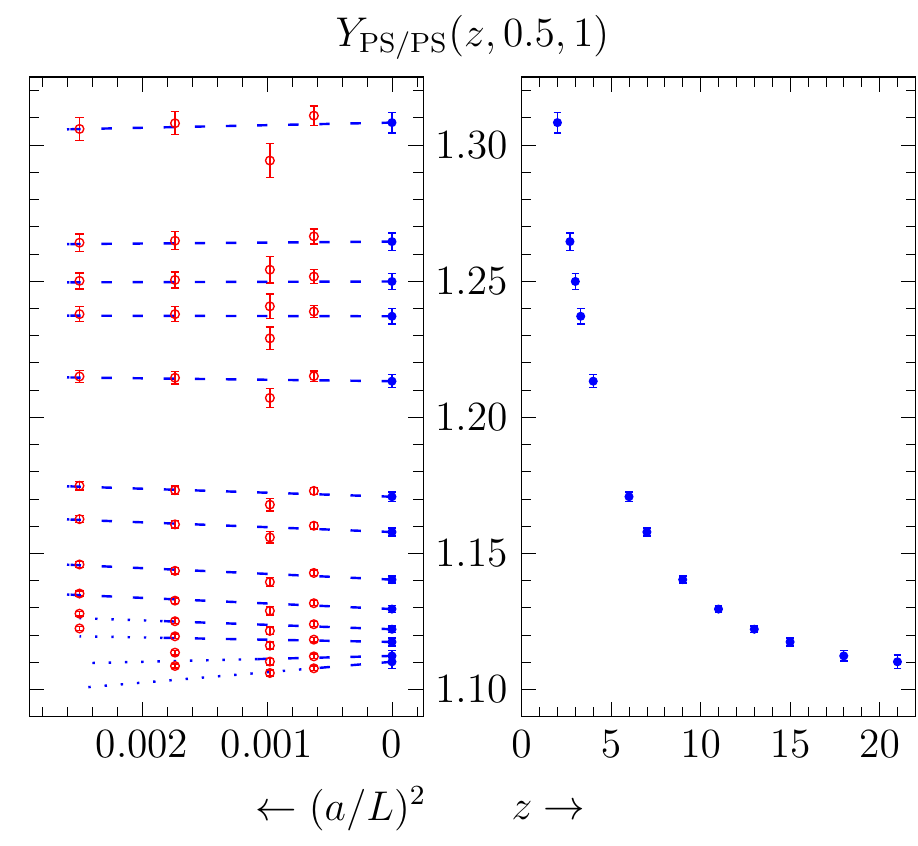}\\
  \vskip-0.25cm
  \caption{Continuum extrapolation and resulting continuum $z$-dependence of 
           $\Yps(\theta_0,z)$ (left) and $\Ypsps(\theta_1,\theta_2,z)$ (right).
           Dashed lines of the continuum extrapolation cover data points that
           enter the global fit, while the dotted part extends to all points.
           The data points used and shown have been tree-level improved
           in advance, cf.~appendix~\ref{app:TLI}.
          }
  \label{fig:CL-QCD}
\end{figure}

To assure stability in our CL results, different fit ansaetze have also been
studied (e.g., allowing for a cubic term in $a/L$ with mass-dependent
coefficient or omitting some of the lightest masses, say $z\le 4$);
these lead to consistent results. 
As the general scaling behaviour towards the continuum limit looks rather
similar among the different observables, we only show two representative 
examples here.
In the left panel of figure~\ref{fig:CL-QCD} we present the result for the
effective pseudoscalar decay constant at $\theta_0=0.5$ and in the right the
outcome for a ratio of the same quantity evaluated at different kinematical
parameters, namely at $(\theta_1,\theta_2)=(0.5,1)$. 
As yet we have not taken into account the error stemming from the tuning of
the heavy quark mass~\eqref{eq:def_z} at finite lattice spacing. 
The uncertainty of the continuum heavy quark mass $M=z/L_1$ is 
$\Delta M/M=\Delta z/z=1.01\%$~\cite{Fritzsch:2010aw}. We add its error
\begin{align}
    \Delta_{M}\Omega^{\rm QCD}(L,M,0) &= 
    \frac{\partial\Omega^{\rm QCD}(L,M,0)}{\partial M}\,\Delta M 
    \label{eq:add-z-error}
\end{align}
quadratically, before performing any extrapolations to the static limit as
they are presented in the following subsections.
The derivative is estimated numerically from the data at hand.
Its contribution to the total error budget is actually negligible, as can be
inferred from the continuum mass dependence displayed in
figure~\ref{fig:CL-QCD}, for instance. In appendix~\ref{app:CL} we list a
representative selection of results at finite lattice spacing and its
continuum limit.

\begin{figure}[t]
  \centering
     \includegraphics[width=0.49\textwidth]{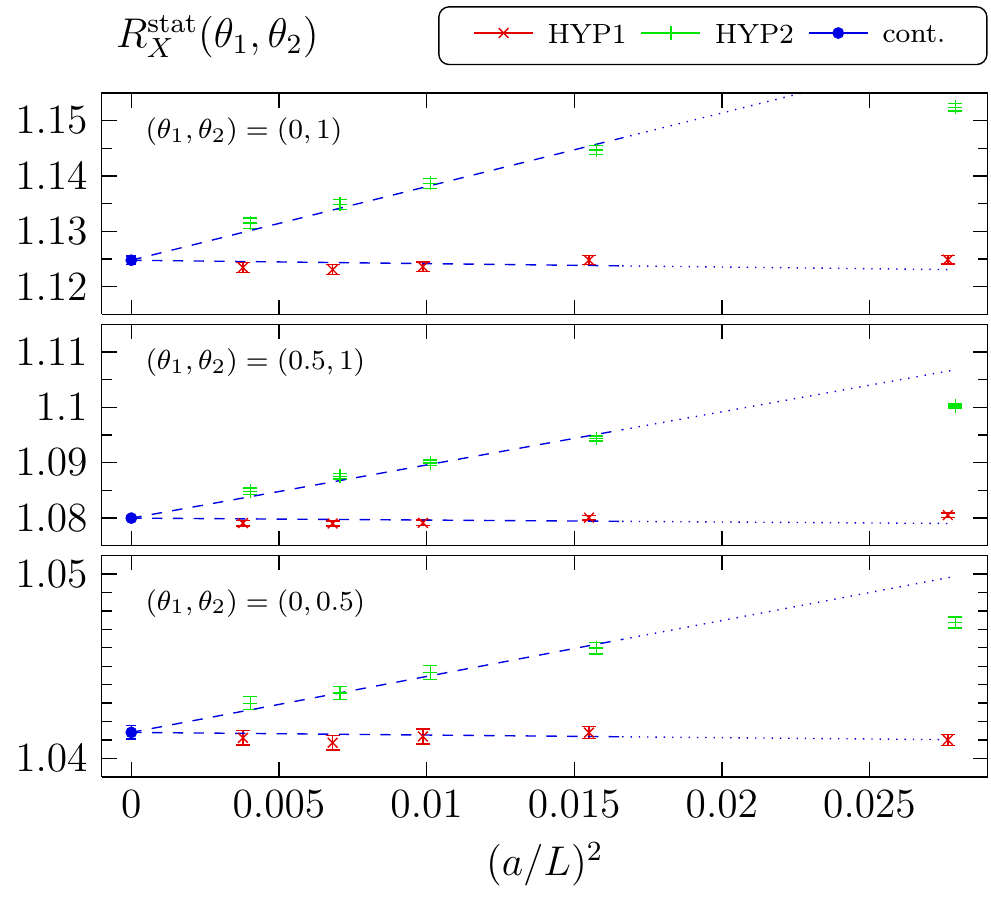}\hfill
     \includegraphics[width=0.49\textwidth]{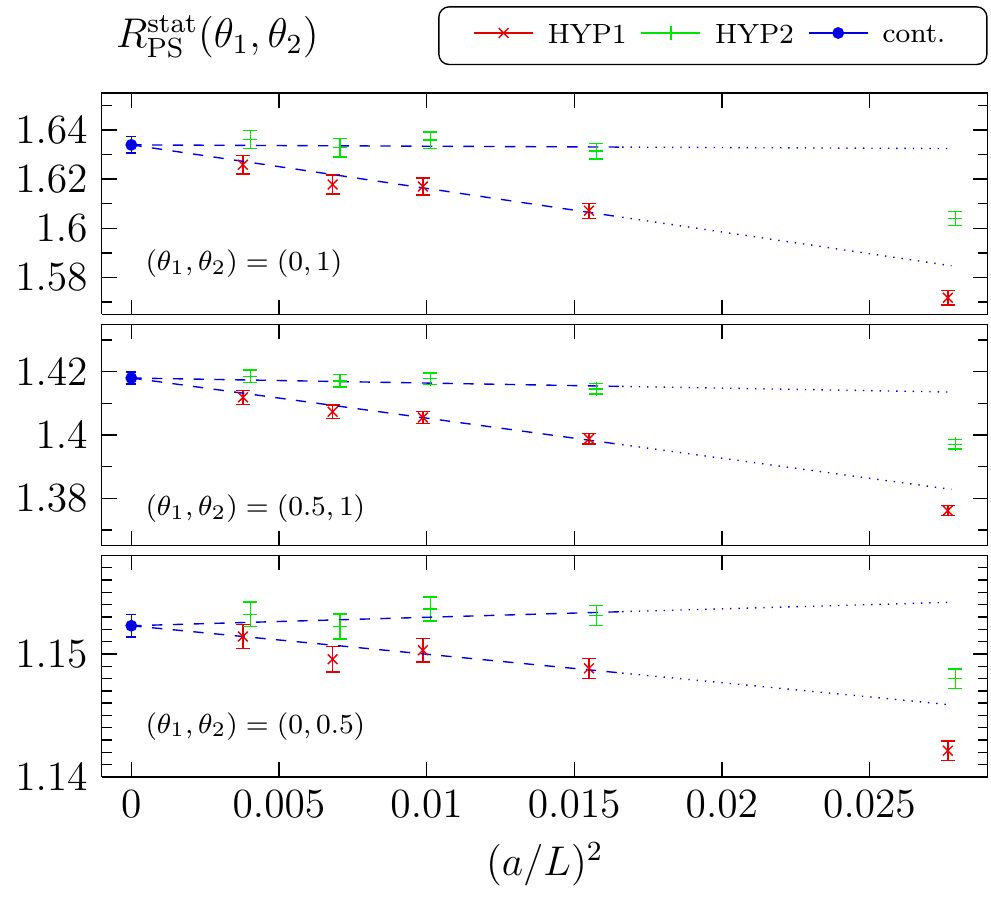}
  \vskip-0.4cm
  \caption{Continuum extrapolations of two representative static observables
  at different kinematical parameters.}
  \label{fig:CL-HQET}
\end{figure}
In some cases, continuum extrapolations of associated observables in HQET
have also to be performed.
Taking the continuum limit of observables in the static theory is much more
straightforward, because there is obsviously no dependence on $z$ and thus
no mass-dependent cutoff effect that needs to be controlled in addition.
However, they depend on the two static actions employed here (HYP$i$,
$i=1,2$), and hence this suggests to adopt a joint continuum extrapolation of
a given HQET observable according to
\begin{align}
        \Omega^{\rm HQET}_i(L,a) &=  
        \Omega^{\rm HQET}(L) \left[ 1 + (a/L)^2\cdot A_i \right] \;.
        \label{eq:HQET-CLfit}
\end{align}
As two explicit examples we show in figure~\ref{fig:CL-HQET} the continuum
extrapolation of $R_{X}^{\rm stat}(\theta_1,\theta_2)$ and 
$R_{\rm PS}^{\rm stat}(\theta_1,\theta_2)$. The numerical results are listed in 
table~\ref{tab:stat-decay-rat} and~\ref{tab:stat-ratios}, respectively.

Before turning our attention to the main results, we want to add some
general remarks on their presentation and the analysis underlying them.

\subsection{Results including perturbative conversion functions}
\label{sec:CX}

As emphasized before, for certain (continuum-extrapolated) observables
$\Omega^{\rm QCD}(L,M,0)\equiv\Omega^{\rm QCD}(L,z)$ we need to take into
account logarithmic corrections when comparing HQET with QCD in the heavy
quark mass limit.  Hence, we divide by the corresponding conversion function
$C_{\Omega}(M/\lMSbar)$ and, in a few cases where convenient, cancel the
leading mass dependence by an appropriate multiplication with some power
$\ell$ of $z$, such as $\ell=-1$ in the case of $\Omega=L\Gamma$,
for instance.  
An exactly --- or even non-perturbatively --- known  conversion function
by definition removes all logarithmic contributions, while the remainder
can then be assumed to be organized as a ``power series'' in $1/z$,
resembling continuum HQET in the asymptotic regime of large $z$.
We thus perform extrapolations to the static limit of HQET of quantities with
perturbative conversion functions according to the fit ansatz
\begin{align}
       \left[z^{\ell}\cdot\Omega^{\rm QCD}(L,M,0)\big/C_{\Omega}(M/\lMSbar)\right] 
       &= \Omega^{[0]} + \Omega^{[1]}z^{-1} + \Omega^{[2]} z^{-2} \;,\quad 
       z^{-1}\lesssim 0.26 \;,
       \label{eq:global-fitfunc-Cx}
\end{align}
where $\Omega^{[0]}$ represents the static limit of the observable in question
as extracted through this fit from the relativistic QCD data.
(The choice $1/z\lesssim 0.26$ is motivated below.)
In practice, we must rely on a \emph{perturbative} evaluation of the
conversion functions up to limited order that have uncertainties decreasing
only logarithmically (see, e.g., ref.~\cite{Sommer:2010ic}).
Since those have to be combined with our \emph{non-perturbative} lattice data,
we generically cannot disentangle logarithmic and power-like contributions
exactly up to some definite value of $1/z$.
A safe statement that can be made, however, is that one asymptotically expects
eq.~\eqref{eq:global-fitfunc-Cx} to yield a good description of the data in the
sense that the l.h.s.~of (\ref{eq:global-fitfunc-Cx}) approaches $\Omega^{[0]}$
in the limit $z\to\infty$. To what extent a quantity $\Omega^{[0]}$ agrees
with its HQET counterpart is subject of the discussion below.

To represent our data, we employed several \emph{unconstrained} (and weighted)
polynomial fits in $1/z$ along (\ref{eq:global-fitfunc-Cx}) by varying the
range of points that enter the fit with the degree of the polynomial.  
Even though the full data set --- down to masses of $\sim 70\%$ of the charm
quark mass --- is well described by a quadratic interpolation within our 
statistical accuracy, it is not surprising that this variation, depending on
the observable under consideration, has a visible effect on the result of the
static extrapolation.
More precisely, we observe the clear general trend (motivating the $z$-range
choice in~(\ref{eq:global-fitfunc-Cx})) that a fit of degree $n$ should
include data in the interval $[0,n\cdot\Delta z^{-1}]$, with 
$\Delta z^{-1}\approx 0.13$, in order to keep the static limit unchanged 
(within errors) and thereby independent of the employed polynomial fit ansatz.
This statement holds separately true for the data incorporating $C_\Omega$
evaluated at two- or three-loop order in perturbation theory.  
However, as will become evident in the sequel, the dominantly significant,
systematic effect on the static extrapolation results originates from the
only perturbative knowledge of the conversion functions that are required to
relate QCD observables at finite $z$ to their HQET counterparts, before the
static limit is taken. 

\subsubsection*{\it Asymptotic behaviour of effective meson masses}
\begin{figure}[t]
  \centering
     \includegraphics[width=0.49\textwidth]{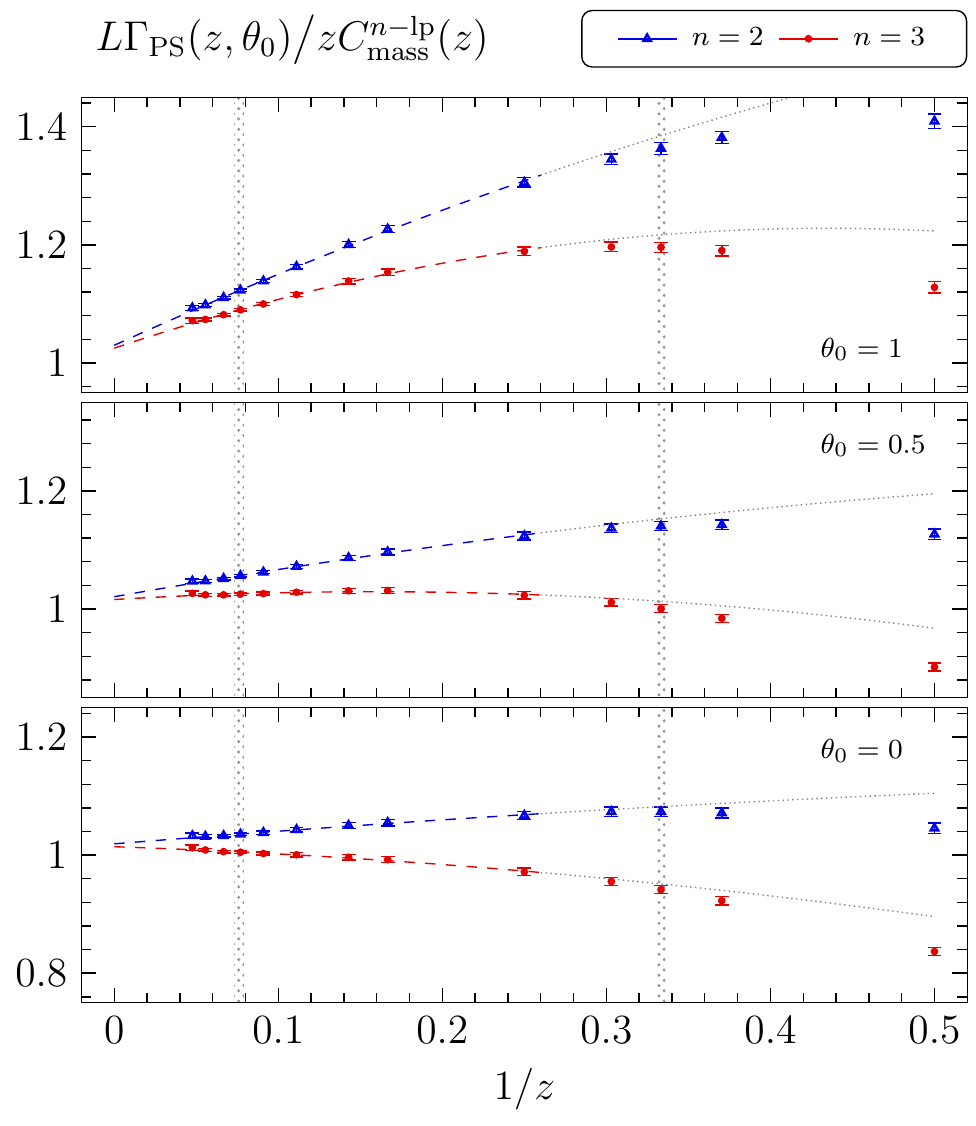}\hfill
     \includegraphics[width=0.49\textwidth]{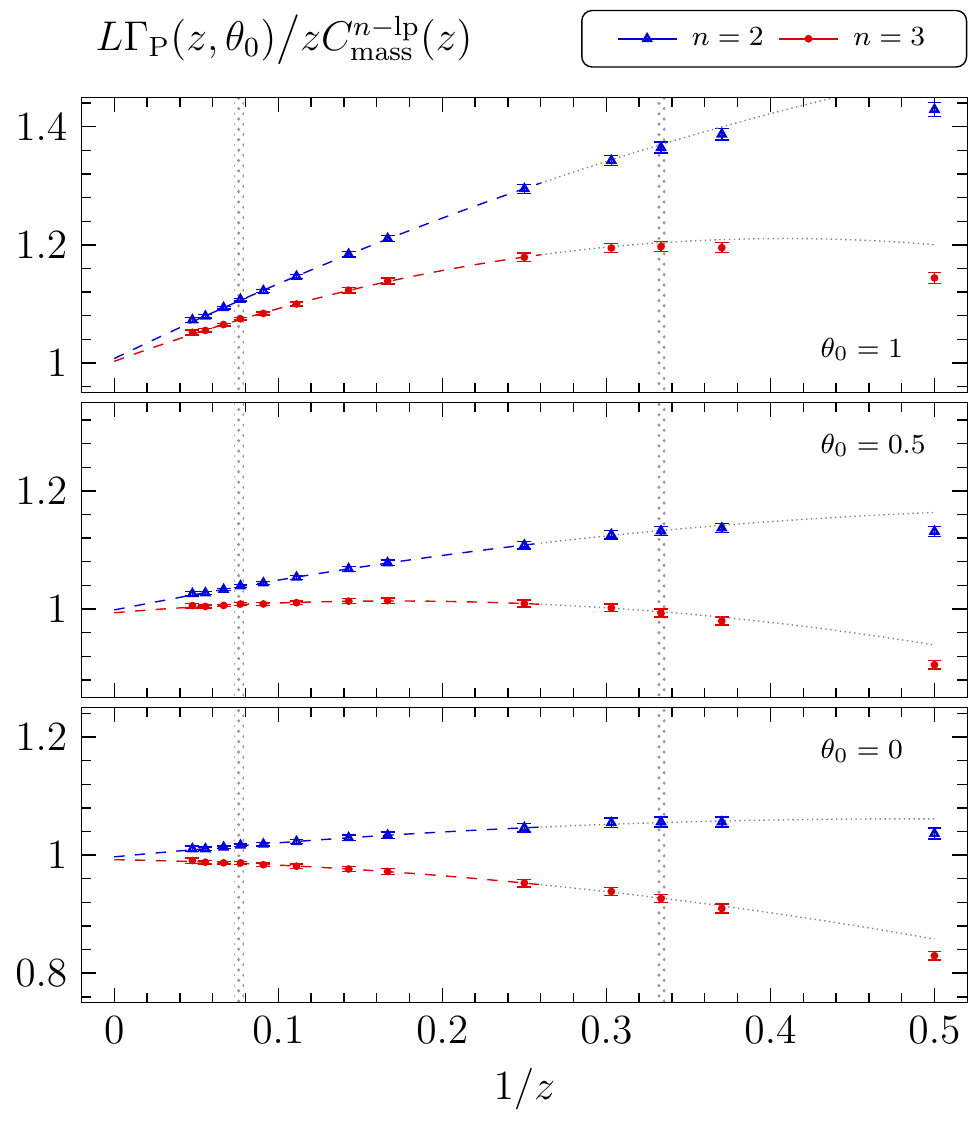}\\
     \includegraphics[width=0.49\textwidth]{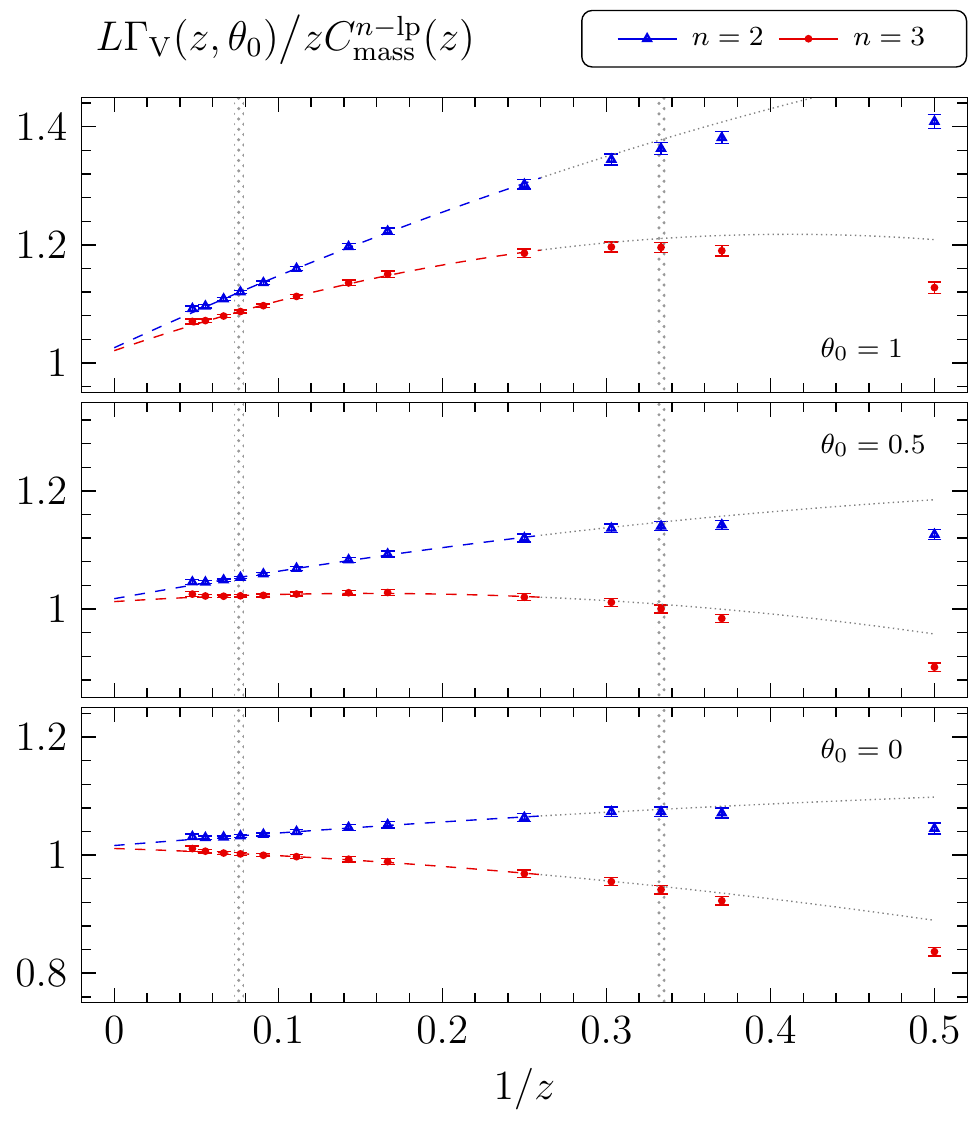}\hfill
     \includegraphics[width=0.49\textwidth]{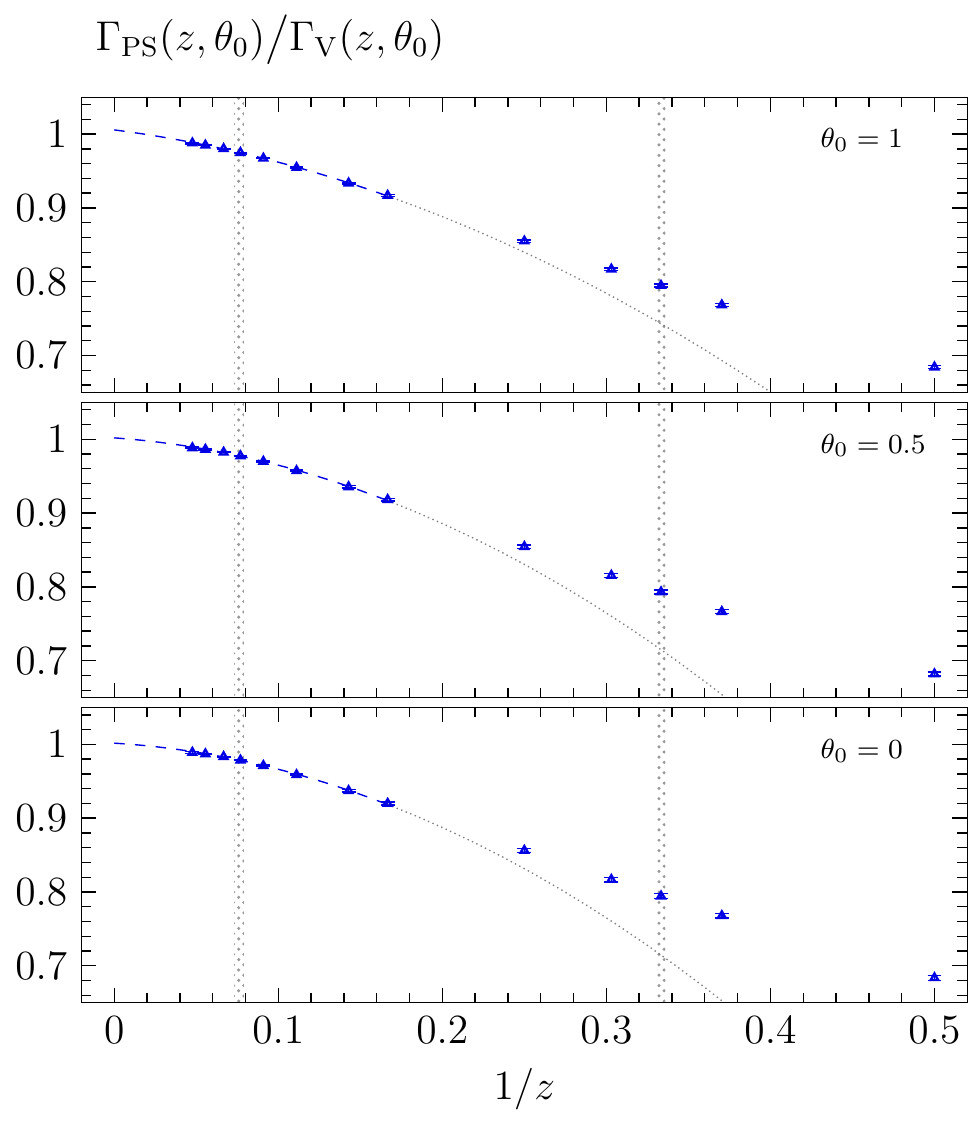}
  \vskip-0.25cm
  \caption{Subleading asymptotic behaviour of effective meson masses at 
           different kinematical parameters using two- and three-loop 
           conversion functions. In comparison, the lower right panel
           shows the asymptotic behaviour of the ratio of effective
           pseudoscalar and vector meson masses. In this ratio, QCD-HQET
           conversion functions and the leading power in $z$ cancel.}
  \label{fig:AL-masses}
\end{figure}
First, we study the large quark mass behaviour of the effective meson
masses~\eqref{eq:def_eff_mass} according to its leading asymptotics in this
regime that is given by eq.~\eqref{eq:LGam-asympt}.  The raw data and the
continuum-extrapolated values for representative $\theta$-values are collected
in tables~\ref{tab:LGamPS} and \ref{tab:LGamVP} of appendix~\ref{app:CL}.
Following eq.~\eqref{eq:global-fitfunc-Cx}, we combine the effective meson
masses with their associated conversion function $C_{\rm mass}$ and remove the
leading heavy quark mass dependence. The remaining finite piece of
$\left[L\Gamma_{\rm X}(L,M,0)\big/\big(zC_{\rm mass}(M/\lMSbar)\big)\right]$
then should approach $1$ in the static (i.e., $1/z\to 0$) limit for all 
${\rm X}={\rm PS},{\rm V},{\rm P}$.

Figure~\ref{fig:AL-masses} confirms this expectation, where for each individual
observable and $\theta_0\in\{0,0.5,1\}$ we depict the remaining (i.e.,
subleading) asymptotic behaviour of our continuum data points for the effective
meson masses using $C_{\rm mass}$ evaluated at two- and three-loop order,
cf.~eq.~\eqref{eq:Cmass-fit}.  For ease of presentation, only the statistical
errors of $L\Gamma_{\rm X}(z,\theta_0)$ are included in the figures.  With the
fit ansatz~\eqref{eq:global-fitfunc-Cx} and all errors taken into account, the
results in the static limit, $\Omega^{[0]}_{\rm X}$, agree for all 
${\rm X}={\rm PS},{\rm V},{\rm P}$ with $1$ within errors indeed.
In order to better judge the validity range of the asymptotic $1/z$-expansion
reflected by the extrapolating fits to the static limit --- as well as the
size of possible particular systematic effects at the physical scale of the
b- and c-quark --- here and in subsequent figures we add vertical error bands
corresponding to $z_{\rm b}=L_1M_{\rm b}\approx13.25$~\cite{Bernardoni:2013xba}
and $z_{\rm c}=L_1M_{\rm c}\approx 3.04$~\cite{Heitger:2013oaa}.

Another non-trivial test consists in studying ratios of masses, in which the
leading asymptotics drops out completely. Such a case is displayed in the
bottom-right panel of figure~\ref{fig:AL-masses}. This is a first explicit
example with the heavy-quark spin symmetry at work, according to which HQET
predicts $\lim_{1/z\to 0}[\GamPS/\GamV]=1$. The deviation from one at the
b-quark scale ($z=z_{\rm b}$), dominated by the spin-splitting term, is about
$2\%$. With increasing $1/z$, higher-order terms appear to become relevant and
contribute with opposite signs, leading to a deviation of about $20\%$ at the
charm quark mass scale.  At least for $\GamPS/\GamV$, this supports the general
expectation that HQET does not provide an all too good description for charm
physics any more.

\subsubsection*{\it Asymptotic behaviour of ratios of heavy-light currents}
\begin{figure}[t]
  \centering
     \includegraphics[width=0.49\textwidth]{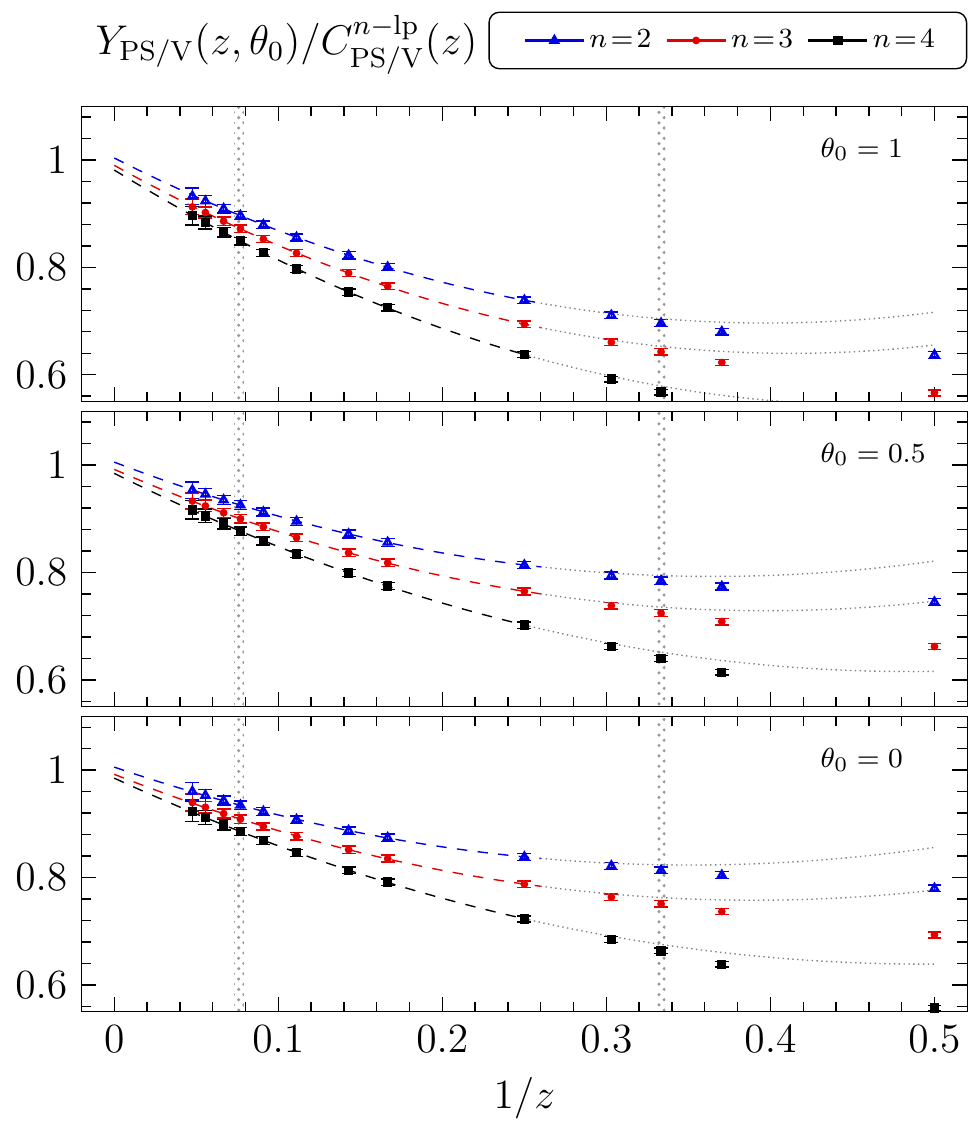}\hfill
     \includegraphics[width=0.49\textwidth]{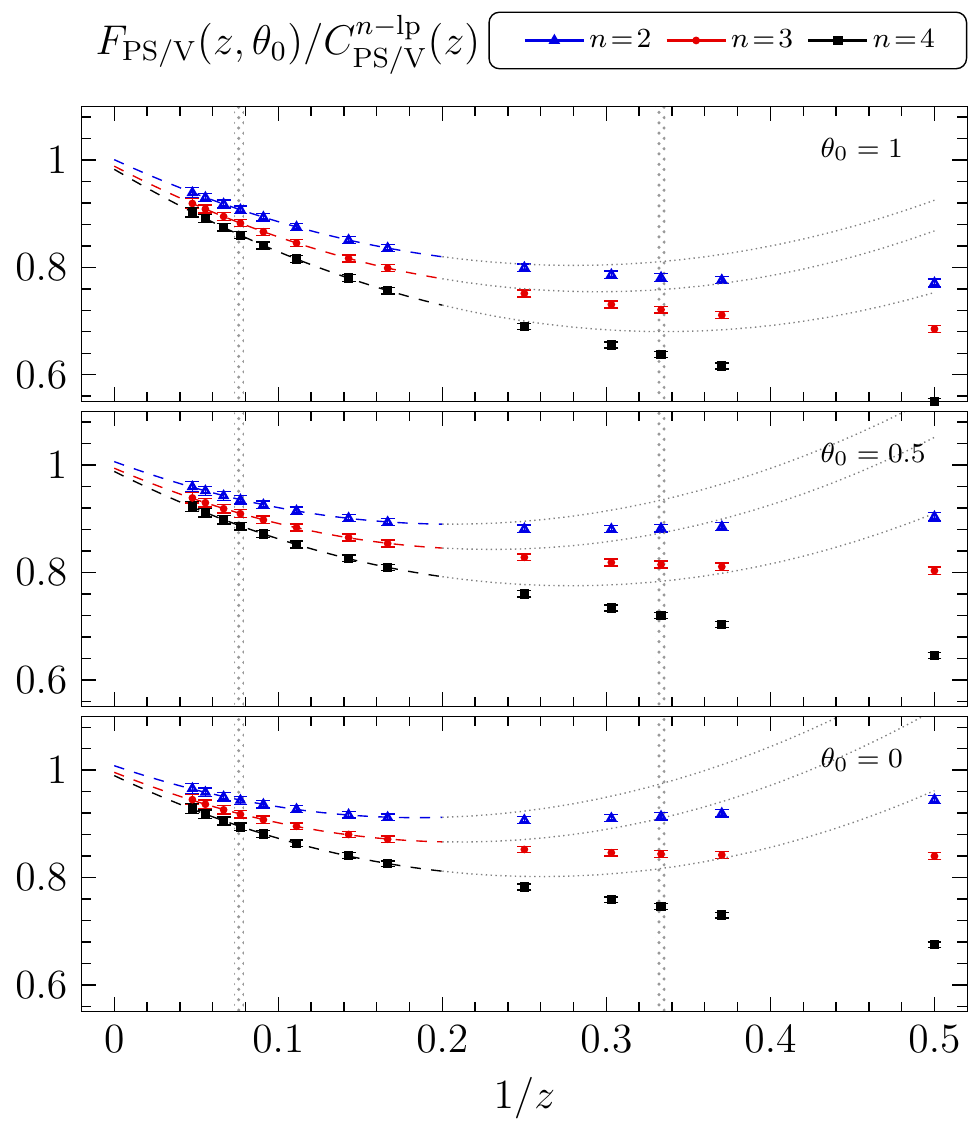}
  \vskip-0.25cm
  \caption{Asymptotic behaviour of ratios of effective decay constants at 
           different kinematical parameters involving two-, three- and
           four-loop conversion functions.}
  \label{fig:Ypsv-extrapol}
\end{figure}
Owing to the definitions in subsections~\ref{sec:eff_masses} and
\ref{sec:DecayConst}, the finite-volume continuum QCD observable
$\Ypsv=\Yps/\Yv$ approaches in the large-volume limit a combination of ratios
of pseudoscalar to vector heavy-light meson decay constants and masses, which
becomes of phenomenological relevance at the b-quark mass scale:
\begin{align}\label{eq:Ypsv-to-LV}
       \lim_{L\to\infty}\Ypsv(z,\theta_0)\big|^{z=z_{\rm b}}_{\theta_0=0}  
       &= \frac{\fB}{\fBstar}\sqrt{\frac{\mB}{\mBstar}} \;.
\end{align}
Thus it is interesting to also inspect the $1/z$-dependence of $\Ypsv$.  As
before, after extrapolating to the continuum limit and accounting for the
proper full (logarithmic) mass dependence via attaching the function $C_{\rm
PS/V}(z)$ to two-, three- and even four-loop accuracy,
cf.~eq.~\eqref{eq:Cpsv-fit}, we obtain $\Ypsv(z,\theta_0)$ and the
corresponding extrapolations to the static limit of HQET reproduced in the left
panel of figure~\ref{fig:Ypsv-extrapol}.  Again, we find that for every
$\theta_0$ the static HQET prediction ($=1$) is reached within errors, with an
almost linear approach as $1/z\to 0$ for $z>10$; its slope grows with the
flavour-twisted momentum $\theta_0$.

As discussed in appendix~\ref{sec:ConvFuncs}, the conversion formula for the
ratio of decay constants ${\fB}/{\fBstar}$, $\Cpsv$, is even known to four-loop
accuracy~\cite{Bekavac2010}, because in the entering difference of anomalous
dimensions the unknown four-loop anomalous dimensions of the currents
themselves drop out.  It was already noted in~\cite{Bekavac2010,Sommer:2010ic}
(and is also reflected in the middle-right panel of figure~\ref{fig:C_X} in
appendix~\ref{sec:ConvFuncs}) that the perturbative expansion of $\Cpsv$
exhibits a bad behaviour, since the perturbative coefficients grow further with
the loop order such that the concept of asymptotic convergence of the
perturbative series appears to be meaningful only for rather small couplings or
masses far above the mass of the b-quark. At the scale of the b-quark mass, for
instance, every known perturbative order approximately contributes by an equal
amount.%
\footnote{In~\cite{Sommer:2010ic} it was also demonstrated that a
rearrangement of the perturbative series (i.e., re-expanding the relevant
anomalous dimension function in the coupling at a different scale such as to
obtain smaller perturbative coefficients) does not lead to a substantially
more stable perturbative prediction.}
The worrying behaviour of perturbation theory for $\Cpsv$ may also be read off
from the data points in figure~\ref{fig:Ypsv-extrapol}, where in fact no signs
of an asymptotic convergence with the loop order in $\Cpsv$ from about the
b-quark mass scale on of any of the curves through the points can be stated.
This is even more so for  
\begin{align}\label{eq:Ypsv-to-LV}
        \Fpsv(z,\theta_0) &\equiv 
        \Ypsv(z,\theta_0) 
        \left({\frac{\GamPS(z,\theta_0)}{\GamV(z,\theta_0)}} \right)^{-1/2} \;, 
\end{align}
in the right panel of figure~\ref{fig:Ypsv-extrapol}, as a direct effective
finite-volume estimates of ${\fB}/{\fBstar}$ involving $\Cpsv$, too, where
$\GamPS/\GamV$ cancels the ratio of meson masses in $\Ypsv$.  The growing
deviations of the data points from the polynomial fits in $1/z$ towards the
charm quark scale for this quantity are inherited from the corresponding
behaviour of the higher-order terms in $\GamPS/\GamV$ (see previous paragraph).

\subsubsection*{\it Asymptotic behaviour of effective decay constants}
\begin{figure}[t]
  \centering
     \includegraphics[width=\textwidth]{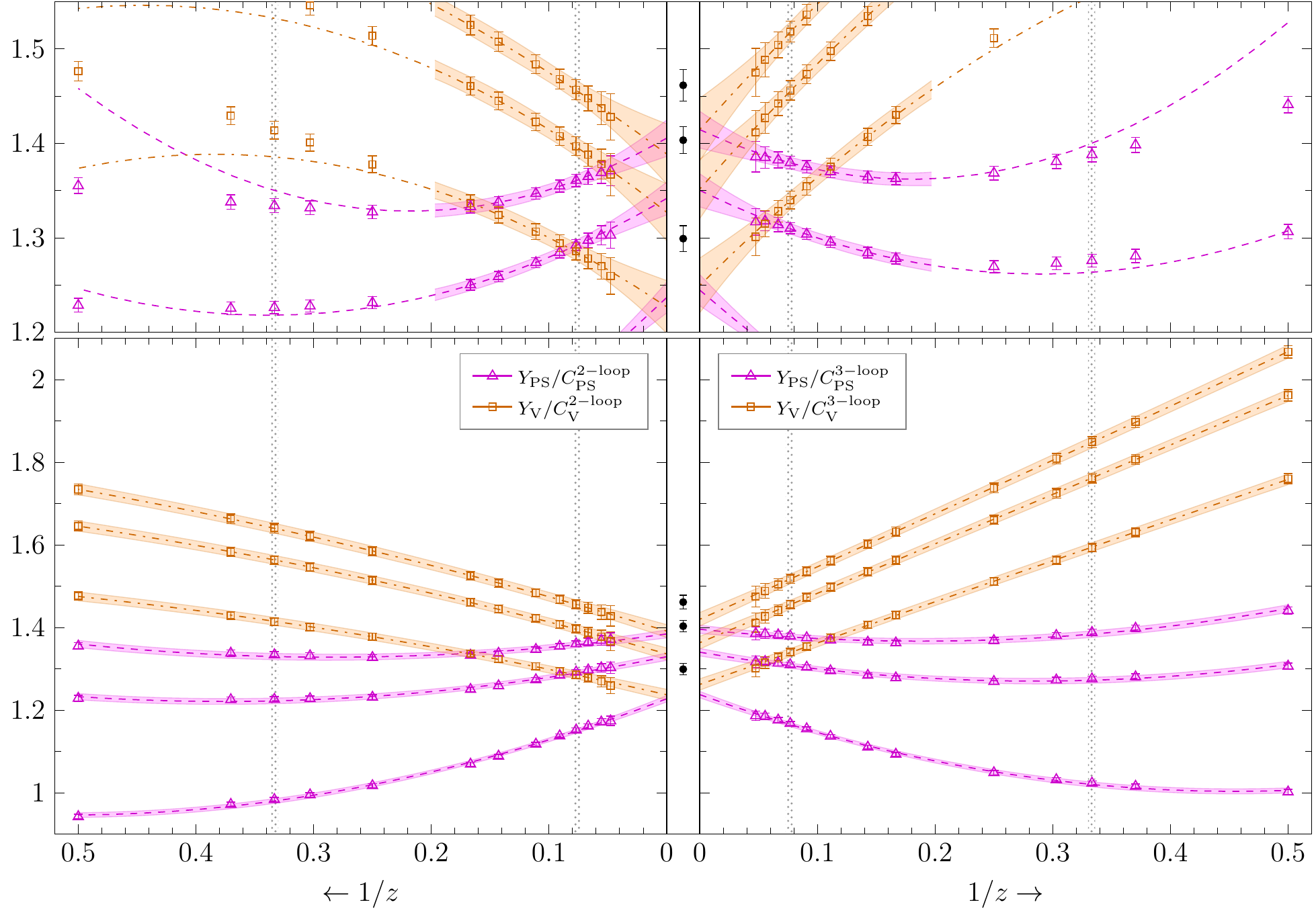}
  \vskip-0.5cm
  \caption{Comparison of the static extrapolations of $\Yps(\theta_0,z)$ and 
           $\Yv(\theta_0,z)$ to the non-perturbative HQET results in the 
           continuum (indicated as the data points in the middle bewteen the
           panels). In the left panels we use conversion functions $\Cx$,
           $\rmX={\rm PS},{\rm V}$, evaluated at two-loop, while in the
           right panels they are evaluated at three-loop order of
           perturbation theory.
           Additionally, the lower panels show (weighted) quadratic fits to
           all data points, while the upper panels only include data points
           in the expected applicability domain of HQET ($1/z<0.2$).
           A linear extrapolation over the further restricted range 
           $1/z\lesssim 0.13$ would lead to compatible results.}
  \label{fig:decay-const-vs-hqet}
\end{figure}
\begin{table}[t]
        \small
        \centering
        \begin{tabular}{CCCCCCCCCCC}\toprule
                   & \multicolumn{2}{c}{\text{static limit in QCD}} & {\text{static HQET}} 
                   & \multicolumn{2}{c}{\text{static limit in QCD}}                        \\
                   & \multicolumn{2}{c}{for $1/z<0.2$} & & \multicolumn{2}{c}{for $1/z\le 0.5$}  \\\cmidrule(lr){2-3}\cmidrule(lr){4-4}\cmidrule(lr){5-6}
         \theta_0  & \Yps^{[0]} & \Yv^{[0]}  &  X_{\rm RGI} & \Yps^{[0]} & \Yv^{[0]}       \\\cmidrule(lr){1-1}\cmidrule(lr){2-3}\cmidrule(lr){4-4}\cmidrule(lr){5-6}
           0.0     & 1.413(20) & 1.413(33) & 1.461(17)    & 1.3978(57) & 1.4206(50)      \\           
           0.5     & 1.348(18) & 1.350(32) & 1.403(11)    & 1.3427(48) & 1.3642(45)      \\            
           1.0     & 1.242(16) & 1.248(29) & 1.299(14)    & 1.2400(41) & 1.2630(45)      \\\bottomrule 
        \end{tabular}
        \caption{Selected results of static extrapolations using the 
                 three-loop $C_{\rm X}$, $\rmX={\rm PS},{\rm V}$, and the 
                 associated non-perturbative result computed in
                 static effective theory; all numbers refer to the 
                 continuum limit.}
        \label{tab:YpsYv-X}
\end{table}
We now turn to an example, where the serious concerns about the usefulness of
perturbation theory for the evaluation of the conversion functions raised in
the foregoing discussion yet becomes evident in a mismatch between the 
large- mass asymptotics on the QCD side and a non-trivial, non-perturbative 
HQET prediction itself. These are the effective finite-volume pseudoscalar and
vector meson decay constants $\Yps$ and $\Yv$, which according to
subsection~\ref{sec:HQETobs} have to obey the predictions
\begin{align}
       \frac{\Yps(\theta_0,z)}{\Cps(z)} &= \XRGI(\theta_0)+\rmO(1/z) \;, &
       \frac{\Yv(\theta_0,z)}{\Cv(z)}   &= \XRGI(\theta_0)+\rmO(1/z) \;,
\end{align}
in the asymptotic regime of $1/z\to 0$.  Recall that $\XRGI$ is the
renormalization group invariant matrix element of the static axial current,
eq.~(\ref{eq:Xrgi}), and its occurrence as the static limit of both QCD
observables is a consequence of the degeneracy of pseudoscalar and vector
channels at static order of HQET owing to the heavy-quark spin symmetry.  As
outlined around \eqref{eq:Xrgi} and in appendix~\ref{app:AxialC}, the
renormalization factors entering in $\XRGI$ were determined non-perturbatively
in the Schr\"odinger functional renormalization scheme~\cite{DellaMorte:2006sv}
so that $\XRGI$ is numerically available without perturbative uncertainties at
an overall precision of about 1\%, see table~\ref{tab:YpsYv-X}.

The comparison between the $z$-dependence of the small-volume pseudoscalar and
vector meson decay constants, together with its static extrapolations, and the
non-perturbative HQET results, after prior continuum limit extrapolations of
all individual pieces involved, is presented in
figure~\ref{fig:decay-const-vs-hqet}.  The various panels distinguish between
two- and three-loop perturbative evaluations of the conversion functions
$\Cps,\Cv$, respectively, as well as between extrapolations quadratic in $1/z$
including data with $1/z<0.2$ only and over the whole range.  
We also remark that by linear fits over the further restricted range 
$1/z\lesssim 0.13$ we arrive at compatible extrapolations.
All errors from our numerical simulations and extrapolations were taken into
account as explained earlier, but we obviously can not do so for any 
systematic error from the conversion functions because of their perturbative
nature.  
While the degeneracy of pseudoscalar and vector channels at static order is
nicely reproduced by the unconstrained fits via the coincidence of the
respective $1/z=0$ limits of $\Yps/\Cps$ and $\Yv/\Cv$, the agreement of the
extrapolation of the relativistic QCD results with the associated predictions
at static order of HQET does not look very convincing. As can be inferred from
table~\ref{tab:YpsYv-X} and figure~\ref{fig:decay-const-vs-hqet}, the results
obtained in the static effective theory (black data points in the center of the
figure) differ systematically for each value of $\theta_0$ from the results of
the static extrapolations using unconstrained quadratic fits that represent the
data very well.  Although these differences tend to decrease when going from
the two-loop to the three-loop evaluation of the $C_{\rm X}$, $\rmX={\rm
PS},{\rm V}$, the disagreement still remains at the 1$-$2$\sigma$ level of the
statistical errors.  One thus may speculate whether this is just an unfortunate
statistical effect, but given the previously discussed doubts on the
reliability of perturbation when matching the quark-mass dependent QCD results
to HQET via the conversion functions, it may also very well be attributed to
the \emph{only perturbative approximation} of $\Cps$ and $\Cv$.

In order to provide further support for the conjecture that the use of
perturbation theory for the $C_{\rm X}$ is actually responsible for the
observed disagreement between the results of the $1/z\to 0$ extrapolations of
the QCD data and the genuinely non-perturbative static HQET predictions, let us
go back to the very definition of $\XRGI$ in eq.~\eqref{eq:Xrgi} and
appendix~\ref{app:AxialC}.  It is an example for a renormalization group
invariant, which is independent of schemes and scales, allowing a clean
factorization of observables into a non-perturbative matrix element of some
composite field operator and a multiplicative matching (resp. conversion)
function that possesses a perturbative expansion.  In the situation at hand, we
always refer to the axial vector current $A_0$ in the lowest-order (i.e.,
static) effective theory, where its multiplicative renormalization factor is
\emph{not} protected against a scale dependence by a suitable axial Ward
identity, as it is the case for the axial current in QCD.  Along the lines of
ref.~\cite{Sommer:2010ic} one can express the lowest-order HQET approximation
of the QCD matrix element of $A_0$, in slight adaption of our notation
introduced before, as
\begin{align}
       \Yps(m_\ast) &= \Cps(M/\Lambda)\times\XRGI+\rmO(1/\mh) \;, \notag\\
       {\rm with}\quad
       \Cps(M/\Lambda) &= \exp\left\{\int^{g_\ast}\rmd x\; 
                          \frac{\gamma_{\rm match}(x)}{\beta(x)}\right\} \;,
       \label{eq:Cps}
\end{align}
such that at leading order in the inverse heavy quark mass, $1/\mh$, the
conversion function $\Cps$ defines a RGI-mass scaling function that contains
the full (logarithmic) mass dependence, whereas the non-perturbative matrix
element in the static effective theory, $\XRGI$, becomes a pure mass
independent number.  Here, the renormalization scale $\mu$ of the static
current is identified with $\mu=m_\ast$, where $m_\ast$ is implicitly defined
by the solution of $m_\ast=\mbar(m_\ast)$, $\mbar$ being the renormalized
(running) heavy quark mass.
The mass dependence is induced via the renormalized coupling 
$g_\ast\equiv\gbar(m_\ast)$ that is understood as a function of the ratio of
renormalization group invariants, $g_\ast=g_\ast(M/\Lambda)$, and can be 
determined in perturbation theory for any value of $M/\Lambda$ from an 
integral expression for $\Lambda/M$, analogous to eq.~\eqref{eq:Cps}, but in
terms of the beta-function and the quark mass anomalous dimension.  
Moreover, $\gamma_{\rm match}$ in~\eqref{eq:Cps} denotes
the anomalous dimension in a renormalization scheme for the static axial
current, $\Astat$, called the \emph{matching scheme}, which is defined by the
condition that --- in this scheme and at $\mu=m_\ast$ --- matrix elements of
$\Astat$ are equal to the QCD ones up to $\rmO(1/\mh)$
(cf.~appendix~\ref{sec:ConvFuncs}, and
refs.~\cite{Heitger:2003xg,Heitger:2004gb,Sommer:2010ic} for more details).%
\footnote{The particular choice of renormalization scheme for the running 
coupling and (heavy) quark mass, $\gbar,\mbar$, and the QCD $\Lambda$-parameter 
is not relevant here, but one may typically think of the $\msbar$-scheme.}
The crucial observation at this point is now that, in the
transition to renormalization group invariants leading to~\eqref{eq:Cps}, the
perturbative running enters in $\Cps(M/\Lambda)=\Yps(m_\ast)/\XRGI$ through the
only perturbative knowledge of the beta-function and the anomalous dimensions
of the quark mass and $\Astat$.  Hence, it does not come as a too big surprise
that the large-mass extrapolation of $\Yps(m_\ast)/\Cps(M/\Lambda)$, which
combines \emph{fully non-perturbative QCD results} with a \emph{perturbative
matching function}, fails to meet the expected static order HQET limit
$\XRGI=X_{\rm R}(\mu=L_1^{-1})/\big[X_{\rm R}(\mu=L_1^{-1})/\XRGI\big]$, a
\emph{fully non-perturbative} number, both factors of which are precisely known
for our setup via the data from the non-perturbative computation
in~\cite{DellaMorte:2006sv}, see eq.~\eqref{eq:Xrgi} and
appendix~\ref{app:AxialC}.  On the other hand, replacing in the calculation of
$\XRGI$ the \emph{non-perturbative} value $\XRGI/X_{\rm R}(\mu=L_1^{-1})=
Z^{\rm stat}_{\rm A,RGI}/Z^{\rm stat}_{\rm A}(\mu=L_1^{-1})=0.875(7)$
of~\eqref{eq:Xrgi} by the corresponding value at $\mu=L_1^{-1}$ obtained from
\emph{perturbative} running (using the available three-loop beta-function and
two-loop anomalous dimension of $\Astat$ in the SF renormalization scheme), the
black HQET data points in the center of figure~\ref{fig:decay-const-vs-hqet}
receive a downward shift of about 5\% to finally coincide with the polynomial
extrapolation results of $Y_{\rm X}/C_{\rm X}$, $\rmX={\rm PS},{\rm V}$, in the
static limit.  This finding clearly suggests that the quark mass dependence of
the conversion function $\Cps$, which employs inputs from the
\emph{perturbative} matching between HQET and QCD only, is then also only
enough to reproduce the \emph{perturbative} prediction for the matrix elements
in the static effective theory.  However, it is not able to reproduce the
non-perturbative result, since it does not comprise the full non-perturbative
mass dependence that is required for a fully consistent matching between HQET
and QCD. 

All in all we therefore conclude that with only perturbative knowledge of the
conversion functions one is not automatically guaranteed to extrapolate
relativistic QCD data at finite values of the (heavy) quark mass to the correct
static limit of HQET as $z\to\infty$ (resp. $M\to\infty$).  In fact, $\Cps$
(and $\Cv$) at three-loop accuracy do not seem to be qualified for use in
conjunction with non-perturbative QCD results on the small-volume meson decay
constants to extrapolate their heavy quark mass dependence to the genuinely
non-perturbative prediction of the effective theory.  The distinct mismatch
between the QCD extrapolations and the expected HQET results in
figure~\ref{fig:decay-const-vs-hqet} clearly illustrates this.
A possible reason for the use of three-loop conversion functions such as $\Cps$
not to correctly recover the expected static limit can be traced back to the
not that well behaved perturbative expansion of the underlying anomalous
dimension $\gamma_{\rm match}$ in the aforementioned matching scheme, which
enters in $\Cps$ in addition to the well behaved perturbative series of the
beta-function and the quark mass anomalous dimension.  Although the overall
mass dependence of, say, $\Cps$ in figure~\ref{fig:C_X} of
appendix~\ref{sec:ConvFuncs} evaluated for different perturbative orders looks
quite innocent, a more careful estimation of the coefficients of $\gamma_{\rm
match}$ for the different known orders (up to three loops) still gives rise to
worries about neglecting higher-order terms at values of the renormalized
coupling around the b-quark mass scale or even below~\cite{Sommer:2010ic}.

Our observations on the small-volume decay constants should also be taken as a
warning that the method of extracting, e.g., the B-meson decay constant via an
interpolation of large-volume lattice data in the charm region and HQET data in
the static limit (as sometimes adopted when applying lattice QCD to B-physics
phenomenology) can easily yield misleading results, as long as only
perturbation theory is employed in the matching step of relating the HQET
numbers to the quark mass-dependent ones in QCD.

In general, of course, these statements on the influence of the conversion
functions on the validity of HQET as an effective theory of QCD may depend on
the individual observable in question and should rather be investigated on a
case-by-case basis as we do it in the present study.  For instance, in the
discussion of the meson masses and the heavy-light current ratios above we have
already seen that their asymptotic $1/z\to 0$ behaviour together with the
perturbative conversion functions meets the corresponding HQET predictions very
well.  Even more convincing tests of the effective theory will follow shortly
in subsection~\ref{sec:No-CX}, when we come to consider observables in which
perturbative factors such as $\Cps,\Cv$ drop out completely. 

\subsubsection*{\it Asymptotic behaviour of the spin splitting}
\begin{figure}[t]
        \small
     \centering
     \includegraphics[width=0.9\textwidth]{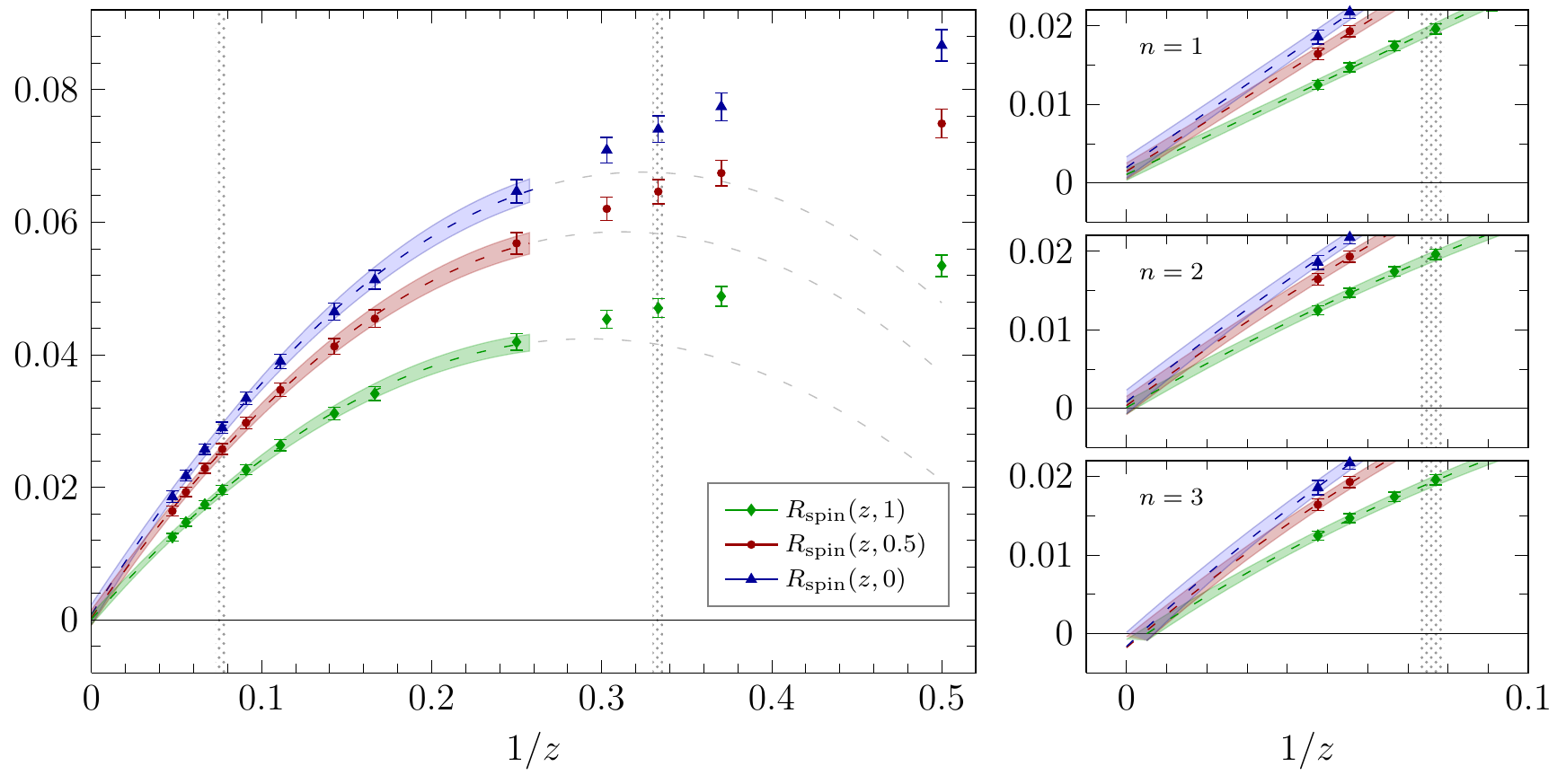}\hfill
     \vskip-1em
     \caption{Asymptotic behaviour of the spin splitting observable 
              $R_{\rm spin}(z,\theta_0)$ together with a quadratic
              unconstrained (and weighted) fit including data at $1/z\le 1/4$.
              In the right panels we compare the static extrapolation in
              the asymptotic scaling region for a linear ($n=1$),
              quadratic ($n=2$) and cubic ($n=3$) polynomial fit ansatz.
             }
     \label{fig:Rspin}
\end{figure}
As can be inferred from figure~\ref{fig:Rspin}, our spin splitting observable
$R_{\rm spin}(z,\theta_0)$ shows the expected asymptotic behaviour towards the
$1/z\to 0$ limit, where it has to vanish due to the heavy-quark spin symmetry.
Opposed to the case of the static axial current, data for the corresponding
RGI matrix element, $X^{\rm spin}_{\rm RGI}$, are not available for the
two-flavour theory. Hence, we refrain from studying the static limit of our
data in the form of applying eq.~\eqref{eq:global-fitfunc-Cx} via
\eqref{eq:asym-spin} to $R_{\rm spin}/\Cspin$, as we did for the axial and
vector meson decay constants in the previous paragraph.  Rather, we studied
different static extrapolations using a linear ($n=1$, $z\ge 13$),
quadratic ($n=2$, $z\ge 4$) and cubic ($n=3$, $z\ge 3$) fit ansatz for the
static extrapolation.  They are presented for better visibility in the
asymptotic region only (right panel of figure~\ref{fig:Rspin}). The left panel
shows the $n=2$ case with all available data points; the fit ansaetze are
able to describe the data very well, and its behaviour confirms the
HQET expectation.

\subsection{Results without perturbative conversion functions}\label{sec:No-CX}
\begin{figure}
  \small
  \centering
  \includegraphics[width=\textwidth]{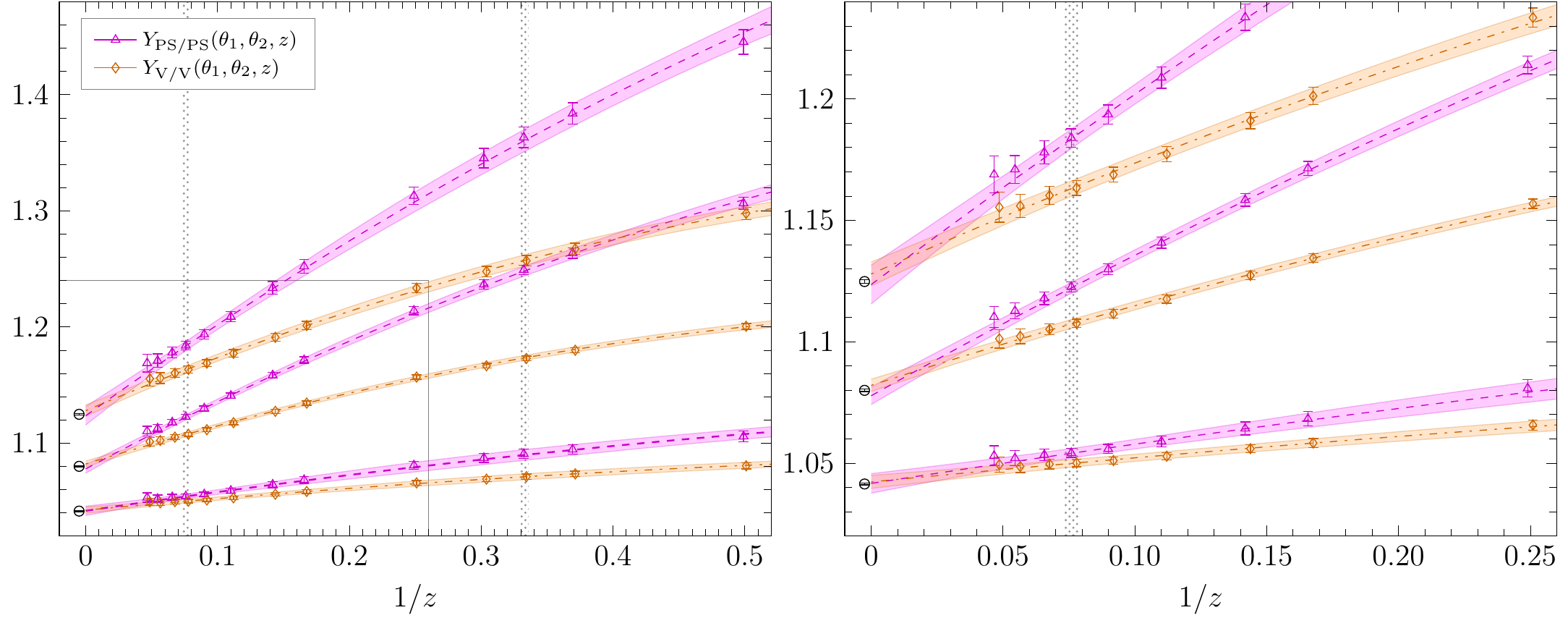}
  \includegraphics[width=0.5\textwidth]{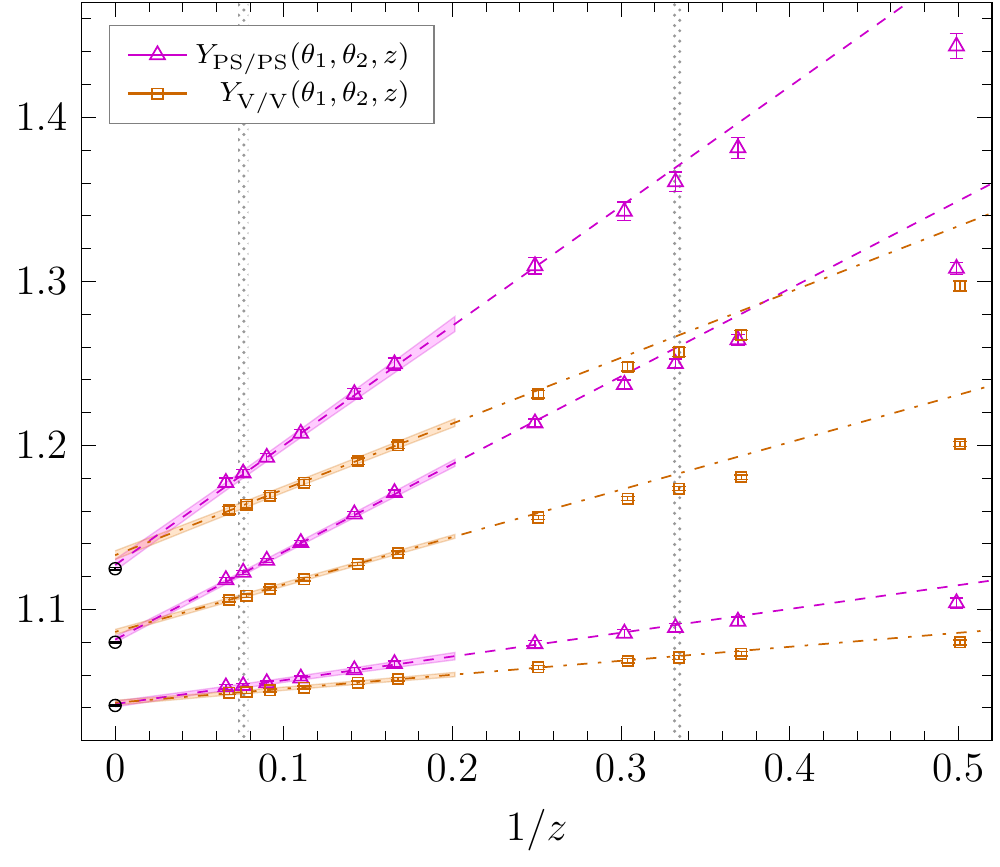}
  \vskip-1em
  \caption{Extrapolations of $Y_{\rm PS/PS}(z,\theta_1,\theta_2)$ and 
           $Y_{\rm V/V}(z,\theta_1,\theta_2)$ to the static limit for all three
           combinations of $(\theta_1,\theta_2)$.
           The results are listed in table~\ref{tab:stat-decay-rat}. 
           The right panel shows a scaled excerpt of the heavy quark mass
           region.
           Black data points indicate the continuum results of the
           corresponding quantity at static order of HQET. 
           Its continuum extrapolation is displayed in
           figure~\ref{fig:CL-HQET}. 
           For comparison, the lower panel displays an extrapolation with
           just a linear function in $1/z$ for $1/z\leq 0.2$ that leads to
           an equally well confirmation of the HQET expectation.}
  \label{fig:extrapol-decay-const}
\end{figure}

\begin{table}
        \small
        \centering
        \begin{tabular}{CCCCCCCCCCC}\toprule
                & \multicolumn{2}{c}{\text{static limit in QCD}} & {\text{static HQET}} 
                & \multicolumn{2}{c}{\text{static limit in QCD}} & \text{static HQET} \\\cmidrule(lr){2-3}\cmidrule(lr){4-4}\cmidrule(lr){5-6}\cmidrule(lr){7-7}
  (\theta_1,\theta_2) &    \Ypsps  &  \Yvv  & R_X^{\rm stat}  &    R_f    &  R_k      &  R_f^{\rm stat}  \\\cmidrule(lr){1-1}\cmidrule(lr){2-4}\cmidrule(lr){5-7}
           (0,0.5)    & 1.0417(20) & 1.0424(15) & 1.0414(4)   & 1.222(11) & 1.220(11) & 1.2245(16)        \\
           (0.5,1)    & 1.0796(22) & 1.0841(16) & 1.0800(5)   & 1.760(21) & 1.759(21) & 1.7234(34)        \\
           (0.0,1)    & 1.1247(42) & 1.1300(29) & 1.1248(8)   & 2.143(46) & 2.142(46) & 2.1107(64)        \\\bottomrule
        \end{tabular}
        \caption{Results of static extrapolations of decay constant ratios 
                 in continuum QCD and its corresponding non-perturbative 
                 continuum extrapolation results in the static effective 
                 theory.}
        \label{tab:stat-decay-rat}
\end{table}

\begin{figure}
  \centering
  \small
  \includegraphics[width=\textwidth]{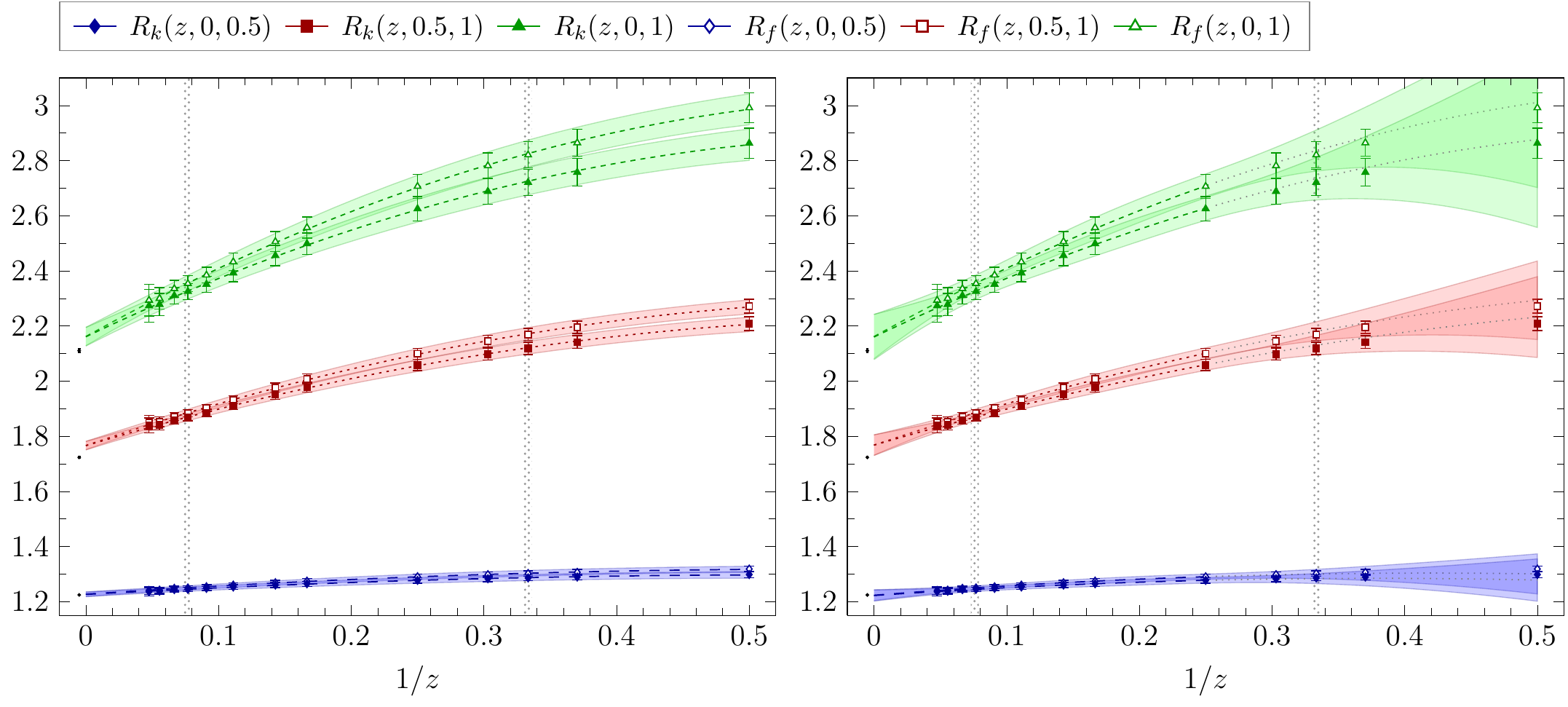}
  \vskip-1em
  \caption{Static extrapolations of $R_f(z,\theta_1,\theta_2)$ and 
           $R_k(z,\theta_1,\theta_2)$ for all three combinations of
           $(\theta_1,\theta_2)$.
           In the left panel, all data points enter the static 
           extrapolations, whereas in the right panel only those with
           $1/z\le 1/4$ contribute.
           The results are listed in table~\ref{tab:stat-decay-rat}. 
           Black data points indicate the continuum results of the
           corresponding quantity at static order of HQET, 
           cf.~eq.~\eqref{eq:stat-extrap2}.
           }
  \label{fig:extrapol-f1k1-ratio}
\end{figure}

We now turn our attention to quantities that do not depend on any conversion
functions and as such are free of any influence of perturbative uncertainties;
they thus are expected to exhibit an unambiguous static extrapolation
compatible with their heavy quark mass expansion. In most cases we find that a
fit function such as
\begin{align}
        \Omega^{\rm QCD}(L,M,0) &= 
        \upsilon_0 + \upsilon_1 z^{-1} + \upsilon_2 z^{-2}     
        \label{eq:global-fitfunc}
\end{align}
models their $z$-dependence in the continuum limit very well over the whole
region of available heavy quark mass values.

\subsubsection*{\it Ratios of currents}
As prototype observables, we first consider and probe 
eq.~\eqref{eq:stat-extrap2-C}, viz.
\begin{align}
       \Ypsps(z,\theta_1,\theta_2) &= R_{X}^{\rm stat}(\theta_1,\theta_2)
                                      +\rmO(1/z) \;, \notag\\
       \Yvv(z,\theta_1,\theta_2)   &= R_{X}^{\rm stat}(\theta_1,\theta_2)
                                      +\rmO(1/z) \;,
\end{align}
where --- again owing to the heavy-quark spin symmetry --- the static
extrapolation of the ratio of effective vector current matrix elements computed
at different kinematical parameters, $\Yvv(z,\theta_1,\theta_2)$, is expected
to agree in the limit $1/z\to 0$ with the associated ratio in the pseudoscalar
channel, $\Ypsps(z,\theta_1,\theta_2)$, \emph{and} to approach the common,
non-trivial leading-order HQET prediction $R_{X}^{\rm stat}(\theta_1,\theta_2)$,
which denotes the corresponding ratio of matrix
elements computed by replacing the relativistic fields by the static ones.

Performing an unconstrained extrapolating fit of all data points
($z\ge 2$), according to the fit ansatz~\eqref{eq:global-fitfunc}, gives an
asymptotic behaviour as depicted in figure~\ref{fig:extrapol-decay-const} for
the three available $\theta$-combinations.  The black circles (slightly moved
to the left towards negative $1/z$ for ease of presentation) represent the
results for the ratio of static-order HQET matrix elements, 
$R_{X}^{\rm stat}(\theta_1,\theta_2)$.
As usual throughout this work, all data points were
extrapolated to the continuum limit first; in particular, the continuum
extrapolation of $R_{X}^{\rm stat}(\theta_1,\theta_2)$ is presented in the left
panel of figure~\ref{fig:CL-HQET} and its numerical values can be read off
together with the results of the static extrapolations from
table~\ref{tab:stat-decay-rat}.

In contrast to the decay constant discussed in the previous subsection, the
comparison in figure~\ref{fig:extrapol-decay-const} illustrates a very good
agreement between the static extrapolations of the results on $\Yvv,\Ypsps$ and
the numbers for $R_{X}^{\rm stat}$, computed to even much higher statistical
accuracy directly in the static approximation.  
Moreover, from the fits in the lower panel of the figure one reads off an
equally excellent agreement with the HQET expectation from an extrapolation
with just a linear function in $1/z$ for $z\geq 5$ ($1/z\leq 0.2$).
These findings not only demonstrate the correctness of the effective theory
itself, but also strongly advocate such observables, which are not affected
by any perturbative imperfections deriving from conversion functions,
as most promising candidates for observables for use
in a non-perturbative strategy for matching HQET to QCD in the spirit
of~\cite{Heitger:2003nj}.  Two additional observations worth to be mentioned
are: a) the error for all three combinations of $(\theta_1,\theta_2)$ grows
with increasing $\theta_2^2-\theta_1^2$ in a similar way for all three
observables; b) also the $1/z$-terms, i.e., the slope at $1/z=0$, grows with
that difference.  These features also hold true for the quantities considered
next and in principle can serve as further helpful criteria for a sensible
choice of matching observables.

\subsubsection*{\it Ratios of boundary-to-boundary matrix elements}
At next, we look at the static extrapolation of the quantities entering the
prediction~\eqref{eq:stat-extrap2-A}.  Compared to observables studied before,
the QCD data points obtained in the vector ($R_k$) and pseudoscalar channel
($R_f$) lie quite close to each other already at finite quark mass.  From
figure~\ref{fig:extrapol-f1k1-ratio} (left panel) one concludes that 
superficially the quadratic fit ansatz~\eqref{eq:global-fitfunc} very well
represents the data points, which approach a (due to spin symmetry) common HQET
limit for the three $\theta$-combinations.  Note that the static HQET results
for $R_k$ and $R_f$, independently computed in the continuum limit from
different simulations, have not been constrained to be equal, but their
agreement is just excellent though.  However, only the extrapolation for the
combination $(\theta_1,\theta_2)=(0,0.5)$ leads to the correct result in the
static limit, $R_k=R_f$, represented in the figure by the leftmost black data
points.  The results for this particular static extrapolation and the
corresponding HQET results are given in table~\ref{tab:stat-decay-rat}. 

This partial mismatch of the results from extrapolations of the QCD data
over the whole available $z$-range down to $z=2$ ($1/z=0.5$) with the HQET
predictions should be taken as a warning that the validity of the
HQET-inspired $1/\mh$-expansion of heavy-light QCD obeservables in the 
large-mass regime can not be tacitly trusted down to the charm quark scale
or below.
In fact, if we restrict the fits to include data points for $1/z<0.26$ only,
as done in the previous section, we obtain a static extrapolation as shown
in the right panel of figure~\ref{fig:extrapol-f1k1-ratio}, where the static
numbers can be seen to be covered by the error of the extrapolation results.
An even further restriction of the fit interval to $1/z<0.1$ would easily
allow for a feasible linear interpolating fit including the HQET result as a
constraint.
Here, extending those fits to the lower $z$'s that were excluded yields only
mild deviations of about 1$\sigma$ between the fit functions and the data
points, though, but this may also be more pronounced for other observables.
Turning the argument around, an interpolating fit somewhat arbitrarily
extended, at fixed polynomial degree, to include data below the charm region
($z=2$), which appears to give a good description of the data, can lead to
an underestimation of the error of the static extrapolation and thereby 
pretend to miss the HQET prediction.

\begin{figure}
  \small
  \centering
  \includegraphics[width=0.95\textwidth]{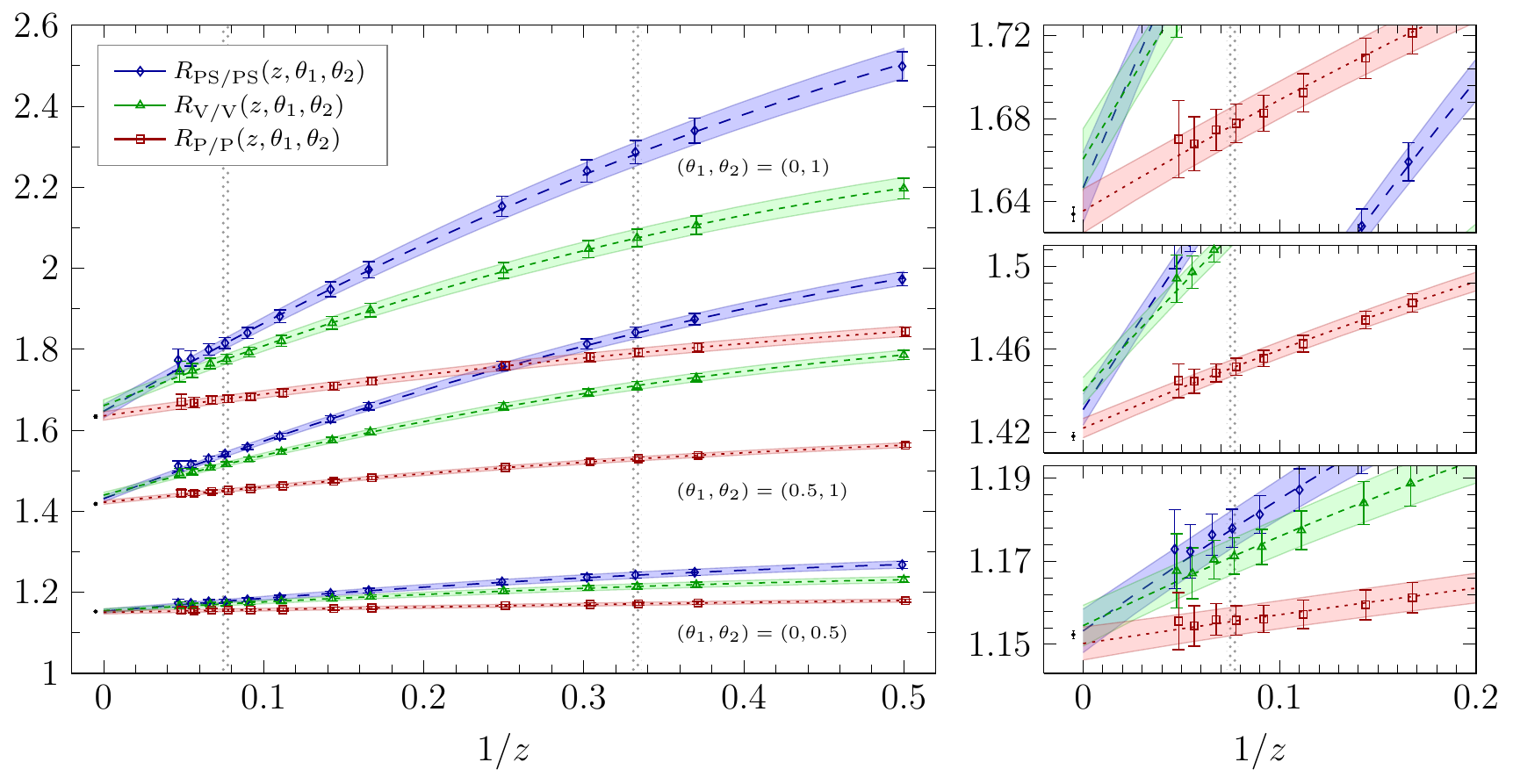}
  \vskip-1em
  \caption{Static extrapolations of $R_{\rm PS/PS}(z,\theta_1,\theta_2)$, 
           $R_{\rm V/V}(z,\theta_1,\theta_2)$ and 
           $R_{\rm P/P}(z,\theta_1,\theta_2)$ for all three combinations of 
           $(\theta_1,\theta_2)$ in comparison to their HQET counterparts
           after taking the continuum limit in the static effective theory 
           (these data points are slightly shifted for better readability).
           The results are listed in table~\ref{tab:stat-decay-rat}. 
           The right panel shows a scaled excerpt of the heavy quark mass
           region individually for the three available 
           $(\theta_1,\theta_2)$-combinations such that error bars of the
           HQET results become visible.
           }
  \label{fig:extrapol-curr-ratio}
\end{figure}
\begin{table}
        \small
        \centering
        \begin{tabular}{CCCCCCCCCCC}\toprule
                & \multicolumn{3}{c}{\text{static limit in QCD}} & {\text{static HQET}}\\\cmidrule(lr){2-4}\cmidrule(lr){5-5}
  (\theta_1,\theta_2) &    \Rpsps  &  \Rvv      &  \Rpp      & R_{\rm PS}^{\rm stat}   \\\cmidrule(lr){1-1}\cmidrule(lr){2-5}
           (0,0.5)    & 1.1532(52) & 1.1545(49) & 1.1502(40) & 1.1523(9)               \\ 
           (0.5,1)    & 1.4309(76) & 1.4399(65) & 1.4219(47) & 1.4180(18)              \\ 
           (0.0,1)    & 1.647(17)  & 1.660(15)  & 1.635(11)  & 1.6339(34)              \\\bottomrule
        \end{tabular}
        \caption{Results for static extrapolations of $\Rpsps$, $\Rvv$ and 
                 $\Rpp$ in continuum QCD as displayed in 
                 figure~\ref{fig:extrapol-curr-ratio} and its corresponding
                 non-perturbative continuum extrapolation results in the
                 static effective theory.}
        \label{tab:stat-ratios}
\end{table}

\subsubsection*{\it Ratios of boundary-to-bulk matrix elements}
Finally, we check the static extrapolation of~eq.~\eqref{eq:stat-extrap2-B}.
The HQET results for $R_{\rm PS}^{\rm stat}(\theta_1,\theta_2)$ follow from the
continuum extrapolation presented in figure~\ref{fig:CL-HQET}. 
Their values are given in table~\ref{tab:stat-ratios}, together with the 
results that stem from a static extrapolation of the (continuum) QCD
observables $\Rpsps$, $\Rvv$ and $\Rpp$ as defined in
eq.~\eqref{eq:all-good-obs}. 
Once more, the data sets themselves are very well represented by a quadratic
polynomial fit ansatz over the whole range of available data points,
whereas the static HQET result (common to all three QCD observables,
again due to spin symmetry) is only met for the $\theta$-combination 
$(\theta_1,\theta_2)=(0,0.5)$ as in the foregoing case of ratios of
boundary-to-boundary matrix elements.
Accordingly, the same discussion (and warning) carries over literally here,
except for $\Rpp$ that very well extrapolates to the HQET result for all 
$\theta$-combinations studied.
Since the errors on the data points for $\Rpp$ stay roughly the same 
when going from the scale of the b-quark mass down to $z=2$, the slight
mismatch in the other cases may likely be attributed partly to statistical
effects.

\section{Conclusions}\label{sec:conclude}
We have studied the asymptotic large-mass behaviour of heavy-light meson
observables in lattice QCD with two massless dynamical sea quarks, in a 
small volume of linear extent $L_1\approx 0.4\,\fm$ and for heavy quark mass
values within a range from beyond the b- to below the c-quark scale, 
in order to confront them with their HQET predictions.

Having taken the continuum limit in all parts of our entirely
non-perturbative calculations on both the QCD and the HQET side
and subsequently performed unconstrained static extrapolations along the
limit $z=L_1\Mh\to\infty$, in most cases we generically find (within the
numerical precision on our results with errors at the few-\% level) a very
satisfactory --- sometimes an even excellent --- agreement between the
large-mass asymptotics of the QCD observables and their expected 
leading-order HQET limits.

Moreover, the overall quality of the polynomial fits in $1/z$ to the heavy
quark mass dependence of the (continuum) heavy-light QCD observables 
convincingly demonstrates that the theory is very well described by simple
$1/\mh$-corrections to the static limit of the effective theory.
In particular, for $1/z\lesssim 0.1$ our extrapolating fits to the HQET
predictions can consistently be modeled by functions linear in $1/z$.
We are thus led to conclude that the effective theory is very well tested
and that the regime with $1/z\lesssim 0.1$, which is a key ingredient to the
finite-volume matching part of the ALPHA Collaboration's B-physics
programme based on HQET non-perturbatively renormalized and matched to QCD
at $\rmO(1/\mh)$~\cite{Heitger:2003nj,Blossier2010,Blossier2010a,Blossier:2010mk,Blossier:2012qu,Bernardoni:2013xba,Bernardoni:2014fva,Bernardoni:2015nqa},
lies very well within the applicability domain of HQET.
This is, for instance, also in line with the onset of the linear behaviour
reported in the tree-level study~\cite{DellaMorte:2013ega} of the 
$1/z$-dependence of the HQET parameters contributing to the full set of
heavy-light flavour currents in HQET at $\rmO(1/\mh)$.
Alltogether, these findings are very reassuring, since they imply that in
the finite-volume setup of the aforementioned non-perturbative matching
strategy between HQET and QCD, higher-order corrections beyond the 
$\rmO(1/\mh)$ ones already included can be expected to be suppressed by a
factor of about~$10$.

A prominent exception to this favourable outcome of our tests of HQET
consists in the large-mass asymptotics of the small-volume heavy-light
axial vector and vector meson decay constants, which fails to meet the
leading-order HQET prediction in the static limit.
But, as argued in section~\ref{sec:CX}, rather than interpreting this as
an inherent shortcoming of HQET being not an appropriate and predictive
effective theory of QCD, one should take this as an advice that it is
generally safer \emph{not} to use conversion functions such as $\Cps,\Cv$
(which inevitably enter in a consistent comparison of QCD and HQET decay
constants) from \emph{perturbation theory}, even if they are determined at
three-loop accuracy, i.e., with relative errors of order $\gbar^6(m_\ast)$ 
at some intrinsic mass scale $m_\ast$; e.g., their perturbative convergence
appears to be relatively poor still at the b-quark scale.
Otherwise, the combination of non-perturbative QCD data with the 
mass dependence via perturbative matching functions (encoding the running 
in HQET) to recover the correct HQET limit can easily lead to
inconsistencies between QCD and non-perturbatively computed matrix elements
in the effective theory.
For the decay constant, we encounter a systematic effect of up to
$\sim 5\%$ from relying on perturbative running in the effective theory. 
Therefore, we consider this as a warning when, e.g., the physical 
(i.e., large-volume) B-meson decay constant is being extracted involving 
interpolations between QCD data below the b-scale and the static limit,
and advocate to perform the matching entirely \emph{non-perturbatively}.

As already noted earlier in~\cite{Sommer:2010ic}, a more quantitative
understanding of this deficiency can be gained from the relative error that
results from a truncation of the perturbative matching expression at $l$-loop
order, viz.
\begin{align}
{\Delta(\Cps)\over\Cps} &\;\propto\;
\big[\gbar^2(m_\ast)\big]^{l}\;\sim\;
\left[{1\over 2b_0\ln\big(m_\ast/\Lambda\big)}\right]^{l}
\;\;\stackrel{m_\ast\gg\Lambda}{\gg}\;\;
{\Lambda\over m_\ast} \;.
\end{align}
I.e., since this perturbative uncertainty only decreases logarithmically
as $m_\ast$ becomes large, at some point it starts to dominate over the
power correction that one needs to include at next-to-leading order in
the HQET expansion for precision physics at the b-quark mass scale.
This underlines once more that with an only perturbative conversion
function, a consistent next-to-leading order expansion with errors 
decreasing as $1/\mh^2$ can not be achieved. 

Another interesting aspect of our work is that the continuum extrapolations
of the HQET observables, such as those presented in 
figure~\ref{fig:CL-HQET} of section~\ref{sec:rep-CL} for two discretizations
of the static quark, provide strong numerical evidence that an universal
continuum limit of the static effective theory exists and that the
non-perturbative renormalizability of HQET along the finite-volume matching
strategy of ref.~\cite{Heitger:2003nj} can be established indeed.

Finally, let us emphasize that exploring the size of the higher-order 
corrections in our QCD observables to test HQET non-perturbatively may also
readily serve as a guide so single out preferred choices among observables,
suitable for a specific HQET-QCD matching problem in question, that have only
small $\rmO(1/\mh^2)$ contributions. 
In addition to that, any flexibility in having different matching equations
made of different observables to determine the same (set of) HQET parameters
enables further useful checks in actual computations, because the final 
results should be independent of any specific but sensible choice of
matching equations and kinematical parameters (such as $T/L$, $x_0$ and the
$\theta$'s) entering them, up to small $\rmO(1/\mh^2)$ corrections.  

\acknowledgments
%
%
We are grateful to M.~Della~Morte, D.~Hesse and M.~Serritiello for
providing the tree-level improvement coefficients entering into some of our
results.
We also would like to thank R.~Sommer for fruitful discussions and a critical 
reading of an earlier version of this manuscript.
This work is part of the ALPHA Collaboration research programme.
We thank NIC/DESY, INFN and the ZIV of the University of M\"{u}nster for
allocating computer time on the APE computers, the PAX Cluster and the PALMA 
HPC Cluster to this project, respectively, as well as the staff of the 
computer centers at Zeuthen, Rome and M\"{u}nster for their support.
We acknowledge partial support by the European Community through 
EU Contract No.~MRTN-CT-2006-035482, ``FLAVIAnet''.
This work was also supported by the grants HE~4517/2-1 (P.~F. and J.~H.)
and HE~4517/3-1 (J.~H.) of the Deutsche Forschungsgemeinschaft.
N.~G. is funded by the Leverhulme Trust, research grant RPG-2014-118.
Finally, P.~F. thanks the organizers and participants of the workshop
``High-precision QCD at low energy'' at the
``Centro de Ciencias de Benasque Pedro Pascual'' for stimulating discussions
and comments.

\appendix
\section{Definitions}\label{app:defs}
Here we provide the definitions of the observables of this study in terms of
the traditional notation for Schr\"odinger functional (SF) correlation 
functions.
We do not repeat all details here; for explicit expression for SF heavy-light
meson correlators in lattice QCD as well as in the static limit of lattice 
HQET, the reader may, e.g., 
consult~\cite{Heitger:2003nj,DellaMorte:2006cb,Kurth:2000ki,DellaMorte:2013ega}.  
\subsection*{QCD observables}
\noindent%
For the effective masses we use SF correlation functions $f_{\rm X}$ with
bulk insertions of $\Or(a)$ improved heavy-light currents
(cf.~section~\ref{sec:obs}) in the pseudoscalar and vector channel and 
define
\begin{align}
   \GamPS &= \left. -\drvsym0 \ln \fAimpr(x_0) \right|_{x_0=\tfrac{T}{2}}  \;, &
   \GamP  &= \left. -\drvsym0 \ln \fP(x_0)     \right|_{x_0=\tfrac{T}{2}}  \;, \notag\\ 
   \GamV  &= \left. -\drvsym0 \ln \kVimpr(x_0) \right|_{x_0=\tfrac{T}{2}}  \;.
\end{align}
Effective decay constants associated with these correlators as well as
suitable ratios thereof require renormalization and are given by
\begin{align}
\label{eq:all-Y}
  \Yps  &\equiv + \za (1+ \tfrac{1}{2}\ba am_{\rm q,h}) \dfrac{\fAimpr(T/2)}{\sqrt{f_1}}  \;, & 
  \Yv   &\equiv - \zv (1+ \tfrac{1}{2}\bv am_{\rm q,h}) \dfrac{\kVimpr(T/2)}{\sqrt{k_1}} \;, \\[0.5em]
\label{eq:all-R}
  \Rpsp &\equiv - \dfrac{\za (1+ \tfrac{1}{2}\ba am_{\rm q,h})}{Z_{{\rm P},{\rm RGI}} (1+ \tfrac{1}{2}\bp am_{\rm q,h})} \dfrac{ \fAimpr(T/2) }{ \fP(T/2) } \;, &
  \Rpsv &\equiv - \dfrac{\za (1+ \tfrac{1}{2}\ba am_{\rm q,h})}{\zv (1+ \tfrac{1}{2}\bv am_{\rm q,h})} \dfrac{ \fAimpr(T/2) }{ \kVimpr(T/2)}  \;.
\end{align}
The renormalization constants $Z_{\rm X}$, $\rmX={\rm A},{\rm V},{\rm P}$, are 
known non-perturbatively in two-flavour QCD;
as in our earlier work~\cite{Fritzsch:2010aw}, $\za$ and $\zv$ have been 
taken from~\cite{DellaMorte:2008xb}, while $Z_{{\rm P},{\rm RGI}}$ is available
through the scale dependent renormalization factor $\zp$ computed 
in~refs.~\cite{DellaMorte:2005kg,Blossier:2012qu}. 
The improvement coefficients $b_{\rm X}$ (multiplying the bare subtracted heavy
quark mass) are known to one-loop order of perturbation theory and can be
found in~\cite{Sint:1997jx}.

The spin splitting observable, however, is constructed from a ratio of SF
boundary-to-boundary correlators with pseudoscalar and vector channel 
composite fields, free of any improvement coefficient and renormalization 
factor:
\begin{align}
  \Rspin &\equiv \tfrac{3}{4}\ln \left[ \fone/ \kone  \right] \;.
\end{align}

As already explained in the main text, the foregoing finite-volume observables 
have been computed for one fermionic phase angle out of 
$\theta_0\in\{0,0.5,1\}$. 
To enlarge the variety in probing QCD and HQET at different kinematics,
it remains to specify the observables that depend on two such angles.
Here we build them from SF correlation functions with different
$\theta$'s, i.e.,
\begin{align} \label{eq:kinQ-ps}
        \Rpsps(\theta_1,\theta_2) &= \left.\frac{ \fAimpr(x_0,\theta_1) }{ \fAimpr(x_0,\theta_2) }\right|_{x_0=T/2}  \;,   &
        R_f   (\theta_1,\theta_2) &= \frac{ \fone(\theta_1) }{ \fone(\theta_2) }                                     \;, \notag \\[0.5em]
        \Rvv  (\theta_1,\theta_2) &= \left.\frac{ \kVimpr(x_0,\theta_1) }{ \kVimpr(x_0,\theta_2) }\right|_{x_0=T/2}  \;,   &
        R_k   (\theta_1,\theta_2) &= \frac{ \kone(\theta_1) }{ \kone(\theta_2) }                                     \;, \notag \\[0.5em]
        \Rpp  (\theta_1,\theta_2) &= \left.\frac{ \fP(x_0,\theta_1)     }{ \fP(x_0,\theta_2)  }\right|_{x_0=T/2}     \;,
\end{align}
where in practice we have chosen to extract them from our simulations for 
the non-trivial combinations $(\theta_1,\theta_2)\in\{(0,0.5),(0.5,1),(0,1)\}$.
Again, renormalization factors drop out in these ratios.
\subsection*{HQET observables}
\noindent%
Among the QCD observables defined above depending on a single $\theta$-angle,
only the effective decay constants in eq.~\eqref{eq:all-Y} possess a
(common) non-trivial static limit, which is given by
\begin{align}
X(\theta) = 
\left. \frac{\fastat(x_0,\theta)}{\sqrt{\fonestat(\theta)}} \right|_{x_0=T/2}
\end{align}
up to renormalization factors (cf.~subsection~\ref{sec:HQETobs}) to be 
discussed in appendix~\ref{app:AxialC}, while the ratios in (\ref{eq:all-R}) 
approach $1$ as a consequence of the heavy-quark spin symmetry that entails
pseudoscalar and vector channels to coincide.
The ratios in eq.~\eqref{eq:kinQ-ps} depending two $\theta$-angles,
on the other hand, approach in the static limit one of the following 
observables at static order of HQET:
\begin{align}
    R_{\rm PS}^{\rm stat}(\theta_1,\theta_2) &= \left. \frac{\fastat(x_0,\theta_1)}{\fastat(x_0,\theta_2)} \right|_{x_0=T/2} \;, &
         R_{f}^{\rm stat}(\theta_1,\theta_2) &= \frac{\fonestat(\theta_1)}{\fonestat(\theta_2)}  \;, \\
         R_{X}^{\rm stat}(\theta_1,\theta_2) &= \frac{X(\theta_1)}{X(\theta_2)} \;. &
\label{}
\end{align}
The corresponding static-light correlation functions $\fastat$ and 
$\fonestat$ entering in these expressions have first been defined 
in~\cite{Kurth:2000ki}.

\section{Conversion functions}\label{sec:ConvFuncs}
\begin{figure}[t]
        \small
  \centering
  \includegraphics[width=0.95\textwidth]{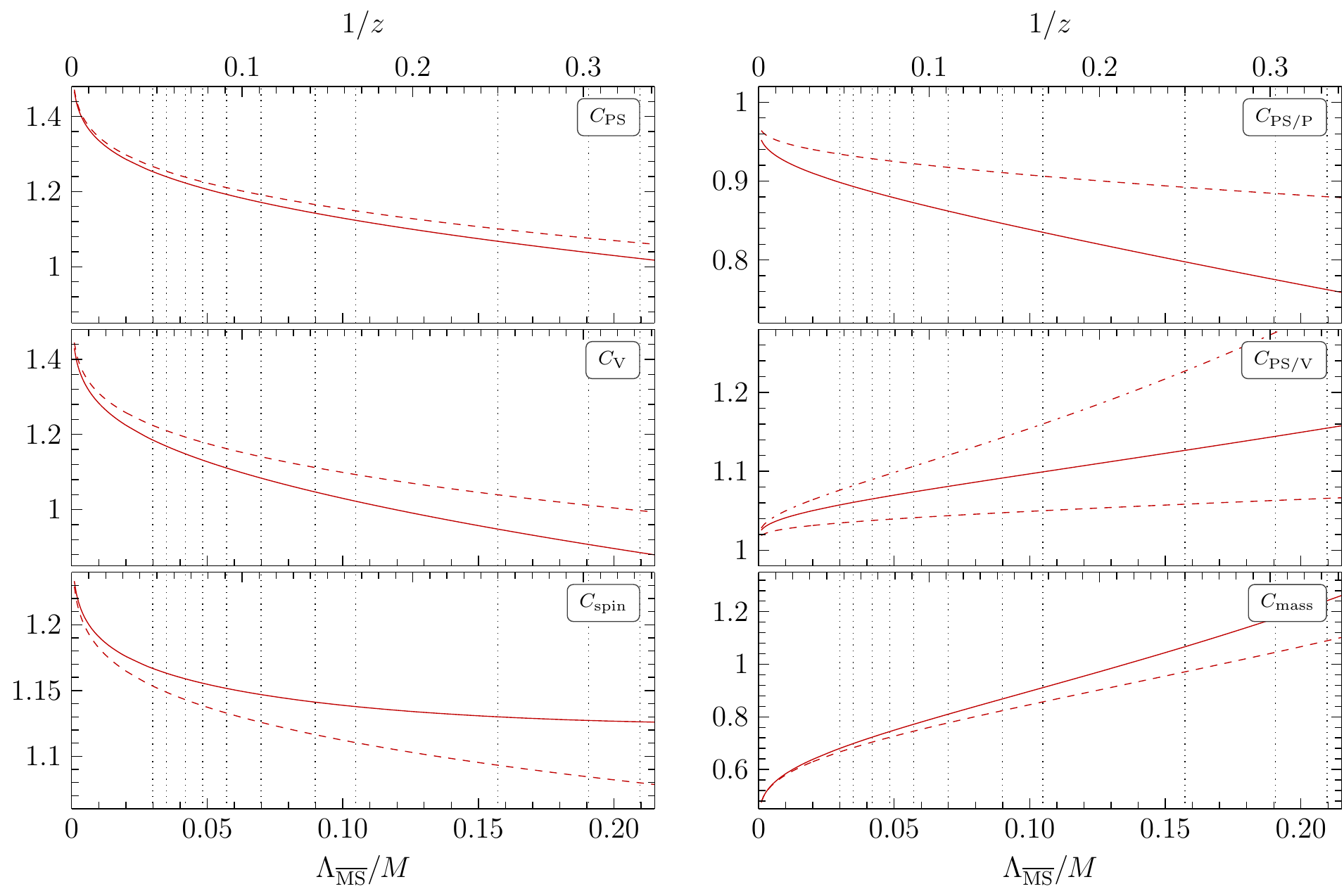}
  \caption{HQET-QCD conversion functions $\Cx$ for heavy-light currents as
  described in the text. The solid and dashed curves correspond to the 
  three-loop and two-loop order anomalous dimensions in the matching scheme, 
  respectively.
  In the case of $\Cpsv$ (middle panel on the right), the four-loop
  expression, which is known from~\cite{Bekavac2010}, is also included as
  dash-dotted curve.
  Comparing the curves for different loop orders suggests that
  perturbation theory converges rather slowly.
  Vertical dotted lines indicate the fixed values of $z=L_1M$ used in our 
  study.}
  \label{fig:C_X}
\end{figure}

In this section we summarize the expressions of the perturbative conversion
functions $\Cx$, $\rmX\in\{{\rm PS},{\rm V},{\rm spin},{\rm PS}/{\rm P},
{\rm PS}/{\rm V},{\rm mass}\}$, that have
been employed to compare our non-perturbatively renormalized observables
in QCD to their counterparts in HQET. They are computed and parameterized in
the so-called \emph{matching scheme}, which has been introduced 
in~\cite{Heitger:2003xg} and specified in appendix B of~\cite{Heitger:2004gb}
(see also~\cite{Sommer:2010ic} for another detailed discussion). 
Our formulae for all relevant operators below refer to the two-flavour theory
and are based on their anomalous dimensions known up to three-loop order
in continuum perturbation
theory~\cite{Tarrach:1980up,Shifman:1987sm,Politzer:1988wp,Gray:1990yh,
Eichten:1990vp,Falk:1991pz,Ji:1991pr,Broadhurst:1991fz,Gimenez:1991bf,
HQET:sigmabI,HQET:sigmabII,Chetyrkin:2003vi,Grozin:2007fh},
to be combined with the matching coefficients between QCD and the effective
theory up to two loops~\cite{Eichten:1989zv,Eichten:1989kb,Eichten:1990vp,
Broadhurst:1994se,Grozin:1998kf},
see appendix B of~\cite{Heitger:2004gb}.
In order to judge the impact of the order of the perturbative expansion on
this comparison between QCD and HQET, introduced by the conversion functions
$\Cx$ as far as they enter the observables under study, we evaluate the $\Cx$
including the two-loop and three-loop anomalous dimensions (together with the
respective matching coefficients in one- and two-loop accuracy) separately.
In both cases we use the four-loop beta-function for the
coupling~\cite{'tHooft:1973mm,Gross:1973id,Caswell:1974gg,Tarasov:1980au,
vanRitbergen:1997va,Czakon:2004bu}, while the conversion function for
effective masses involves the quark mass anomalous dimension $\tau$ at
four-loop~\cite{Chetyrkin:1997dh,Vermaseren:1997fq,Czakon:2004bu}. 

Moreover, thanks to the three-loop calculation of the matching coefficients
for the heavy-light currents available from~\cite{Bekavac2010}, the anomalous
dimensions of \emph{ratios} of currents become effectively known to
four-loop order in the matching scheme, because the unknown four-loop
anomalous dimensions of the currents themselves cancel out in the 
\emph{difference} of anomalous dimensions contributing to these ratios.
As an example, we therefore also include the conversion function
$\Cpsv$ of the ratio of axial vector ($A_0$) to vector ($V_k$) currents at
four-loop accuracy in our study.

Since so far only $\Cps$ for the $\nf=2$ theory was given earlier 
in~\cite{DellaMorte:2006sv}, we here list the parametrization of all
conversion functions $\Cx$ entering this study that were obtained along the
lines detailed in appendix B of~\cite{Heitger:2004gb}.
They are expressed as smooth functions in terms of the variable
\begin{align} \label{eqn:fit-parametr}
        x &\equiv 1\left/\ln\left[M/\lMSbar\right] \right.
        \;, &
        \dfrac{M}{\lMSbar} &= \dfrac{z}{L_1\lMSbar} \;,
\end{align}
with the product $L_1\lMSbar=0.629(36)$ in the two-flavour theory taken 
from~\cite{Fritzsch:2012wq} and $M$ denoting the renormalization group
invariant (RGI) heavy quark mass as in the main text.
The functions decompose into a pre-factor which encodes the leading
logarithmic asymptotics (if any) as $x\rightarrow 0$, multiplied by a
polynomial in $x$ which reflects the perturbative order of the underlying
anomalous dimension in conjunction with the associated matching coefficients.
\begin{align}
 \label{eq:Cps-fit}
 \Cps(x) &= 
     \begin{cases}
             \begin{array}{ll}
     x^{\gamps_0/(2b_0)}\left\{1-0.107x+0.093x^2          \right\}  & \;\text{: 2-loop } \gamps  \\[0.25em]
     x^{\gamps_0/(2b_0)}\left\{1-0.118x-0.010x^2+0.043x^3 \right\}  & \;\text{: 3-loop } \gamps
             \end{array}
     \end{cases} \;, \\[0.5em]
 \label{eq:Cv-fit}
 \Cv(x) &= 
     \begin{cases}
             \begin{array}{ll}
     x^{\gamv_0/(2b_0)}\left\{1-0.239x+0.153x^2          \right\} \hspace*{.25cm}& \text{: 2-loop } \gamv  \\[0.25em]
     x^{\gamv_0/(2b_0)}\left\{1-0.266x-0.178x^2+0.193x^3 \right\} \hspace*{.25cm}& \text{: 3-loop } \gamv
             \end{array}
     \end{cases} \;, \\[0.5em]
 \label{eq:Cspin-fit}
 \Cspin(x) &= 
     \begin{cases}
             \begin{array}{ll}
     x^{\gamspin_0/(2b_0)}\left\{1+0.043x+0.09x^2           \right\} & \text{: 2-loop } \gamspin  \\[0.25em]
     x^{\gamspin_0/(2b_0)}\left\{1+0.044x+0.179x^2-0.099x^3 \right\} & \text{: 3-loop } \gamspin
             \end{array}
     \end{cases} \;, \\[0.5em]
 \label{eq:Cpsps-fit}
 \Cpsp(x) &= 
     \begin{cases}
             \begin{array}{ll}
         1-0.266x+0.123x^2           \hspace*{2.15cm} & \,\text{: 2-loop } \gampsp  \\[0.25em]
         1-0.293x-0.304x^2+0.284x^3  \hspace*{2.15cm} & \,\text{: 3-loop } \gampsp
             \end{array}
     \end{cases} \;,\\[0.5em]
 \label{eq:Cpsv-fit}
 \Cpsv(x) &= 
     \begin{cases}
             \begin{array}{ll}
             1+0.136x-0.052x^2                    & \quad\;\text{: 2-loop } \gampsv  \\[0.25em]
             1+0.142x+0.250x^2-0.148x^3           & \quad\;\text{: 3-loop } \gampsv \\[0.25em]
             1+0.135x+0.323x^2+0.614x^3-0.436x^4  & \quad\;\text{: 4-loop } \gampsv
             \end{array}
     \end{cases} \;, \\[0.5em]
 \label{eq:Cmass-fit}
 \Cmass(x) &= 
     \begin{cases}
             \begin{array}{ll}
     x^{\,d_0/(2b_0)}\left\{1+0.373x+0.176x^2          \right\} \hspace*{0.255cm}& \,\text{: 2-loop } \tau  \\[0.25em]
     x^{\,d_0/(2b_0)}\left\{1+0.287x+0.752x^2+0.011x^3 \right\} \hspace*{0.255cm}& \,\text{: 3-loop } \tau
             \end{array}
     \end{cases} \;. 
\end{align}

The functional dependence of the $\Cx$ on $\lMSbar/M$ is shown in 
figure~\ref{fig:C_X}.
The solid curves represent the matching functions to (in most cases) highest
available perturbative accuracy, i.e., corresponding to three-loop anomalous
dimensions of the heavy-light currents involved and the four-loop 
beta-function.
The dashed curves are those obtained with two-loop anomalous dimensions only.
For illustration, we plot as vertical dotted lines the values of $z$ which
have been fixed in order to non-perturbatively compute our test observables.
In the exemplary case of the ratio of pseudoscalar to vector currents,
$\Cpsv$, which we are able to consider even up to the maximally available
four-loop precision (dash-dotted curve in the middle-right panel of 
figure~\ref{fig:C_X}), one observes an increase in the size of correction 
when going from two- to three- and three- to four-loop order. 
This is in accordance with the conclusion of the authors of~\cite{Bekavac2010}
that the perturbative series for this ratio of currents ``converges very 
slowly at best''.
We finally remark that, since the previous expressions derive from continuum
perturbation theory, the functions $\Cx$ must be properly attached as factors
to the HQET-QCD test observables in question \emph{after} the lattice results
on them have been extrapolated to the continuum limit,
cf.~section~\ref{sec:HQETobs}.

\section{Further details and tables}\label{app:CL}

\subsection*{Tree-level improvement}\label{app:TLI}
Even though our simulations are performed at rather fine lattice spacings, one
ultimately encounters mass-dependent cutoff effects, which essentially grow
with the heavy quark mass, i.e., to some power of $z\times a/L=aM$ in the
non-perturbatively $\Or$ improved theory.  To attenuate these effects, we also
apply perturbative improvement to some of our lattice observables under
consideration.  Here we restrict ourselves to \emph{tree-level improvement}
(TLI), which for a generic test observable $\Omega$ amounts to the replacement
\begin{align}
        \Omega(\gbsq,a/L,z) \;\to\; 
        \Omega^{(0)}(\gbsq,a/L,z) &= 
        \frac{\Omega(\gbsq,a/L,z)}{1+\delta^{(0)}_{\Omega}(a/L,z)} \;, \notag \\
        \delta^{(0)}_{\Omega}(a/L,z) &=
        \frac{\Omega^{\rm tree}(a/L,z)-\Omega^{\rm tree}(0,z)}
        {\Omega^{\rm tree}(0,z)} \;,
        \label{eq:def-TLI}
\end{align}
and thereby removes all the $\rmO\left((\frac{a}{L})^n\right)$ effects to
produce classically perfect observables.  For additional details on the actual
extraction of the improvement terms $\delta^{(0)}_{\Omega}$, see appendix D
of~\cite{Blossier2010}.  Besides its obvious dependence on $z$ and $a/L$,
$\delta^{(0)}_{\Omega}$ also depends on the kinematical setup of the
Schr\"odinger functional, inherited from the observable itself.  These are the
choice of boundary gauge field, the ratio $T/L$ and the fermionic periodicity
angles involved in the definition of $\Omega$.  Except for
$\Omega\in\{\Rpsps,\Rvv,\Rpp,R_1,R_f,R_k\}$, where TLI has not been accounted
for, and observables such as $\Rspin$, where it vanishes exactly, the presented
results always correspond to the tree-level improved quantities.

We furthermore remark that before any continuum limit in QCD and HQET is taken
numerically, we use the data from appendix~B of~\cite{Blossier:2012qu} to
correct for small deviations in our observables from the renormalized (i.e.,
constant physics) trajectory at $(L_1m_{\rm l},\gbsq(L_1))=(0,4.484)$.
The associated errors are propagated quadratically.

\subsection*{Special case: Axial current renormalization}\label{app:AxialC}
We add some remarks about the continuum extrapolation of the RGI static decay
constant as defined in eq.~\eqref{eq:Xrgi}, because we need to fully specify
the renormalization scheme applied.

The \emph{total} renormalization factor to obtain the RGI matrix element of the
static axial current, $\XRGI$, from the bare matrix element, $\Xbare$,
decomposes into 
\begin{align}
   Z^{\rm stat}_{\rm A,RGI}(g_0,\theta)  
      &\equiv \left[\frac{Z^{\rm stat}_{\rm A,RGI}}{Z^{\rm stat}_{\rm A}(\mu)} \cdot
                   \zastat(g_0,a\mu) \right]_{\mu=L_1^{-1}}^{(\theta,\rho)}
       = \left[\frac{Z^{\rm stat}_{\rm A,RGI}}{Z^{\rm stat}_{\rm A}(\mu)}  
         \frac{X(0,a\mu)}{\Xbare(g_0,a\mu)}\right]_{\mu=L_1^{-1}}^{(\theta,\rho)} \;.
   \label{eq:Zrgi}
\end{align}
As indicated here, the universal part $Z^{\rm stat}_{\rm A,RGI}/Z^{\rm
stat}_{\rm A}(\mu)$, which links a renormalized matrix element of the static
axial current at a renormalization scale $\mu$ to the RGI one, is defined
within the massless SF scheme for vanishing boundary field and
$(\theta,\rho=T/L)=(0.5,1)$ non-perturbatively and in the continuum limit.  For
the particular scale $\mu=L_1^{-1}$ corresponding to the physical volume
employed in the present work, $Z^{\rm stat}_{\rm A,RGI}/Z^{\rm stat}_{\rm
A}(\mu)$ has been re-computed from the two-flavour data of
ref.~\cite{DellaMorte:2006sv} to yield the estimate quoted in~\eqref{eq:Xrgi}.
(Note that in~\cite{DellaMorte:2006sv}, the universal ratio was denoted as
$\Phirgi/\Phi(\mu)$).  In the definition of the scale dependent renormalization
factor itself, $\zastat(g_0,a\mu)$, the tree-level normalization factor
$X(0,L/a)$ enters at the respective values of $(\theta,\rho)$ such that
$\zastat$ approaches one in the limit $g_0\to 0$, cf.~\cite{Heitger:2003xg}.
In fact, as we always have $\rho=T/L=1$ in the present work, too, we can take
the continuum limit according to
\begin{align}
        \XRGI(\theta_0) &= \lim_{L_1/a\to 0}\,Z^{\rm stat}_{\rm A,RGI}(g_0,\theta)
                           \,\Xbare(g_0,L_1/a,\theta_0) \notag \\   
                        &\equiv \lim_{L_1/a\to 0}\, 
                           \frac{Z^{\rm stat}_{\rm A,RGI}}{Z^{\rm stat}_{\rm A}(\mu)}
                           \,X_{\rm R}(\mu) \;,
        \label{eq:Xrgi-cont}
\end{align}
with the particular values $\theta_0\in\{0,0.5,1\}$ and $\theta=0.5$.%
\footnote{In case of $\theta_0=\theta=0.5$, the bare matrix element
in eq.~\eqref{eq:Zrgi} cancels exactly and only the tree-level matrix element
contributes in place of $X_{\rm R}$ in the second line of 
(\ref{eq:Xrgi-cont}).} 
The second equality in (\ref{eq:Xrgi-cont}) picks up the notation of
eq.~(\ref{eq:Xrgi}), and the results are listed in table~\ref{tab:YpsYv-X}.
Finally, let us point out that its total error is dominated by the error of the
universal continuum factor for the non-perturbative running, $Z^{\rm stat}_{\rm
A,RGI}\big/Z^{\rm stat}_{\rm A}(\mu)$ (as quoted in~\eqref{eq:Xrgi}), and that
this part of the error is only to be accounted for \emph{after} $\XRGI$ has
been extrapolated to the continuum limit.

\subsection*{Tables}

\begin{table}[t]
        \small
        \centering
        \begin{tabular}{cccccc}\toprule
                & $L_1/a$      & $20$           & $24$           & $32$           & $40$           \\
          $z$   & $\beta$      & $6.1569   $    & $6.2483   $    & $6.4574   $    & $6.6380   $    \\\midrule
     $\approx0$ & $\kapl\approx\hopc$      & $0.1360536$    & $0.1359104$    & $0.1355210$    & $0.1351923$    \\
          $2.0$ & $\kappa(z,\beta)$ & $0.1344153854$ & $0.1345754522$ & $0.1345486577$ & $0.1344246956$ \\
          $2.7$ & $\kappa(z,\beta)$ & $0.1338267897$ & $0.1340982540$ & $0.1342031414$ & $0.1341528164$ \\
          $3.0$ & $\kappa(z,\beta)$ & $0.1335718447$ & $0.1338920152$ & $0.1340541848$ & $0.1340357592$ \\
          $3.3$ & $\kappa(z,\beta)$ & $0.1333152045$ & $0.1336846995$ & $0.1339046863$ & $0.1339183724$ \\\midrule
            &$\zp(L_1,\gosq)$& $0.5310(22)$   & $0.5182(17)$ & $0.5161(16)$ & $0.5166(44)$      \\\bottomrule
        \end{tabular}
        \caption{Bare parameters for the four additional ensembles 
        (supplementing table~3 of~\cite{Blossier:2012qu}) used to compute
        QCD observables in $L_1$, including the heavy valence quark hopping
        parameters for the respective $z$-values, as well as the 
        pseudoscalar renormalization constant, $\zp$. The gauge field
        ensembles have been produced with vanishing SF boundary fields
        and a kinematical setup specified by $(T/L,\theta_{\rm sea})=(1,0.5)$.}
        \label{tab:QCD_run_details}
\end{table}
\begin{table}[t]
        \small
        \centering
        \begin{tabular}{cccccc}\toprule
             $L_1/a$ & $6$        & $8$        & $10$       & $12$       & $16$ \\
             $\beta$ & $5.2638 $  & $5.4689 $  & $5.6190 $  & $5.7580 $  & $5.9631 $ \\
             $\hopc$ & $0.135985$ & $0.136700$ & $0.136785$ & $0.136623$ & $0.136422$ \\\bottomrule
        \end{tabular}
        \caption{Bare parameters for the five ensembles used to compute HQET
        observables in $L_1$. 
        The gauge field ensembles have been produced with vanishing SF
        boundary fields and a kinematical setup specified by 
        $(T/L,\theta_{\rm sea})=(1,0.5)$.}
        \label{tab:HQET_run_details}
\end{table}

In table~\ref{tab:QCD_run_details} we collect addition details concerning the
measurements of heavy-light QCD observables in the present publication, thereby
extending the parameter set of table~3 in~\cite{Blossier:2012qu}, also referred
to as set ``QCD$[L_1]$''. Table~\ref{tab:HQET_run_details} repeats the bare
parameters relevant for the measurements in the static theory, defining
ensemble set ``HQET$[L_1]$''.

Tables~\ref{tab:LGamPS}~--~\ref{tab:decay-ratios} list the results at
finite lattice spacing and the corresponding continuum limits of some
selected observables. Listing all results
in full detail would be beyond the scope of this paper%
\footnote{They may be obtained from the authors upon request.},
since the qualitative and quantitative behaviour can also be well inferred
from the plots shown in the main text.
As an example for the $\theta$-dependence of an observable,
we reproduce $L\GamPS$ for all values of
$z\in\{2,2.7,3,3.3,4,6,7,9,11,13,15,18,21\}$ and
$\theta=\theta_0\in\{0,0.5,1\}$ in table~\ref{tab:LGamPS}. For all other
observables we only list values for $\theta_0=0.5$ or the combination
$(\theta_1,\theta_2)=(0.5,1)$. Note that values in squared brackets have not
been taken into account in the continuum extrapolations as detailed in
section~\ref{sec:rep-CL} and that the phase angle 
$\theta\equiv\theta_{\rm sea}=0.5$ has been used for the doublet of sea
quarks in the production runs to generate the underlying two-flavour
gauge field configuration ensembles.

\begin{table}[thb]
        \small
        \centering
        \begin{tabular}{CCCCC>{\it}C}
        \toprule
 L/a & 20 & 24 & 32 & 40 & CL \\
        \midrule
   z & \multicolumn{5}{c}{$L\GamPS^{(0)}(z,\theta_0=0.0)$}\\\cmidrule(lr){1-1}\cmidrule(lr){2-6} 
   2 &   1.934(12) &   1.912(09) &   1.949(11) &   1.944(10) &   1.952(10)  \\
 2.7 &   2.508(12) &   2.494(11) &   2.533(13) &   2.529(11) &   2.541(12)  \\
   3 &   2.746(13) &   2.735(11) &   2.774(14) &   2.771(12) &   2.784(13)  \\
 3.3 &   2.980(13) &   2.972(12) &   3.010(14) &   3.008(12) &   3.023(14)  \\
   4 &   3.476(14) &   3.477(12) &   3.522(15) &   3.524(13) &   3.537(15)  \\
   6 &   4.879(15) &   4.895(14) &   4.946(18) &   4.954(15) &   4.974(17)  \\
   7 &   5.547(15) &   5.568(14) &   5.622(18) &   5.633(15) &   5.658(18)  \\
   9 &   6.835(15) &   6.864(15) &   6.924(19) &   6.944(16) &   6.978(18)  \\
  11 &   8.081(15) &   8.113(15) &   8.179(19) &   8.208(16) &   8.254(18)  \\
  13 &  [9.308(15)]&   9.330(15) &   9.400(19) &   9.438(16) &   9.490(19)  \\
  15 & [10.559(15)]&  10.529(15) &  10.596(19) &  10.645(16) &  10.710(25)  \\
  18 &  \text{---} & [12.341(14)]&  12.359(19) &  12.421(16) &  12.499(38)  \\
  21 &  \text{---} & [14.688(14)]& [14.108(19)]&  14.171(16) &  14.274(62)  \\
        \midrule
   z & \multicolumn{5}{c}{$L\GamPS^{(0)}(z,\theta_0=0.5)$}\\\cmidrule(lr){1-1}\cmidrule(lr){2-6} 
   2 &   2.061(11) &   2.053(11) &   2.082(10) &   2.088(11) &   2.098(11)  \\
 2.7 &   2.650(11) &   2.653(12) &   2.681(11) &   2.689(13) &   2.703(14)  \\
   3 &   2.892(12) &   2.899(12) &   2.927(12) &   2.935(13) &   2.951(14)  \\
 3.3 &   3.130(12) &   3.139(13) &   3.167(12) &   3.175(13) &   3.193(15)  \\
   4 &   3.635(12) &   3.653(13) &   3.686(13) &   3.698(14) &   3.714(16)  \\
   6 &   5.048(13) &   5.081(15) &   5.118(15) &   5.135(15) &   5.158(18)  \\
   7 &   5.718(13) &   5.756(15) &   5.795(16) &   5.815(15) &   5.841(18)  \\
   9 &   7.012(13) &   7.056(15) &   7.100(16) &   7.127(16) &   7.159(18)  \\
  11 &   8.264(13) &   8.309(15) &   8.356(17) &   8.392(16) &   8.433(18)  \\
  13 &  [9.498(13)]&   9.529(15) &   9.579(17) &   9.625(16) &   9.669(18)  \\
  15 & [10.759(12)]&  10.733(15) &  10.778(17) &  10.832(16) &  10.887(24)  \\
  18 &  \text{---} & [12.555(14)]&  12.545(17) &  12.611(16) &  12.673(36)  \\
  21 &  \text{---} & [14.922(14)]& [14.300(16)]&  14.364(15) &  14.455(56)  \\
        \midrule
   z & \multicolumn{5}{c}{$L\GamPS^{(0)}(z,\theta_0=1.0)$}\\\cmidrule(lr){1-1}\cmidrule(lr){2-6} 
   2 &   2.573(18) &   2.578(20) &   2.578(22) &   2.638(22) &   2.634(21)  \\
 2.7 &   3.201(18) &   3.219(21) &   3.217(22) &   3.279(23) &   3.278(23)  \\
   3 &   3.454(18) &   3.476(21) &   3.473(22) &   3.535(23) &   3.535(24)  \\
 3.3 &   3.699(18) &   3.724(21) &   3.721(22) &   3.782(23) &   3.784(24)  \\
   4 &   4.226(18) &   4.257(21) &   4.256(22) &   4.317(23) &   4.320(25)  \\
   6 &   5.652(17) &   5.696(20) &   5.697(21) &   5.758(23) &   5.762(25)  \\
   7 &   6.325(16) &   6.371(20) &   6.374(20) &   6.436(22) &   6.440(24)  \\
   9 &   7.623(16) &   7.672(19) &   7.678(20) &   7.742(22) &   7.747(23)  \\
  11 &   8.882(15) &   8.926(19) &   8.934(19) &   9.003(21) &   9.014(22)  \\
  13 & [10.126(14)]&  10.152(18) &  10.158(19) &  10.234(21) &  10.250(22)  \\
  15 & [11.403(14)]&  11.362(17) &  11.359(19) &  11.440(20) &  11.474(29)  \\
  18 &  \text{---} & [13.201(16)]&  13.132(18) &  13.220(20) &  13.266(44)  \\
  21 &  \text{---} & [15.606(15)]& [14.895(17)]&  14.975(19) &  15.084(70)  \\
        \bottomrule
        \end{tabular}
        \caption{Tree-level improved pseudoscalar effective mass
                 $L\GamPS^{(0)}(x_0=T/2)$ as defined in eq.~\eqref{eq:GamPS} 
                 for all available values of $z$, $a/L$ and $\theta_0$, 
                 together with the continuum limit (CL) result obtained 
                 according to eq.~\eqref{eq:global-CLfit}.}
        \label{tab:LGamPS}
\end{table}

\begin{table}[thb]
        \small
        \centering
        \begin{tabular}{CCCCC>{\it}C}
        \toprule
 L/a & 20 & 24 & 32 & 40 & CL \\
        \midrule
   z & \multicolumn{5}{c}{$L\GamV^{(0)}(z,\theta_0)$}\\\cmidrule(lr){1-1}\cmidrule(lr){2-6} 
   2 &   2.907(20) &   2.976(26) &   3.026(26) &   3.048(25) &   3.084(25)  \\
 2.7 &   3.354(19) &   3.420(25) &   3.470(25) &   3.487(24) &   3.535(26)  \\
   3 &   3.546(19) &   3.612(24) &   3.662(25) &   3.677(23) &   3.729(26)  \\
 3.3 &   3.740(19) &   3.805(24) &   3.854(25) &   3.868(23) &   3.923(26)  \\
   4 &   4.164(18) &   4.231(23) &   4.283(24) &   4.298(22) &   4.353(26)  \\
   6 &   5.425(17) &   5.493(22) &   5.545(23) &   5.561(21) &   5.619(25)  \\
   7 &   6.047(17) &   6.116(21) &   6.169(22) &   6.187(20) &   6.243(24)  \\
   9 &   7.274(16) &   7.344(20) &   7.400(21) &   7.425(19) &   7.474(22)  \\
  11 &   8.482(15) &   8.549(19) &   8.608(21) &   8.642(19) &   8.688(21)  \\
  13 &  [9.682(15)]&   9.734(18) &   9.797(20) &   9.840(18) &   9.891(21)  \\
  15 & [10.914(14)]&  10.910(18) &  10.969(19) &  11.022(18) &  11.089(27)  \\
  18 &  \text{---} & [12.699(17)]&  12.705(19) &  12.772(17) &  12.868(42)  \\
  21 &  \text{---} & [15.032(15)]& [14.436(18)]&  14.503(17) &  14.662(66)  \\
        \midrule
   z & \multicolumn{5}{c}{$L\GamP^{(0)}(z,\theta_0)$}\\\cmidrule(lr){1-1}\cmidrule(lr){2-6} 
   2 &   2.939(18) &   3.003(25) &   3.043(25) &   3.061(28) &   3.096(25)  \\
 2.7 &   3.359(18) &   3.420(24) &   3.460(24) &   3.474(26) &   3.516(25)  \\
   3 &   3.543(18) &   3.602(24) &   3.642(24) &   3.655(25) &   3.699(25)  \\
 3.3 &   3.729(17) &   3.787(23) &   3.826(23) &   3.837(25) &   3.884(25)  \\
   4 &   4.156(17) &   4.209(23) &   4.247(23) &   4.257(24) &   4.305(25)  \\
   6 &   5.434(16) &   5.474(21) &   5.500(21) &   5.506(21) &   5.544(25)  \\
   7 &   6.085(16) &   6.114(20) &   6.131(21) &   6.135(21) &   6.159(24)  \\
   9 &   7.411(15) &   7.405(19) &   7.393(20) &   7.391(19) &   7.376(22)  \\
  11 &   8.792(15) &   8.716(19) &   8.655(19) &   8.641(19) &   8.578(21)  \\
  13 & [10.278(15) &  10.065(18) &   9.922(19) &   9.886(18) &   9.760(21)  \\
  15 & [11.990(14) &  11.485(18) &  11.203(19) &  11.133(18) &  10.942(28)  \\
  18 &  \text{---} & [13.893(17)]&  13.181(18) &  13.018(17) &  12.670(42)  \\
  21 &  \text{---} & [17.646(17)]& [15.283(18)]&  14.945(17) &  14.413(66)  \\
        \midrule
   z & \multicolumn{5}{c}{$\GamPS^{(0)}(z,\theta_0)/\GamV^{(0)}(z,\theta_0)$}\\\cmidrule(lr){1-1}\cmidrule(lr){2-6} 
   2 &  0.7089(35) &  0.6900(36) &  0.6881(34) &  0.6850(33) &  0.6820(28)  \\
 2.7 &  0.7902(30) &  0.7757(31) &  0.7727(29) &  0.7711(29) &  0.7664(28)  \\
   3 &  0.8156(28) &  0.8025(28) &  0.7994(27) &  0.7981(27) &  0.7930(27)  \\
 3.3 &  0.8368(26) &  0.8250(26) &  0.8218(26) &  0.8209(25) &  0.8155(26)  \\
   4 &  0.8728(22) &  0.8634(22) &  0.8606(22) &  0.8605(21) &  0.8544(24)  \\
   6 &  0.9305(14) &  0.9250(14) &  0.9229(14) &  0.9234(14) &  0.9181(18)  \\
   7 &  0.9456(11) &  0.9411(12) &  0.9393(11) &  0.9399(11) &  0.9354(15)  \\
   9 & 0.96391(79) & 0.96077(81) & 0.95937(77) & 0.95991(81) & 0.95720(98)  \\
  11 & 0.97435(58) & 0.97191(60) & 0.97072(57) & 0.97114(60) & 0.96989(65)  \\
  13 & 0.98103(43) & 0.97895(45) & 0.97782(44) & 0.97812(47) & 0.97736(48)  \\
  15 & 0.98576(32) & 0.98375(35) & 0.98262(35) & 0.98280(38) & 0.98228(56)  \\
  18 &  \text{---} & 0.98871(24) & 0.98740(25) & 0.98741(28) & 0.98621(82)  \\
  21 &  \text{---} & 0.99269(15) & 0.99056(19) & 0.99041(22) &  0.9883(12)  \\
        \bottomrule
        \end{tabular}
        \caption{Tree-level improved vector and pseudoscalar effective masses
                 $L\GamV^{(0)}(x_0=T/2)$ and $L\GamP^{(0)}(x_0=T/2)$ and the
                 ratio $[\GamPS^{(0)}(x_0)/\GamV^{(0)}(x_0)]_{x_0=T/2}$
                 for all available values of $z$ and $a/L$ with $\theta_0=0.5$.}
        \label{tab:LGamVP}
\end{table}

%
%
\begin{table}[thb]
        \small
        \centering
        \begin{tabular}{CCCCC>{\it}C}\toprule
   L/a  &      20     &        24     &      32      &     40      & CL   \\\midrule
    z   & \multicolumn{5}{c}{$\Yps^{(0)}(z,\theta_0)$}      \\\cmidrule(lr){1-1}\cmidrule(lr){2-6}
      2 &  1.2653(58) &  1.2519(65) &  1.2514(61) &  1.2417(65) &  1.2367(74)  \\
    2.7 &  1.3088(57) &  1.2978(63) &  1.2973(59) &  1.2889(62) &  1.2850(72)  \\
      3 &  1.3253(57) &  1.3153(62) &  1.3149(59) &  1.3069(62) &  1.3035(71)  \\
    3.3 &  1.3407(57) &  1.3316(62) &  1.3313(58) &  1.3236(61) &  1.3209(70)  \\
      4 &  1.3696(57) &  1.3630(61) &  1.3640(58) &  1.3573(60) &  1.3552(69)  \\
      6 &  1.4380(57) &  1.4362(59) &  1.4396(56) &  1.4343(58) &  1.4359(66)  \\
      7 &  1.4641(57) &  1.4642(59) &  1.4688(56) &  1.4639(57) &  1.4670(65)  \\
      9 &  1.5062(58) &  1.5092(58) &  1.5158(56) &  1.5115(58) &  1.5170(64)  \\
     11 &  1.5396(59) &  1.5443(59) &  1.5524(57) &  1.5484(58) &  1.5547(69)  \\
     13 & [1.5682(59)]&  1.5731(59) &  1.5819(58) &  1.5781(59) &  1.5841(73)  \\
     15 & [1.5953(61)]&  1.5979(60) &  1.6067(59) &  1.6028(60) &  1.6067(93)  \\
     18 &  \text{---} & [1.6317(61)]&  1.6376(61) &  1.6331(61) &  1.633(13)  \\
     21 &  \text{---} & [1.6782(63)]& [1.6639(63)]&  1.6580(63) &  1.650(18)  \\
    \midrule
    z   & \multicolumn{5}{c}{$\Yv^{(0)}(z,\theta_0)$}                             \\\cmidrule(lr){1-1}\cmidrule(lr){2-6}
     2  &  1.539(10) &  1.539(10) &  1.542(10) &  1.538(10) &  1.539(11)  \\
   2.7  &  1.551(10) &  1.551(10) &  1.555(10) &  1.551(10) &  1.554(11)  \\
     3  &  1.555(10) &  1.556(10) &  1.560(10) &  1.556(10) &  1.559(11)  \\
   3.3  &  1.559(10) &  1.560(10) &  1.565(10) &  1.561(10) &  1.564(10)  \\
     4  &  1.565(10) &  1.568(10) &  1.573(10) &  1.569(10) &  1.573(10)  \\
     6  &  1.583(10) &  1.587(10) &  1.595(10) &  1.590(10) &  1.597(10)  \\
     7  &  1.590(10) &  1.595(10) &  1.604(10) &  1.599(10) &  1.606(11)  \\
     9  &  1.605(10) &  1.611(10) &  1.620(10) &  1.615(10) &  1.623(11)  \\
    11  &  1.619(10) &  1.625(10) &  1.634(10) &  1.629(10) &  1.636(11)  \\
    13  & [1.633(10)]&  1.638(10) &  1.647(10) &  1.642(10) &  1.647(12)  \\
    15  & [1.650(11)]&  1.652(10) &  1.659(11) &  1.653(10) &  1.656(14)  \\
    18  & \text{---} & [1.673(11)]&  1.676(11) &  1.669(11) &  1.668(19)  \\
    21  & \text{---} & [1.711(11)]& [1.693(11)]&  1.683(11) &  1.673(27)  \\
    \midrule
    z   & \multicolumn{5}{c}{$\Ypsv^{(0)}(z,\theta_0)$}                             \\\cmidrule(lr){1-1}\cmidrule(lr){2-6}
     2  &  0.8220(61) &  0.8136(63) &  0.8115(61) &  0.8072(62) &  0.8034(72)  \\
   2.7  &  0.8440(62) &  0.8367(63) &  0.8342(61) &  0.8308(62) &  0.8270(72)  \\
     3  &  0.8523(63) &  0.8455(63) &  0.8429(62) &  0.8398(62) &  0.8360(72)  \\
   3.3  &  0.8601(63) &  0.8536(63) &  0.8509(62) &  0.8482(63) &  0.8444(72)  \\
     4  &  0.8751(64) &  0.8695(64) &  0.8670(63) &  0.8649(63) &  0.8612(72)  \\
     6  &  0.9086(65) &  0.9049(65) &  0.9028(64) &  0.9019(64) &  0.8991(74)  \\
     7  &  0.9206(66) &  0.9177(66) &  0.9159(65) &  0.9154(64) &  0.9131(75)  \\
     9  &  0.9387(67) &  0.9369(67) &  0.9357(65) &  0.9359(65) &  0.9347(77)  \\
    11  &  0.9512(68) &  0.9504(67) &  0.9499(66) &  0.9505(66) &  0.9503(83)  \\
    13  & [0.9603(68)]&  0.9602(68) &  0.9603(67) &  0.9613(66) &  0.9616(86)  \\
    15  & [0.9671(69)]&  0.9675(68) &  0.9682(67) &  0.9696(67) &  0.970(10)  \\
    18  &  \text{---} & [0.9753(68)]&  0.9769(67) &  0.9786(67) &  0.980(14)  \\
    21  &  \text{---} & [0.9811(69)]& [0.9829(68)]&  0.9850(67) &  0.987(19)  \\
    \bottomrule
        \end{tabular}
        \caption{Tree-level improved effective decay constants $\Yps$, $\Yv$
                 and and their ratio $\Ypsv$ for all available values of $z$, $a/L$ at 
                 $\theta_0=0.5$. Continuum limit (CL) has been
                 taken according to eq.~\eqref{eq:global-CLfit}.}
        \label{tab:decay-const}
\end{table}

%
%
\begin{table}[thb]
        \small
        \centering
        \begin{tabular}{CCCCC>{\it}C}\toprule
   L/a  &      20     &        24     &      32      &     40      & CL   \\\midrule
   z   & \multicolumn{5}{c}{$F_{\rm PS/V}^{(0)}(z,\theta_0)$}      \\\cmidrule(lr){1-1}\cmidrule(lr){2-6}
     2 &  0.9763(74) &  0.9794(72) &  0.9784(71) &  0.9753(73) &  0.9750(83)  \\
   2.7 &  0.9494(70) &  0.9500(69) &  0.9490(68) &  0.9462(70) &  0.9463(77)  \\
     3 &  0.9438(69) &  0.9438(69) &  0.9427(67) &  0.9400(69) &  0.9403(75)  \\
   3.3 &  0.9402(69) &  0.9398(68) &  0.9387(66) &  0.9361(68) &  0.9364(74)  \\
     4 &  0.9369(68) &  0.9359(67) &  0.9347(66) &  0.9324(67) &  0.9327(71)  \\
     6 &  0.9425(68) &  0.9413(67) &  0.9400(66) &  0.9387(66) &  0.9386(69)  \\
     7 &  0.9476(68) &  0.9465(67) &  0.9453(66) &  0.9444(66) &  0.9440(69)  \\
     9 &  0.9573(68) &  0.9567(68) &  0.9558(67) &  0.9555(66) &  0.9548(72)  \\
    11 &  0.9654(69) &  0.9652(68) &  0.9647(67) &  0.9649(67) &  0.9641(76)  \\
    13 &  0.9719(69) &  0.9719(68) &  0.9719(67) &  0.9725(67) &  0.9719(80)  \\
    15 &  0.9771(69) &  0.9773(69) &  0.9777(68) &  0.9786(67) &  0.9783(86)  \\
    18 &  \text{---} &  0.9834(69) &  0.9843(68) &  0.9855(67) &  0.9862(87)  \\
    21 &  \text{---} &  0.9883(69) &  0.9891(68) &  0.9907(68) &  0.993(10)  \\
    \midrule
    z   & \multicolumn{5}{c}{$\Rpsv^{(0)}(z,\theta_0)$}                             \\\cmidrule(lr){1-1}\cmidrule(lr){2-6}
     2  &  0.8557(64) &  0.8503(64) &  0.8505(62) &  0.8463(65) &  0.8447(73)  \\
   2.7  &  0.8748(64) &  0.8704(64) &  0.8702(63) &  0.8668(65) &  0.8653(72)  \\
     3  &  0.8820(64) &  0.8780(64) &  0.8776(63) &  0.8746(65) &  0.8731(71)  \\
   3.3  &  0.8887(65) &  0.8850(65) &  0.8846(63) &  0.8818(65) &  0.8803(71)  \\
     4  &  0.9015(65) &  0.8987(65) &  0.8983(64) &  0.8961(65) &  0.8947(71)  \\
     6  &  0.9301(67) &  0.9288(66) &  0.9287(65) &  0.9276(66) &  0.9269(71)  \\
     7  &  0.9402(67) &  0.9396(67) &  0.9396(65) &  0.9389(66) &  0.9386(72)  \\
     9  &  0.9551(68) &  0.9555(67) &  0.9561(66) &  0.9560(66) &  0.9565(73)  \\
    11  &  0.9652(69) &  0.9664(68) &  0.9676(67) &  0.9680(67) &  0.9692(79)  \\
    13  & [0.9723(69)]&  0.9741(68) &  0.9759(67) &  0.9767(67) &  0.9783(82)  \\
    15  & [0.9772(69)]&  0.9797(69) &  0.9820(68) &  0.9833(67) &  0.985(10)  \\
    18  &  \text{---} & [0.9852(69)]&  0.9886(68) &  0.9903(68) &  0.992(13)  \\
    21  &  \text{---} & [0.9885(69)]& [0.9929(68)]&  0.9951(68) &  0.997(19)  \\
    \midrule
    z   & \multicolumn{5}{c}{$\Rspin(z,\theta_0)$}                             \\\cmidrule(lr){1-1}\cmidrule(lr){2-6}
     2  & 0.0603(17) & 0.0661(16) &  0.0703(21)  &  0.0710(16) &  0.0750(21)  \\
   2.7  & 0.0538(15) & 0.0592(15) &  0.0632(19)  &  0.0636(14) &  0.0675(19)  \\
     3  & 0.0514(15) & 0.0566(14) &  0.0606(18)  &  0.0608(14) &  0.0647(18)  \\
   3.3  & 0.0492(14) & 0.0542(14) &  0.0581(18)  &  0.0583(13) &  0.0621(18)  \\
     4  & 0.0447(13) & 0.0494(12) &  0.0532(16)  &  0.0532(12) &  0.0568(16)  \\
     6  & 0.0351(10) & 0.0391(10) &  0.0424(13)  &  0.0421(10) &  0.0454(13)  \\
     7  & 0.0315(9)  & 0.0353(9)  &  0.0384(12)  &  0.0380(9)  &  0.0411(12)  \\
     9  & 0.0260(8)  & 0.0294(7)  &  0.0322(10)  &  0.0318(9)  &  0.0345(10)  \\
    11  & 0.0219(6)  & 0.0251(6)  &  0.0277(9)   &  0.0273(7)  &  0.0295(9)   \\
    13  &[0.0185(5)] & 0.0216(5)  &  0.0241(8)   &  0.0238(6)  &  0.0258(8)   \\
    15  &[0.0156(5)] & 0.0188(5)  &  0.0213(7)   &  0.0210(5)  &  0.0225(8)   \\
    18  & \text{---} &[0.0151(4)] &  0.0179(6)   &  0.0178(5)  &  0.0191(9)   \\
    21  & \text{---} &[0.0112(3)] & [0.01511(48)]&  0.0153(4)  &  0.0160(9)   \\
    \bottomrule
        \end{tabular}
        \caption{Tree-level improved ratios $F_{\rm PS/V}$ and $\Rpsv$ together with the
                 spin splitting $\Rspin$ for all available values of $z$, $a/L$ at 
                 $\theta_0=0.5$. Continuum limit (CL) has been
                 taken according to eq.~\eqref{eq:global-CLfit}.}
        \label{tab:Rspin}
\end{table}

%
%
\begin{table}[thb]
        \small
        \centering
        \begin{tabular}{CCCCC>{\it}C}
        \toprule
   L/a  &      20     &        24     &      32      &     40      & CL   \\\midrule
    z   & \multicolumn{5}{c}{$\Rpsps(z,\theta_1,\theta_2)$}      \\\cmidrule(lr){1-1}\cmidrule(lr){2-6}
    2   &  1.941(16)  &  1.956(17)   &  1.912(24)  & 1.975(15)   & 1.970(16)    \\ 
    2.7 &  1.849(14)  &  1.861(14)   &  1.825(20)  & 1.876(12)   & 1.873(14)    \\ 
    3   &  1.818(13)  &  1.830(13)   &  1.796(19)  & 1.843(11)   & 1.841(13)    \\ 
    3.3 &  1.791(12)  &  1.802(12)   &  1.771(17)  & 1.814(11)   & 1.812(13)    \\ 
    4   &  1.741(11)  &  1.750(11)   &  1.723(15)  & 1.7609(93)  & 1.759(12)    \\ 
    6   &  1.6466(82) &  1.6545(81)  &  1.635(12)  & 1.6628(71)  & 1.6611(90)   \\ 
    7   &  1.6164(75) &  1.6239(74)  &  1.606(11)  & 1.6318(64)  & 1.6301(81)   \\ 
    9   &  1.5730(65) &  1.5804(65)  &  1.5660(96) & 1.5882(56)  & 1.5871(68)   \\ 
    11  &  1.5428(59) &  1.5507(59)  &  1.5385(87) & 1.5590(51)  & 1.5594(63)   \\ 
    13  &  1.5199(55) &  1.5288(54)  &  1.5185(81) & 1.5380(47)  & 1.5409(62)   \\ 
    15  &  1.5007(51) &  1.5116(51)  &  1.5032(77) & 1.5220(44)  & 1.5281(79)   \\ 
    18  &  \text{---} &  1.4907(48)  &  1.4856(73) & 1.5040(42)  & 1.513(11)    \\ 
    21  &  \text{---} &  1.4696(45)  &  1.4721(69) & 1.4906(40)  & 1.508(15)    \\ 
    \midrule
    z   & \multicolumn{5}{c}{$\Rvv(z,\theta_1,\theta_2)$}                             \\\cmidrule(lr){1-1}\cmidrule(lr){2-6}
    2   &  1.769(12)  &  1.778(11)    &  1.748(17)  &  1.787(10)  & 1.784(12)    \\ 
    2.7 &  1.714(10)  &  1.7227(99)   &  1.698(14)  &  1.7306(88) & 1.728(10)    \\ 
    3   &  1.6956(96) &  1.7038(94)   &  1.680(14)  &  1.7112(83) & 1.709(10)    \\ 
    3.3 &  1.6791(92) &  1.6870(90)   &  1.665(13)  &  1.6940(79) & 1.6920(96)   \\ 
    4   &  1.6478(84) &  1.6553(82)   &  1.635(12)  &  1.6617(72) & 1.6598(89)   \\ 
    6   &  1.5870(69) &  1.5937(67)   &  1.5779(98) &  1.5993(59) & 1.5977(74)   \\ 
    7   &  1.5665(64) &  1.5732(63)   &  1.5587(92) &  1.5788(55) & 1.5773(69)   \\ 
    9   &  1.5363(58) &  1.5430(57)   &  1.5306(83) &  1.5490(49) & 1.5481(59)   \\ 
    11  &  1.5145(54) &  1.5217(53)   &  1.5108(78) &  1.5283(45) & 1.5288(57)   \\ 
    13  &  1.4975(51) &  1.5055(50)   &  1.4961(74) &  1.5130(43) & 1.5155(57)   \\ 
    15  &  1.4829(48) &  1.4925(48)   &  1.4845(71) &  1.5012(41) & 1.5060(73)   \\ 
    18  &  \text{---} &  1.4764(45)   &  1.4710(68) &  1.4876(39) & 1.4941(97)   \\ 
    21  &  \text{---} &  1.4597(43)   &  1.4604(65) &  1.4773(38) & 1.491(14)    \\
    \midrule
    z   & \multicolumn{5}{c}{$\Rpp(z,\theta_1,\theta_2)$}                          \\\cmidrule(lr){1-1}\cmidrule(lr){2-6}
    2   &  1.5534(57) &  1.5571(56)   &  1.5430(83) &  1.5628(48) & 1.5619(56)  \\ 
    2.7 &  1.5292(51) &  1.5327(50)   &  1.5205(75) &  1.5380(43) & 1.5372(52)  \\ 
    3   &  1.5211(50) &  1.5246(48)   &  1.5129(73) &  1.5297(41) & 1.5290(51)  \\ 
    3.3 &  1.5141(48) &  1.5175(47)   &  1.5063(70) &  1.5225(40) & 1.5218(50)  \\ 
    4   &  1.5009(46) &  1.5044(44)   &  1.4940(67) &  1.5092(38) & 1.5084(47)  \\ 
    6   &  1.4761(41) &  1.4795(40)   &  1.4705(61) &  1.4841(34) & 1.4833(43)  \\ 
    7   &  1.4677(40) &  1.4712(39)   &  1.4626(59) &  1.4759(33) & 1.4751(42)  \\ 
    9   &  1.4553(39) &  1.4590(38)   &  1.4509(57) &  1.4639(32) & 1.4636(39)  \\ 
    11  &  1.4462(38) &  1.4502(37)   &  1.4426(56) &  1.4554(31) & 1.4561(41)  \\ 
    13  &  1.4389(38) &  1.4435(37)   &  1.4363(55) &  1.4491(31) & 1.4508(42)  \\ 
    15  &  1.4327(38) &  1.4380(36)   &  1.4313(55) &  1.4442(30) & 1.4475(54)  \\ 
    18  &  \text{---} &  1.4312(36)   &  1.4255(54) &  1.4385(30) & 1.4426(72)  \\ 
    21  &  \text{---} &  1.4242(36)   &  1.4209(54) &  1.4342(30) & 1.443(10)  \\
    \bottomrule
        \end{tabular}
        \caption{Observables $\Rpsps$, $\Rvv$ and $\Rpp$ for all available values of $z$, 
                 $a/L$ at $(\theta_1,\theta_2)=(0.5,1)$. Continuum limit (CL) has been 
                 taken according to eq.~\eqref{eq:global-CLfit}. All three observables 
                 are expected to approach the same static limit.}
        \label{tab:current-rat}
\end{table}

%
%
\begin{table}[thb]
        \small
        \centering
        \begin{tabular}{CCCCC>{\it}C}
        \toprule
   L/a  &      20     &        24     &      32      &     40      & CL   \\
        \midrule
    z   & \multicolumn{5}{c}{$R_1(z,\theta_1,\theta_2)$}      \\\cmidrule(lr){1-1}\cmidrule(lr){2-6}
    2   &  0.777(11) &  0.788(11) &  0.762(16) &  0.800(10) & 0.799(12)  \\ 
    2.7 &  0.747(10) &  0.758(10) &  0.735(14) &  0.769(9)  & 0.768(11)  \\ 
    3   &  0.736(10) &  0.747(10) &  0.725(14) &  0.758(9)  & 0.756(11)  \\ 
    3.3 &  0.726(10) &  0.737(9)  &  0.716(13) &  0.747(9)  & 0.747(10)  \\ 
    4   &  0.708(9)  &  0.719(9)  &  0.699(13) &  0.728(8)  & 0.727(10)  \\ 
    6   &  0.669(8)  &  0.679(8)  &  0.664(11) &  0.688(7)  & 0.687(9)   \\ 
    7   &  0.655(8)  &  0.665(7)  &  0.651(10) &  0.673(7)  & 0.673(8)   \\ 
    9   &  0.632(7)  &  0.643(7)  &  0.630(10) &  0.652(6)  & 0.652(7)   \\ 
    11  &  0.616(7)  &  0.627(7)  &  0.615(9)  &  0.636(6)  & 0.637(7)   \\ 
    13  &  0.602(6)  &  0.614(6)  &  0.604(9)  &  0.623(6)  & 0.627(7)   \\ 
    15  &  0.590(6)  &  0.604(6)  &  0.595(9)  &  0.614(6)  & 0.618(9)   \\ 
    18  &  \text{---}&  0.590(6)  &  0.583(8)  &  0.602(5)  & 0.607(12)  \\ 
    21  &  \text{---}&  0.576(6)  &  0.575(8)  &  0.593(5)  & 0.605(17)  \\ 
        \midrule
    z   & \multicolumn{5}{c}{$R_f(z,\theta_1,\theta_2)$}             \\\cmidrule(lr){1-1}\cmidrule(lr){2-6}
    2   &  2.209(23) &  2.237(24) &  2.181(34) &  2.271(22) & 2.269(25)  \\ 
    2.7 &  2.139(21) &  2.166(21) &  2.118(30) &  2.195(19) & 2.194(23)  \\ 
    3   &  2.115(20) &  2.141(20) &  2.096(29) &  2.168(18) & 2.168(22)  \\ 
    3.3 &  2.093(19) &  2.119(19) &  2.076(27) &  2.145(18) & 2.145(21)  \\ 
    4   &  2.051(18) &  2.076(18) &  2.038(25) &  2.100(16) & 2.100(20)  \\ 
    6   &  1.967(15) &  1.991(15) &  1.960(21) &  2.010(14) & 2.011(17)  \\ 
    7   &  1.938(14) &  1.961(14) &  1.933(20) &  1.979(13) & 1.980(16)  \\ 
    9   &  1.893(13) &  1.915(13) &  1.891(18) &  1.933(12) & 1.934(14)  \\ 
    11  &  1.859(12) &  1.882(12) &  1.861(17) &  1.900(11) & 1.903(13)  \\ 
    13  &  1.833(11) &  1.857(11) &  1.838(16) &  1.875(10) & 1.882(13)  \\ 
    15  &  1.810(11) &  1.836(11) &  1.820(16) &  1.856(10) & 1.866(17)  \\ 
    18  &  \text{---}&  1.810(10) &  1.799(15) &  1.834(10) & 1.844(22)  \\ 
    21  &  \text{---}&  1.784(10) &  1.782(14) &  1.817(10) & 1.841(32)  \\
        \midrule
    z   & \multicolumn{5}{c}{$R_k(z,\theta_1,\theta_2)$}             \\\cmidrule(lr){1-1}\cmidrule(lr){2-6}
    2   &  2.163(24) &  2.185(24) &  2.129(34) &  2.211(22) & 2.205(25)  \\ 
    2.7 &  2.100(22) &  2.122(22) &  2.074(30) &  2.144(20) & 2.140(23)  \\ 
    3   &  2.079(21) &  2.100(21) &  2.054(29) &  2.121(19) & 2.118(22)  \\ 
    3.3 &  2.059(20) &  2.081(20) &  2.037(28) &  2.101(18) & 2.097(22)  \\ 
    4   &  2.022(18) &  2.043(18) &  2.004(25) &  2.061(17) & 2.059(20)  \\ 
    6   &  1.946(15) &  1.967(15) &  1.935(21) &  1.982(14) & 1.980(17)  \\ 
    7   &  1.920(14) &  1.940(14) &  1.911(20) &  1.955(13) & 1.953(16)  \\ 
    9   &  1.879(13) &  1.899(13) &  1.874(18) &  1.914(12) & 1.913(14)  \\ 
    11  &  1.848(12) &  1.869(12) &  1.847(17) &  1.884(11) & 1.886(14)  \\ 
    13  &  1.824(12) &  1.845(12) &  1.826(16) &  1.862(11) & 1.867(13)  \\ 
    15  &  1.802(11) &  1.826(11) &  1.809(16) &  1.844(10) & 1.853(17)  \\ 
    18  &  \text{---}&  1.803(10) &  1.790(15) &  1.824(10) & 1.833(23)  \\ 
    21  &  \text{---}&  1.778(10) &  1.774(14) &  1.808(10) & 1.830(32)  \\
        \bottomrule
        \end{tabular}
        \caption{Observables $R_1$, $R_f$ and $R_k$ for all available values of $z$, 
                 $a/L$ at $(\theta_1,\theta_2)=(0.5,1)$. Continuum limit 
                 (CL) has been taken according to eq.~\eqref{eq:global-CLfit}. Due to
                 have quark spin symmetry one expects 
                 $\lim_{1/z\to 0}R_f(z,\theta_1,\theta_2)=\lim_{1/z\to 0}R_k(z,\theta_1,\theta_2)$}
        \label{tab:R1fk}
\end{table}

%
%
\begin{table}[thb]
        \small
        \centering
        \begin{tabular}{CCCCC>{\it}C}\toprule
   L/a  &      20     &        24     &      32      &     40      & CL   \\
        \midrule
    z   & \multicolumn{5}{c}{$\Ypsps^{(0)}(z,\theta_1,\theta_2)$}                     \\\cmidrule(lr){1-1}\cmidrule(lr){2-6}
     2  &  1.3059(41) &  1.3080(42) &  1.2943(63) &  1.3108(36) &  1.3082(38)  \\
   2.7  &  1.2641(32) &  1.2648(33) &  1.2542(49) &  1.2664(28) &  1.2645(32)  \\
     3  &  1.2501(29) &  1.2504(30) &  1.2407(45) &  1.2517(25) &  1.2499(30)  \\
   3.3  &  1.2379(27) &  1.2379(27) &  1.2290(41) &  1.2388(23) &  1.2371(28)  \\
     4  &  1.2150(22) &  1.2145(23) &  1.2071(34) &  1.2151(19) &  1.2132(24)  \\
     6  &  1.1748(15) &  1.1732(15) &  1.1679(24) &  1.1729(13) &  1.1708(17)  \\
     7  &  1.1626(13) &  1.1606(14) &  1.1559(21) &  1.1601(12) &  1.1578(15)  \\
     9  &  1.1459(10) &  1.1436(11) &  1.1395(17) &  1.1428(10) &  1.1403(13)  \\
    11  &  1.13522(9) &  1.1326(10) &  1.1289(15) &  1.1317(9)  &  1.1295(12)  \\
    13  &  1.12784(8) &  1.1251(9)  &  1.1215(14) &  1.1240(9)  &  1.1221(12)  \\
    15  &  1.12241(8) &  1.1195(9)  &  1.1161(13) &  1.1183(9)  &  1.1175(15)  \\
    18  &  \text{---} &  1.1135(8)  &  1.1102(12) &  1.1122(9)  &  1.1123(19)  \\
    21  &  \text{---} &  1.1086(8)  &  1.1060(12) &  1.1078(9)  &  1.1102(25)  \\
        \midrule
    z   & \multicolumn{5}{c}{$\Yvv^{(0)}(z,\theta_1,\theta_2)$}                       \\\cmidrule(lr){1-1}\cmidrule(lr){2-6}
     2  &  1.2027(14) &  1.2027(13) &  1.1984(20) &  1.2021(13) &  1.2008(15)  \\
   2.7  &  1.1829(11) &  1.1825(11) &  1.1789(17) &  1.1819(11) &  1.1809(14)  \\
     3  &  1.1761(10) &  1.1756(11) &  1.1722(16) &  1.1749(11) &  1.1740(13)  \\
   3.3  &  1.1701(10) &  1.1695(10) &  1.1663(15) &  1.1688(10) &  1.1680(13)  \\
     4  &  1.1562(9)  &  1.1562(9)  &  1.1541(14) &  1.1567(10) &  1.1550(12)  \\
     6  &  1.1349(7)  &  1.1346(8)  &  1.1332(12) &  1.1353(9)  &  1.1341(11)  \\
     7  &  1.1280(7)  &  1.1277(8)  &  1.1264(11) &  1.1285(9)  &  1.1275(10)  \\
     9  &  1.1181(7)  &  1.1178(8)  &  1.1170(11) &  1.1190(9)  &  1.1185(10)  \\
    11  &  1.1112(6)  &  1.1111(8)  &  1.1106(10) &  1.1127(9)  &  1.1127(11)  \\
    13  &  1.1058(6)  &  1.1061(8)  &  1.1059(10) &  1.1081(9)  &  1.1084(11)  \\
    15  &  1.1011(6)  &  1.1020(8)  &  1.1023(10) &  1.1046(9)  &  1.1055(13)  \\
    18  &  \text{---} &  1.0969(8)  &  1.0981(10) &  1.1006(9)  &  1.1023(17)  \\
    21  &  \text{---} &  1.0914(7)  &  1.0947(10) &  1.0976(9)  &  1.1008(22)  \\
        \bottomrule
        \end{tabular}
        \caption{Observables $\Ypsps$ and $\Yvv$ for all available values of $z$, 
                 $a/L$ at $(\theta_1,\theta_2)=(0.5,1)$. Continuum limit (CL) has been 
                 taken according to eq.~\eqref{eq:global-CLfit}. Both observables 
                 are expected to agree in the static limit.}
        \label{tab:decay-ratios}
\end{table}

\bibliography{biblio}

\providecommand{\href}[2]{#2}\begingroup\raggedright\begin{thebibliography}{10}

\bibitem{stat:eichten}
E.~Eichten, {\it Heavy quarks on the lattice},  {\em Nucl. Phys. Proc. Suppl.}
  {\bf 4} (1988) 170.

\bibitem{Eichten:1989zv}
E.~Eichten and B.~R. Hill, {\it {An Effective Field Theory for the Calculation
  of Matrix Elements Involving Heavy Quarks}},  {\em Phys. Lett.} {\bf B234}
  (1990) 511.

\bibitem{Georgi:1990um}
H.~Georgi, {\it {An effective field theory for heavy quarks at low energies}},
  {\em Phys. Lett.} {\bf B240} (1990) 447.

\bibitem{Grinstein:1990mj}
B.~Grinstein, {\it {The static quark effective theory}},  {\em Nucl. Phys.}
  {\bf B339} (1990) 253.

\bibitem{Heitger:2003nj}
{\bf ALPHA} Collaboration, J.~Heitger and R.~Sommer, {\it {Non-perturbative
  heavy quark effective theory}},  {\em JHEP} {\bf 0402} (2004) 022,
  [\href{http://arxiv.org/abs/hep-lat/0310035}{{\tt hep-lat/0310035}}].

\bibitem{Blossier2010}
{\bf ALPHA} Collaboration, B.~Blossier, M.~Della~Morte, N.~Garron, and
  R.~Sommer, {\it {HQET at order $1/m$: I. Non-perturbative parameters in the
  quenched approximation}},  {\em JHEP} {\bf 1006} (2010) 002,
  [\href{http://arxiv.org/abs/1001.4783}{{\tt arXiv:1001.4783}}].

\bibitem{Blossier2010a}
{\bf ALPHA} Collaboration, B.~Blossier et~al., {\it {HQET at order $1/m$: II.
  Spectroscopy in the quenched approximation}},  {\em JHEP} {\bf 1005} (2010)
  074, [\href{http://arxiv.org/abs/1004.2661}{{\tt arXiv:1004.2661}}].

\bibitem{Blossier:2010mk}
{\bf ALPHA} Collaboration, B.~Blossier et~al., {\it {HQET at order 1/m: III.
  Decay constants in the quenched approximation}},  {\em JHEP} {\bf 1012}
  (2010) 039, [\href{http://arxiv.org/abs/1006.5816}{{\tt arXiv:1006.5816}}].

\bibitem{Blossier:2012qu}
{\bf ALPHA} Collaboration, B.~Blossier et~al., {\it {Parameters of Heavy Quark
  Effective Theory from $N_{\rm f}=2$ lattice QCD}},  {\em JHEP} {\bf 1209}
  (2012) 132, [\href{http://arxiv.org/abs/1203.6516}{{\tt arXiv:1203.6516}}].

\bibitem{Bernardoni:2013xba}
{\bf ALPHA} Collaboration, F.~Bernardoni et~al., {\it {The b-quark mass from
  non-perturbative $N_{\rm f}=2$ Heavy Quark Effective Theory at $O(1/m_{\rm
  h})$}},  {\em Phys. Lett.} {\bf B730} (2014) 171,
  [\href{http://arxiv.org/abs/1311.5498}{{\tt arXiv:1311.5498}}].

\bibitem{Bernardoni:2014fva}
{\bf ALPHA} Collaboration, F.~Bernardoni et~al., {\it {Decay constants of
  B-mesons from non-perturbative HQET with two light dynamical quarks}},  {\em
  Phys. Lett.} {\bf B735} (2014) 349,
  [\href{http://arxiv.org/abs/1404.3590}{{\tt arXiv:1404.3590}}].

\bibitem{Bernardoni:2015nqa}
{\bf ALPHA} Collaboration, F.~Bernardoni et~al., {\it {B-meson spectroscopy in
  HQET at order 1/m}},  {\em {\rm accepted by PRD}} (2015)
  [\href{http://arxiv.org/abs/1505.03360}{{\tt arXiv:1505.03360}}].

\bibitem{DellaMorte:2013ega}
{\bf ALPHA} Collaboration, M.~Della~Morte, S.~Dooling, J.~Heitger, D.~Hesse,
  and H.~Simma, {\it {Matching of heavy-light flavour currents between HQET at
  order 1/$m$ and QCD: I. Strategy and tree-level study}},  {\em JHEP} {\bf
  1405} (2014) 060, [\href{http://arxiv.org/abs/1312.1566}{{\tt
  arXiv:1312.1566}}].

\bibitem{Korcyl:2013ega}
P.~Korcyl, {\it {On one-loop corrections to matching conditions of Lattice HQET
  including $1/m_{\rm b}$ terms}},  {\em PoS} {\bf Lattice2013} (2013) 380,
  [\href{http://arxiv.org/abs/1312.2350}{{\tt arXiv:1312.2350}}].

\bibitem{Heitger:2004gb}
{\bf ALPHA} Collaboration, J.~Heitger, A.~J\"uttner, R.~Sommer, and
  J.~Wennekers, {\it {Non-perturbative tests of heavy quark effective theory}},
   {\em JHEP} {\bf 0411} (2004) 048,
  [\href{http://arxiv.org/abs/hep-ph/0407227}{{\tt hep-ph/0407227}}].

\bibitem{Hesse:2012hb}
{\bf ALPHA} Collaboration, D.~Hesse and R.~Sommer, {\it {A one-loop study of
  matching conditions for static-light flavor currents}},  {\em JHEP} {\bf
  1302} (2013) 115, [\href{http://arxiv.org/abs/1211.0866}{{\tt
  arXiv:1211.0866}}].

\bibitem{DellaMorte:2007qw}
{\bf ALPHA} Collaboration, M.~Della~Morte et~al., {\it {Towards a
  non-perturbative matching of HQET and QCD with dynamical light quarks}},
  {\em PoS} {\bf LAT2007} (2007) 246,
  [\href{http://arxiv.org/abs/0710.1188}{{\tt arXiv:0710.1188}}].

\bibitem{thesis:patrickf}
P.~Fritzsch, {\em {B-meson properties from non-perturbative matching of HQET to
  finite-volume two-flavour QCD}}.
\newblock PhD thesis, {\rm {Westf\"alische Wilhelms-Universit\"at M\"unster}},
  2009.

\bibitem{Fritzsch:2010aw}
{\bf ALPHA} Collaboration, P.~Fritzsch, J.~Heitger, and N.~Tantalo, {\it
  {Non-perturbative improvement of quark mass renormalization in two-flavour
  lattice QCD}},  {\em JHEP} {\bf 1008} (2010) 074,
  [\href{http://arxiv.org/abs/1004.3978}{{\tt arXiv:1004.3978}}].

\bibitem{Fritzsch:2012wq}
{\bf ALPHA} Collaboration, P.~Fritzsch et~al., {\it {The strange quark mass and
  Lambda parameter of two flavor QCD}},  {\em Nucl. Phys.} {\bf B865} (2012)
  397, [\href{http://arxiv.org/abs/1205.5380}{{\tt arXiv:1205.5380}}].

\bibitem{Luscher:1992an}
M.~L{\"u}scher, R.~Narayanan, P.~Weisz, and U.~Wolff, {\it {The Schr{\"o}dinger
  functional: A renormalizable probe for non-Abelian gauge theories}},  {\em
  Nucl. Phys.} {\bf B384} (1992) 168,
  [\href{http://arxiv.org/abs/hep-lat/9207009}{{\tt hep-lat/9207009}}].

\bibitem{Sint:1993un}
S.~Sint, {\it {On the Schr{\"o}dinger functional in QCD}},  {\em Nucl. Phys.}
  {\bf B421} (1994) 135, [\href{http://arxiv.org/abs/hep-lat/9312079}{{\tt
  hep-lat/9312079}}].

\bibitem{DellaMorte:2005se}
{\bf ALPHA} Collaboration, M.~Della~Morte, R.~Hoffmann, and R.~Sommer, {\it
  {Non-perturbative improvement of the axial current for dynamical Wilson
  fermions}},  {\em JHEP} {\bf 0503} (2005) 029,
  [\href{http://arxiv.org/abs/hep-lat/0503003}{{\tt hep-lat/0503003}}].

\bibitem{Sint:1997jx}
S.~Sint and P.~Weisz, {\it {Further results on O(a) improved lattice QCD to
  one-loop order of perturbation theory}},  {\em Nucl. Phys.} {\bf B502} (1997)
  251, [\href{http://arxiv.org/abs/hep-lat/9704001}{{\tt hep-lat/9704001}}].

\bibitem{DellaMorte:2008xb}
{\bf ALPHA} Collaboration, M.~Della~Morte, R.~Sommer, and S.~Takeda, {\it {On
  cutoff effects in lattice QCD from short to long distances}},  {\em Phys.
  Lett.} {\bf B672} (2009) 407, [\href{http://arxiv.org/abs/0807.1120}{{\tt
  arXiv:0807.1120}}].

\bibitem{DellaMorte:2005yc}
{\bf ALPHA} Collaboration, M.~Della~Morte, A.~Shindler, and R.~Sommer, {\it {On
  lattice actions for static quarks}},  {\em JHEP} {\bf 0508} (2005) 051,
  [\href{http://arxiv.org/abs/hep-lat/0506008}{{\tt hep-lat/0506008}}].

\bibitem{DellaMorte:2006sv}
{\bf ALPHA} Collaboration, M.~Della~Morte, P.~Fritzsch, and J.~Heitger, {\it
  {Non-perturbative renormalization of the static axial current in two-flavour
  QCD}},  {\em JHEP} {\bf 0702} (2007) 079,
  [\href{http://arxiv.org/abs/hep-lat/0611036}{{\tt hep-lat/0611036}}].

\bibitem{Kurth:2001yr}
{\bf ALPHA} Collaboration, M.~Kurth and R.~Sommer, {\it {Heavy quark effective
  theory at one-loop order: An explicit example}},  {\em Nucl. Phys.} {\bf
  B623} (2002) 271, [\href{http://arxiv.org/abs/hep-lat/0108018}{{\tt
  hep-lat/0108018}}].

\bibitem{Sommer:2010ic}
R.~Sommer, {\it {Introduction to Non-perturbative Heavy Quark Effective
  Theory}},  {\em {\rm published in} {\it Les Houches Summer School in
  Theoretical Physics, Modern perspectives in lattice QCD, Les Houches, France,
  3 -- 28 August 2009}, Oxford Univ. Pr.} (2010) 517,
  [\href{http://arxiv.org/abs/1008.0710}{{\tt arXiv:1008.0710}}].

\bibitem{Heitger:2013oaa}
{\bf ALPHA} Collaboration, J.~Heitger, G.~M. von Hippel, S.~Schaefer, and
  F.~Virotta, {\it {Charm quark mass and D-meson decay constants from
  two-flavour lattice QCD}},  {\em PoS} {\bf LATTICE2013} (2013) 475,
  [\href{http://arxiv.org/abs/1312.7693}{{\tt arXiv:1312.7693}}].

\bibitem{Bekavac2010}
S.~Bekavac et~al., {\it {Matching QCD and HQET heavy-light currents at three
  loops}},  {\em Nucl. Phys.} {\bf B833} (2010) 46,
  [\href{http://arxiv.org/abs/0911.3356}{{\tt arXiv:0911.3356}}].

\bibitem{Heitger:2003xg}
{\bf ALPHA} Collaboration, J.~Heitger, M.~Kurth, and R.~Sommer, {\it
  {Non-perturbative renormalization of the static axial current in quenched
  QCD}},  {\em Nucl. Phys.} {\bf B669} (2003) 173,
  [\href{http://arxiv.org/abs/hep-lat/0302019}{{\tt hep-lat/0302019}}].

\bibitem{DellaMorte:2006cb}
{\bf ALPHA} Collaboration, M.~Della~Morte, N.~Garron, M.~Papinutto, and
  R.~Sommer, {\it {Heavy quark effective theory computation of the mass of the
  bottom quark}},  {\em JHEP} {\bf 0701} (2007) 007,
  [\href{http://arxiv.org/abs/hep-ph/0609294}{{\tt hep-ph/0609294}}].

\bibitem{Kurth:2000ki}
{\bf ALPHA} Collaboration, M.~Kurth and R.~Sommer, {\it {Renormalization and
  O(a) improvement of the static axial current}},  {\em Nucl. Phys.} {\bf B597}
  (2001) 488, [\href{http://arxiv.org/abs/hep-lat/0007002}{{\tt
  hep-lat/0007002}}].

\bibitem{DellaMorte:2005kg}
{\bf ALPHA} Collaboration, M.~Della~Morte, R.~Hoffmann, F.~Knechtli, J.~Rolf,
  R.~Sommer, I.~Wetzorke, and U.~Wolff, {\it {Non-perturbative quark mass
  renormalization in two-flavor QCD}},  {\em Nucl. Phys.} {\bf B729} (2005)
  117, [\href{http://arxiv.org/abs/hep-lat/0507035}{{\tt hep-lat/0507035}}].

\bibitem{Tarrach:1980up}
R.~Tarrach, {\it {The pole mass in perturbative QCD}},  {\em Nucl. Phys.} {\bf
  B183} (1981) 384.

\bibitem{Shifman:1987sm}
M.~A. Shifman and M.~B. Voloshin, {\it On annihilation of mesons built from
  heavy and light quark and $\bar{B}_0 \leftrightarrow {B}_0$ oscillations},
  {\em Sov. J. Nucl. Phys.} {\bf 45} (1987) 292.

\bibitem{Politzer:1988wp}
H.~D. Politzer and M.~B. Wise, {\it {Leading Logarithms of Heavy Quark Masses
  in Processes with Light and Heavy Quarks}},  {\em Phys. Lett.} {\bf B206}
  (1988) 681.

\bibitem{Gray:1990yh}
N.~Gray, D.~J. Broadhurst, W.~Grafe, and K.~Schilcher, {\it {Three-loop
  relation of quark $\overline{MS}$ and pole masses}},  {\em Z. Phys.} {\bf
  C48} (1990) 673.

\bibitem{Eichten:1990vp}
E.~Eichten and B.~R. Hill, {\it {Static effective field theory: $1/m$
  corrections}},  {\em Phys. Lett.} {\bf B243} (1990) 427.

\bibitem{Falk:1991pz}
A.~F. Falk, B.~Grinstein, and M.~E. Luke, {\it Leading mass corrections to the
  heavy quark effective theory},  {\em Nucl. Phys.} {\bf B357} (1991) 185.

\bibitem{Ji:1991pr}
X.-D. Ji and M.~Musolf, {\it {Subleading logarithmic mass dependence in heavy
  meson form-factors}},  {\em Phys. Lett.} {\bf B257} (1991) 409.

\bibitem{Broadhurst:1991fz}
D.~J. Broadhurst and A.~Grozin, {\it {Two-loop renormalization of the effective
  field theory of a static quark}},  {\em Phys. Lett.} {\bf B267} (1991) 105,
  [\href{http://arxiv.org/abs/hep-ph/9908362}{{\tt hep-ph/9908362}}].

\bibitem{Gimenez:1991bf}
V.~Gim\'enez, {\it {Two-loop calculation of the anomalous dimension of the
  axial current with static heavy quarks}},  {\em Nucl. Phys.} {\bf B375}
  (1992) 582.

\bibitem{HQET:sigmabI}
G.~Amor\'os, M.~Beneke, and M.~Neubert, {\it Two-loop anomalous dimension of
  the chromomagnetic moment of a heavy quark},  {\em Phys. Lett.} {\bf B401}
  (1997) 81, [\href{http://arxiv.org/abs/hep-ph/9701375}{{\tt
  hep-ph/9701375}}].

\bibitem{HQET:sigmabII}
A.~Czarnecki and A.~G. Grozin, {\it {HQET} chromomagnetic interaction at two
  loops},  {\em Phys. Lett.} {\bf B405} (1997) 142,
  [\href{http://arxiv.org/abs/hep-ph/9701415}{{\tt hep-ph/9701415}}].

\bibitem{Chetyrkin:2003vi}
K.~Chetyrkin and A.~Grozin, {\it {Three-loop anomalous dimension of the heavy
  light quark current in HQET}},  {\em Nucl. Phys.} {\bf B666} (2003) 289,
  [\href{http://arxiv.org/abs/hep-ph/0303113}{{\tt hep-ph/0303113}}].

\bibitem{Grozin:2007fh}
A.~Grozin, P.~Marquard, J.~Piclum, and M.~Steinhauser, {\it {Three-Loop
  Chromomagnetic Interaction in HQET}},  {\em Nucl. Phys.} {\bf B789} (2008)
  277, [\href{http://arxiv.org/abs/0707.1388}{{\tt arXiv:0707.1388}}].

\bibitem{Eichten:1989kb}
E.~Eichten and B.~R. Hill, {\it {Renormalization of heavy-light bilinears and
  $f_{\rm B}$ for Wilson fermions}},  {\em Phys. Lett.} {\bf B240} (1990) 193.

\bibitem{Broadhurst:1994se}
D.~J. Broadhurst and A.~Grozin, {\it {Matching QCD and HQET heavy-light
  currents at two loops and beyond}},  {\em Phys. Rev.} {\bf D52} (1995) 4082,
  [\href{http://arxiv.org/abs/hep-ph/9410240}{{\tt hep-ph/9410240}}].

\bibitem{Grozin:1998kf}
A.~Grozin, {\it {Decoupling of heavy quark loops in light-light and heavy-light
  quark currents}},  {\em Phys. Lett.} {\bf B445} (1998) 165,
  [\href{http://arxiv.org/abs/hep-ph/9810358}{{\tt hep-ph/9810358}}].

\bibitem{'tHooft:1973mm}
G.~'t~Hooft, {\it {Dimensional regularization and the renormalization group}},
  {\em Nucl. Phys.} {\bf B61} (1973) 455.

\bibitem{Gross:1973id}
D.~J. Gross and F.~Wilczek, {\it {Ultraviolet Behavior of Nonabelian Gauge
  Theories}},  {\em Phys. Rev. Lett.} {\bf 30} (1973) 1343.

\bibitem{Caswell:1974gg}
W.~E. Caswell, {\it {Asymptotic Behavior of Nonabelian Gauge Theories to
  Two-Loop Order}},  {\em Phys. Rev. Lett.} {\bf 33} (1974) 244.

\bibitem{Tarasov:1980au}
O.~Tarasov, A.~Vladimirov, and A.~Y. Zharkov, {\it {The Gell-Mann-Low Function
  of QCD in the Three-Loop Approximation}},  {\em Phys. Lett.} {\bf B93} (1980)
  429.

\bibitem{vanRitbergen:1997va}
T.~van Ritbergen, J.~Vermaseren, and S.~Larin, {\it {The four-loop
  beta-function in quantum chromodynamics}},  {\em Phys. Lett.} {\bf B400}
  (1997) 379, [\href{http://arxiv.org/abs/hep-ph/9701390}{{\tt
  hep-ph/9701390}}].

\bibitem{Czakon:2004bu}
M.~Czakon, {\it {The four-loop QCD beta-function and anomalous dimensions}},
  {\em Nucl. Phys.} {\bf B710} (2005) 485,
  [\href{http://arxiv.org/abs/hep-ph/0411261}{{\tt hep-ph/0411261}}].

\bibitem{Chetyrkin:1997dh}
K.~Chetyrkin, {\it {Quark mass anomalous dimension to $O(\alpha_s^4)$}},  {\em
  Phys. Lett.} {\bf B404} (1997) 161,
  [\href{http://arxiv.org/abs/hep-ph/9703278}{{\tt hep-ph/9703278}}].

\bibitem{Vermaseren:1997fq}
J.~Vermaseren, S.~Larin, and T.~van Ritbergen, {\it {The four-loop quark mass
  anomalous dimension and the invariant quark mass}},  {\em Phys. Lett.} {\bf
  B405} (1997) 327, [\href{http://arxiv.org/abs/hep-ph/9703284}{{\tt
  hep-ph/9703284}}].

\end{thebibliography}\endgroup
\bibliographystyle{JHEP}
\end{document}